\DocumentMetadata{} 
\pdfoutput=1
\documentclass[a4paper,11pt,parskip=half-,abstract,bibliography=totocnumbered]{scrartcl}
\usepackage{geometry}
\usepackage[para]{threeparttable}
\usepackage{rotating} 
\usepackage{setspace} 
\usepackage{graphics}
\usepackage{amssymb,amsfonts,amsmath,amsbsy}

\usepackage{enumitem}

\usepackage{url}
\usepackage{nameref}
\usepackage{array}
\usepackage{graphicx}
\usepackage{epstopdf}

\renewcaptionname{english}{\refname}{Bibliography} 
\usepackage{natbib}

\makeatletter
\newcommand\bibstyle@comma{\bibpunct(),a,,}
\newcommand\bibstyle@semicolon{\bibpunct();a,,}
\makeatother
\usepackage{etoolbox}
\pretocmd\citet{\citestyle{comma}}\relax\relax
\pretocmd\Citet{\citestyle{comma}}\relax\relax
\pretocmd\citep{\citestyle{semicolon}}\relax\relax
\pretocmd\Citep{\citestyle{semicolon}}\relax\relax
\pretocmd\citealp{\citestyle{semicolon}}\relax\relax
\pretocmd\Citealp{\citestyle{semicolon}}\relax\relax

\usepackage{nameref}
\usepackage{hyperref}
\hypersetup{
  colorlinks = true,
  citecolor=  black,
  linkcolor = blue,
  filecolor = cyan 
}
\usepackage{xurl} 

\newcommand*{\qpref}[1]{\hyperref[{#1}]{\textit{``\nameref*{#1}'', section \ref*{#1}}}}
\newcommand*{\qref}[1]{\hyperref[{#1}]{\textit{``\nameref*{#1}'' (section \ref*{#1})}}}

\newcommand*{\doi}{}
\makeatletter
\newcommand{\doi@}[1]{\href{https://doi.org/#1}{\textcolor{black}{DOI: } https://doi.org/#1}}
\DeclareRobustCommand{\doi}{\hyper@normalise\doi@}
\makeatother

\usepackage[dvipsnames,table]{xcolor}

\newcommand{\blu}[1]{{\textcolor {blue} {#1}}}
\newcommand{\Burl}[1]{\blu{\url{#1}}}

\usepackage{pdflscape} 
\usepackage{authblk} 
\usepackage[iso,english]{isodate}

\usepackage[english]{babel}

\usepackage[utf8]{inputenc} 

\usepackage[]{newtx}

\usepackage[export]{adjustbox}

\usepackage{booktabs,tabularx}
\newcolumntype{i}{>{\footnotesize\centering\arraybackslash\hsize=1.0\hsize}X} 
\newcolumntype{n}{>{\footnotesize\centering\arraybackslash\hsize=0.3\hsize}X} 
\newcolumntype{m}{>{\footnotesize\centering\arraybackslash\hsize=.95\hsize}X} 
\newcolumntype{q}{>{\footnotesize\centering\arraybackslash\hsize=0.30\hsize}X} 
\newcolumntype{w}{>{\footnotesize\centering\arraybackslash\hsize=1.00\hsize}X} 
\newcolumntype{s}{>{\footnotesize\centering\arraybackslash\hsize=1.00\hsize}X} 
\newcolumntype{t}{>{\footnotesize\centering\arraybackslash\hsize=0.45\hsize}X} 
\newcolumntype{u}{>{\footnotesize\centering\arraybackslash\hsize=0.50\hsize}X} 
\newcolumntype{v}{>{\footnotesize\centering\arraybackslash\hsize=0.50\hsize}X}
\newcolumntype{e}{>{\footnotesize\centering\arraybackslash\hsize=3.3\hsize}X}

\date{2024-06-30}
\title{\textbf{A picture guide to cancer progression and monotonic accumulation models: evolutionary assumptions, plausible interpretations, and alternative uses}}

\author[1,*]{Ramon Diaz-Uriarte}
\author[2]{Iain G. Johnston}
\affil[1]{Department of Biochemistry, School of Medicine, Universidad Autónoma de Madrid, and Instituto de Investigaciones Biomédicas Sols-Morreale (IIBM), CSIC-UAM,  Madrid, Spain.}
\affil[2]{Computational Biology Unit and Department of Mathematics, University of Bergen, Norway.}
\affil[*]{Author for correspondence: \texttt{r.diaz@uam.es}}
\ExplSyntaxOn
\AddToHook{env/sidewaystable/end}{\pdfmanagement_add:nnn{ThisPage}{Rotate}{90}}
\ExplSyntaxOff

\begin{document}

\maketitle

\tableofcontents

\begin{abstract}
  Cancer progression and monotonic accumulation models were developed to discover dependencies in the irreversible acquisition of binary traits from cross-sectional data. They have been used in computational oncology and virology but also in widely different problems such as malaria progression. These methods have been applied to predict future states of the system, identify routes of feature acquisition, and improve patient stratification, and they hold promise for evolutionary-based treatments. New methods continue to be developed.

  But these methods have shortcomings, which are yet to be systematically critiqued, regarding key evolutionary assumptions and interpretations. After an overview of the available methods, we focus on why inferences might not be about the processes we intend. Using fitness landscapes, we  highlight difficulties that arise from bulk sequencing and reciprocal sign epistasis, from conflating lines of descent, path of the maximum, and mutational profiles, and from ambiguous use of the idea of exclusivity. We examine how the previous concerns change when bulk sequencing is explicitly considered, and underline opportunities for addressing dependencies due to frequency-dependent selection. This review identifies major standing issues, and should encourage the use of these methods in other areas with a better alignment between entities and model assumptions.

\end{abstract}

\section{Introduction}
\label{sec:introduction}

\begin{figure}[tb!]
 \centering \includegraphics[width=11.0cm,keepaspectratio]{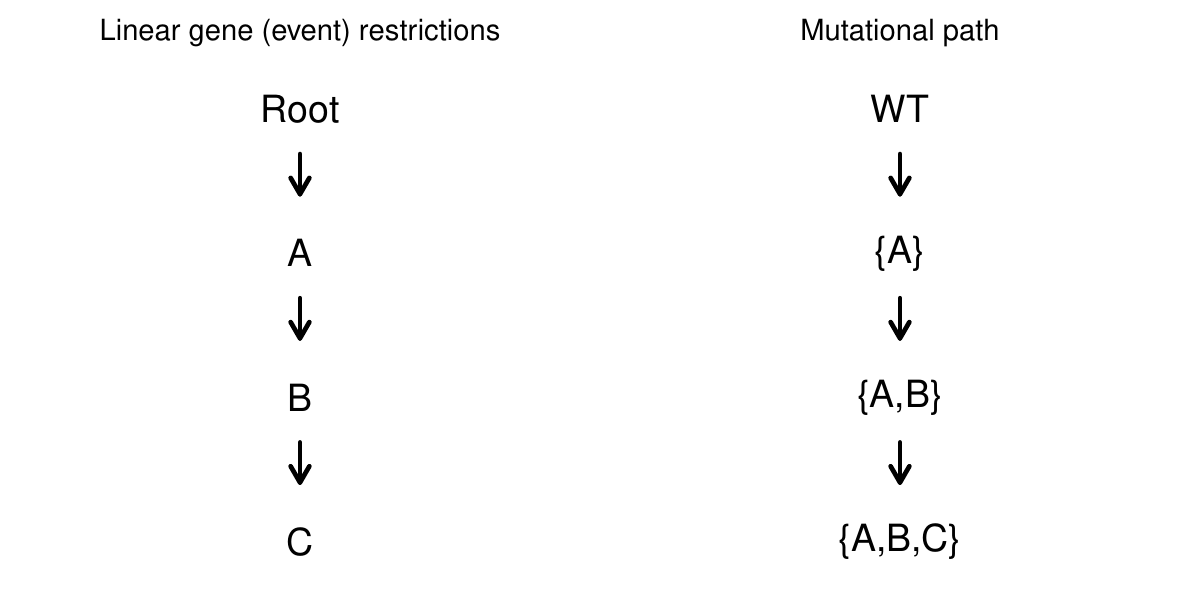}
 \caption{\textbf{Cancer progression model with linear dependencies}. Left: example of a linear model of  dependencies in the acquisition of mutations; gaining a mutation in gene B requires having gained a mutation in A; gaining a mutation in gene C requires having gained a mutation in B (and, thus, A). The only possible tumour states or mutational profiles (or genotypes, if the model applies to within-cell restrictions) under that model are WT (no mutations), \{A\} (only A mutated), \{A,B\} (both A and B mutated), and \{A,B,C\}. Right: mutational path which depicts the possible transitions between the possible mutational profiles (or genotypes, if the model applies to within-cell restrictions).  If our data consisted only of mutational profiles \textit{WT}, \{A\}, \{A,B\}, and \{A,B,C\}, the model on the left would fit this data; in contrast, if our observed data consisted of \textit{WT}, \{B\}, \{A,B\}, and \{A,B,C\}, the best fitting model would have \textit{A} depend on \textit{B}. Note that any of the deterministic models discussed below (section \ref{sec:deterministic-deps}) could fit these linear patterns perfectly. For notation for genes, events, paths, and genotypes see \qref{sec:notation-genots}. }
  \label{fig:linear-deps}
\end{figure}

Cancer progression models (CPMs) were first developed by \citet{Desper1999JCB} to try to understand restrictions in the order of accumulation of events, such as accumulation of chromosomal aberrations, during tumour progression.  Descriptions of cancer progression typically emphasise mutation accumulation over time \citep[e.g.,][]{Hanahan2011,Merlo2006,Weinstein2008}, but not all orders of mutation accumulation need to be equally likely. A mutation in a given gene (say, gene B) might only provide a growth advantage to the cells that harbour it if those cells already have a mutation in another gene, say gene A; this could happen if a mutation in B is deleterious unless A is already mutated (an example of sign epistasis:  \citealp{crona_peaks_2013,weinreich_perspective_2005-1}). A simple model is a linear accumulation one: mutation A comes first, which leads to an initial growth of the tumour, then B is mutated, which depends on A being already present, then a third mutation is acquired, which depends on the previous one, and so on. Under this linear accumulation model, there is a single possible trajectory of mutation accumulation (Figure \ref{fig:linear-deps}).

A linear accumulation model would thus try to find the linear sequence of events that best explains the observed data. These models have been developed further \citep{Hainke2012,Beerenwinkel2015}. The linear accumulation model is a deterministic model that assumes that one event is always directly required for the accumulation of another event; richer deterministic models  (section \ref{sec:deterministic-deps}) allow more complex dependencies (e.g., an event can depend on the occurrence of one among several possible previous events). Stochastic dependencies models (section \ref{sec:stoch-deps}) assume that events have inhibiting or enhancing effects on the probability of acquisition of other events. These models have also been used to address questions in other fields, in particular HIV mutation accumulation, including antiretroviral resistance acquisition \citep{beerenwinkel_evolution_2006, Beerenwinkel2007, montazeri_estimating_2015,posada-cespedes2021}, animal tool use \citep{johnston2020a}, malaria progression in children \citep{johnston2019},  antimicrobial resistance in tuberculosis \citep{greenbury2020}, and gene loss in mitochondria \citep{johnston2016}, and are thus probably better described as ``monotonic accumulation models'' \citep{ramazzotti2019a}. The key idea that has prompted this expansion in expressive power and fields of application is that many phenomena involve the irreversible accumulation (or loss) of certain features or events, and that different events are not independent but can affect the acquisition (or loss) of other events; monotonic accumulation models try to infer the dependencies that give rise to the patterns we observe in the data.

\begin{figure}[tbhp]
  \centering \includegraphics[width=10.5cm,keepaspectratio]{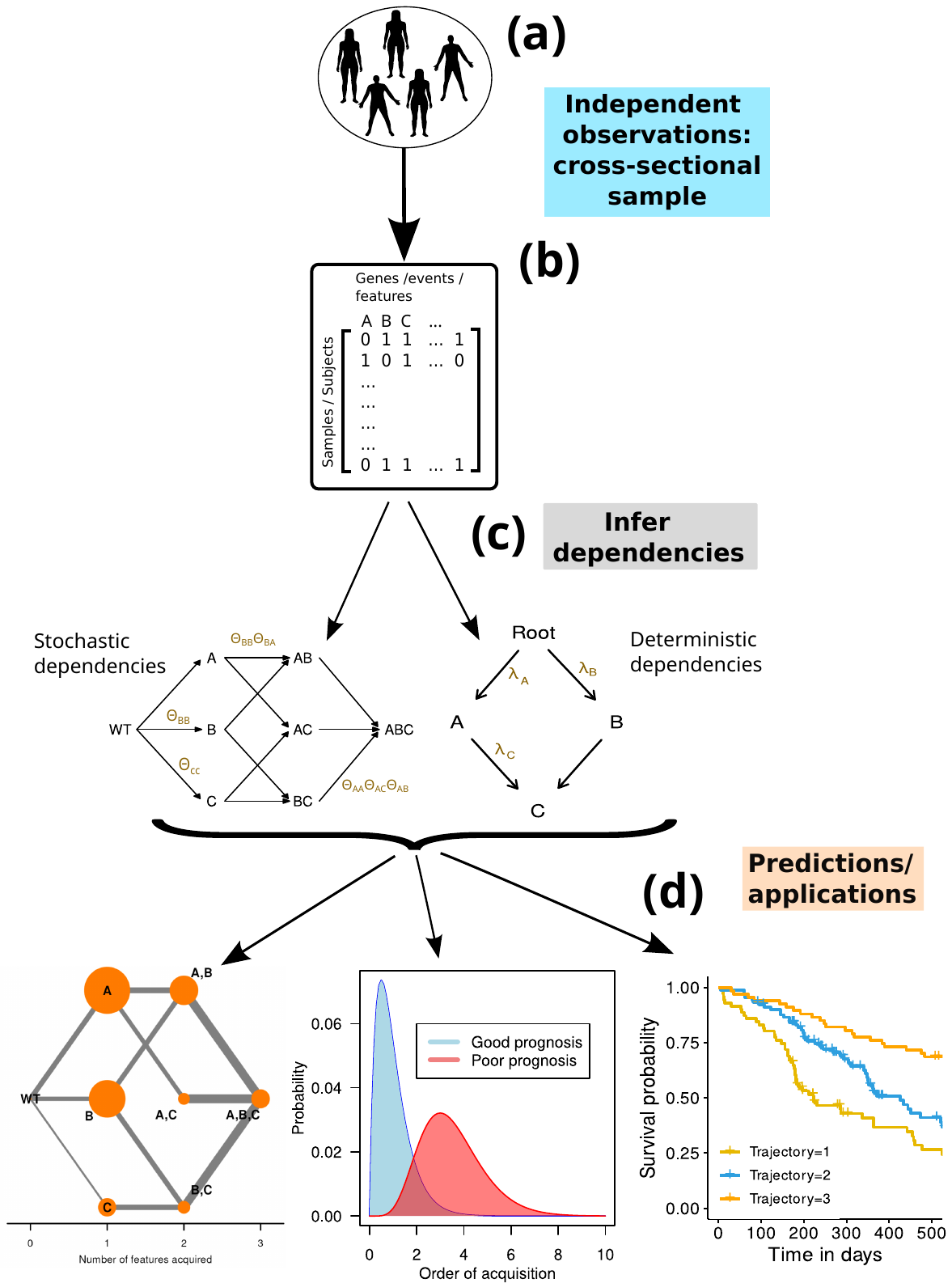}

 \caption{\textbf{Cancer progression and monotonic accumulation models: main steps}.
   (a) Features of relevance (e.g., mutations in genes or presence of malaria symptoms) are measured on some subjects (e.g., patients).
   (b) Data are arranged as a binary matrix of samples by features. For cancer data, each row in (b) is a mutational profile: what mutations or genetic alterations were present in a given subject.
   Here we only focus on data where all samples are assumed independent and each sample provides a single observation: cross-sectional data. This is the   standard scenario for the application of cancer progression models.
     (c) The researcher uses a method to infer the dependencies in the order of accumulation of the events (or mutations) from the data in (b). Some methods (left) model  stochastic dependencies for the transitions between the combinations of events, other methods (right), model deterministic relationships encoded as graphs; in both figures, letters denote events or gene alterations (the column names of the matrix in (b)).
   Each method will find the parameters of the model (interaction and spontaneous rates or transition probabilities for models with stochastic dependencies; trees/graphs, type of dependencies, and rates/conditional probabilities for models with deterministic dependencies) that allow that model to provide the closest match between predicted and observed frequencies of mutational profiles (the data in (b)). As an example, in brown, some of the edges annotated with possible parameters inferred.
   (d) Some uses of cancer progression models. Left: predicting the next genotype and predicting the paths of progression of the disease; edge weights give predicted probabilities.  Centre: distribution of ordering of some feature (e.g., a mutation in a gene) between patients with different prognosis, facilitating stratification. Right: patients are stratified based on evolutionary trajectory, and the survival of the different groups is compared.}
 \label{fig:cpm_cartoon}
\end{figure}

A cartoon depiction of the main steps in the use of cancer progression and monotonic accumulation models is shown in Fig.~\ref{fig:cpm_cartoon}. We are interested in how some features or events that characterise a process accumulate (or are lost) irreversibly. For example, how mutations in driver genes accumulate during cancer progression \citep{Beerenwinkel2015}, or how symptoms (fever, vomiting, anemia, \ldots) accumulate during severe malaria \citep{johnston2019}. Understanding the sequence of events can help identify diagnostic targets or stratify patients for differential treatments.

As shown in Fig.~\ref{fig:cpm_cartoon}(a) and (b), we start by measuring the presence or absence of the events of interest (e.g., mutations in driver genes, malaria symptoms) in a collection of subjects, and arrange the data as a binary matrix of samples by features; for cancer data, each row in Fig.~\ref{fig:cpm_cartoon}(b) would be a mutational profile, whereas for the malaria example each row would be the set of symptoms present in a malaria patient. Most of the cancer progression models assume that each subject or sample provides a single data point and that different subjects are independent: these are cross-sectional data, and this is the standard use-case for these models we will focus on here. But, as we will mention later, some methods for cross-sectional data can also handle phylogenetic (and longitudinal dependencies) in the data \citep{johnston2020a,greenbury2020,peach2021,moen2023};
as well, some cancer progression models have been extended to analyse data sets with multiple observations per subject, where each patient contributes multiple samples, from different tissues or sampling times  \citep{luo2023, caravagna2018}; finally, some models allow missing and uncertain observations \citep{aga2024}.

We then use a specific method to  try to infer the dependencies in the order of accumulation of events that best explains the observed data in Fig.~\ref{fig:cpm_cartoon}(b).  As mentioned above, there is a variety of methods, that we schematise in two groups in Fig.~\ref{fig:cpm_cartoon}(c), as deterministic (section \ref{sec:deterministic-deps}) and stochastic (section \ref{sec:stoch-deps}) dependencies; briefly, models with deterministic dependencies, that we already saw in Fig.~\ref{fig:linear-deps}, assume that some events are always, necessarily required for the accumulation of other events, whereas models with stochastic dependencies assume that events have inhibiting or enhancing effects on the probability of acquisition of other events.

Once models are fitted, they can be used to obtain additional predictions or improved  statistical analyses. The results from the models can be used to make predictions about the next state, conditional on the current observed state, as on the left of side of Fig.~\ref{fig:cpm_cartoon}(d); for example, \citet{johnston2020a} for predictions about future tool use in animals, \citet{diaz-colunga2021} for predicting the next genotype conditional on the currently observed genotype, and \citet{hosseini2019a} and \citet{diaz-uriarte2019a} for estimating the predictability of cancer evolution. These models have also been used  for the closely related problem of  predicting unobserved features in the data set \citep{johnston2020a, johnston2019}. They can also be used to examine how the ordering of some features differs between different groups of subjects;  the centre figure in  Fig.~\ref{fig:cpm_cartoon}(d) is similar to those in Fig.~2b in \citet{johnston2019} where the posterior distribution of the ordering of a feature is compared between patients that eventually died or survived severe malaria; similar analyses have been carried out to show differences in the ordering of tool use acquisition in terrestrial vs.\ aquatic animals \citep{johnston2020a}. Patients can also be stratified according to evolutionary trajectory and ``evolutionary time'', and the survival of the different groups compared \citep{angaroni2021,fontana2023}, as in the rightmost figure in Fig.~\ref{fig:cpm_cartoon}(d).  In fact, denoising patient data by inferring the most probable hidden genotype has been shown to improve survival analysis of cancer data \citep{Gerstung2009}. It is also likely that these models can be used to help identify therapeutic targets and improve evolutionary-based adaptive treatment approaches \citep{fontaneda2023,diaz-uriarte-rios-arroyo-interv}. Moreover, the rich variety of  available models further increases their utility: models with markedly different dependency structures (deterministic restrictions with different operators, stochastic inhibiting/facilitating interactions, direct modelling of state transitions) can be fit and compared on the same data \citep{aga2024,greenbury2020}, which opens the door to a nuanced understanding of dependencies and restrictions in these dynamical systems.

These tools have led to substantial new insights and are valuable ways of reasoning about dependencies in these dynamical processes.  But these methods also have limitations with respect to key evolutionary assumptions and interpretations, and how these are affected by sampling and underlying evolutionary processes, that have not been systematically explored and critiqued. The first objective of this manuscript is to examine  these assumptions, using a generous number of figures  (thus the ``picture guide'' in the title), and the consequences of different scenarios for the interpretations of the methods' output.

After explaining their main assumptions and purpose, we provide an overview of the available models and methods that can be used to analyse cross-sectional data; much of the discussion in the central part of the ms.\ is also relevant to closely related methods, which are briefly mentioned (section \ref{sec:other-methods}). We next highlight the need to be explicit about what empirical entities  are being modelled. The main sections of the paper, \ref{sec:violat} and \ref{sec:what-if-bulk-seq}, examine how and why different methods can fail, and suggest that the limitations and opportunities from bulk sequencing be recognised and exploited explicitly. We finish with a brief discussion about model evaluation, using the data at hand and experimental evolution data, and the exploratory options that using off-the-shelf modelling frameworks for simulation-based inference could offer to monotonic accumulation models.

This overview highlights that we endow certain parameters with a specific interpretation because we make certain assumptions, and how justified we are in making those inferences can be assessed by meticulously examining the different scenarios considered. This manuscript should also promote further theoretical work, and we indicate several standing problems, as well as new research areas; and because some of the problems discussed are related to bulk sequencing, this manuscript should prove useful as more single-cell data become available. Finally, the importance of clarifying the entities being modelled, adjusted to the purpose of the modelling, should also encourage the use of these methods in  other domains with a possibly better fit between entities and model assumptions.

\section{Method overview}\label{meth-overview}

We provide here a quick overview of the existing available methods (see also Table \ref{tab:table-methods}), ordered approximately by increasing complexity of dependencies that can be reflected. We limit ourselves to methods with an existing, working, free software implementation and without dependencies on proprietary software, and focus on methods that can analyse cross-sectional data (closely related methods that can be applied to other types of data are briefly mentioned in section \ref{sec:other-methods}). Before describing each method, we summarise assumptions, type of data, and general purpose that are common to all the methods and models discussed.

\begin{sidewaystable}
  \hspace{-0.5cm}
  \vspace{-10cm}
  \begin{minipage}[t][\dimexpr \textheight + 3cm]{\textwidth}
    \begin{threeparttable}
   \KOMAoption{fontsize}{10pt}
    \renewcommand{\arraystretch}{1.25}
   \rowcolors{2}{white}{gray!25}
   \begin{tabular}{
     p{1.51cm} 
     p{1.6cm} 
     p{2.6cm} 
     p{3.7cm} 
     p{2.5cm} 
     p{1.9cm} 
     p{4cm} 
     p{1.5cm} 
     p{3.5cm} 
     }
     \hline

     Method
     & Timed/
       untimed
     & Type of dependencies
     & Complexity of dependencies
     & Type of input data
     & Number of features\tnote{a}
     & Representation and output\tnote{b}
     & Error model
     & References
     \\   \hline

     OT & Untimed\tnote{c} & Deterministic & Single (each event depends on only one event) & Cross-sectional & $> 2000$ & Tree with edge weights ($\pi$s) & Yes\tnote{g\_OT} & \citet{Desper1999JCB, Szabo2008}\\

     OncoBN (DBN) & Untimed\tnote{c} & Deterministic & Multiple dependencies; OR (DBN version), AND (CBN version)  dependencies. A model can only contain either AND xor OR, not both  & Cross-sectional & $> 100$ & DAG with event thetas ($\theta$s) & Yes\tnote{g\_OBN} & \citet{nicol2021}\\

     CBN & Continuous time  & Deterministic & Multiple dependencies: AND & Cross-sectional & $< 15$; $< 1000$ H/MC-CBN & DAG with event rates ($\lambda$s) & Yes\tnote{g\_CBN}& \citet{Gerstung2009, montazeri_large-scale_2016}\\

     H-ESBCN (PMCE) & Continuous time & Deterministic & Multiple dependencies: AND, OR, XOR  & Cross-sectional & $< 14$ & DAG with event rates ($\lambda$s) & Yes\tnote{g\_HES} & \citet{angaroni2021} \\

     MHN & Continuous time  & Stochastic dependencies: promoting and inhibiting & Pairwise interactions between events & Cross-sectional & $< 25$\tnote{e} & $\Theta$ matrix with baseline hazards and multiplicative effects & None & \citet{schill2020, schill2024a} \\

     HyperTraPS /-CT & Discrete and continuous time & Stochastic dependencies: promoting and inhibiting & Arbitrarily complex\tnote{d} & Cross-sectional, phylogenetically  (and longitudinally) related & $> 120$\tnote{f} & Transition probabilities, average trait ordering, pairs of transition orders 
     & None & \citet{johnston2016,greenbury2020,aga2024} \\

     HyperHMM & Discrete time\tnote{c} & Stochastic dependencies: promoting and inhibiting & Arbitrarily complex dependencies (direct estimation of each hypercubic transition)  & Cross-sectional, phylogenetically (and longitudinally) related & $< 25$ & Transition probabilities, average trait ordering, pairs of transition orders & None & \citet{moen2023}\\

     \hline
   \end{tabular}

   \begin{tablenotes}
     \setstretch{0.75}
{\footnotesize
   \item[a]{Approximate number of features than can be handled by the reference implementation. This can be strongly affected by choice of method's parameters, for example using the dynamic programming instead of a genetic algorithm for OncoBN or using MC-CBN instead of H-CBN (though MC-CBN can scale poorly with number of subjects). Note that ``can be handled by'' does not mean ``can be handled quickly''.
   }\\
   \item[b]{Output directly provided by the reference implementation of each method. Other output, such as predicted probabilities of genotypes, transition rate matrices, and probabilities of paths can be obtained for most methods; see \citet{diaz-uriarte2022a}. Likewise, it is often possible to use a representation from method A with method B, such as average trait ordering and pairs of transitions orders with MHN and OT in \citet[][Fig.~S6]{moen2023}.}\\
   \item[c]{Untimed: for OT and OncoBN parameters of models represent conditional probabilities of observing a given mutation, when the sample is taken, given the parents are observed.  Discrete time for HyperTraPS and HyperHMM: the models estimate transition probabilities, and an event is gained in each step, thus time units need not have the same duration.}\\
   \item[d]{Common rate, \(L, L^2, L^3, L^4\), all interactions.}
   \\
  \item[e]{This limit applies to the MHN implementation. TreeMHN \citep{luo2023} can be used with a larger numbers of features, its runtime and memory usage being limited by sample size.}
       \\
   \item[f]{Depends on complexity of the model.}
       \\
     }
   {\normalfont \item[\textbf{Error models:}]{}}\\
     {\footnotesize
  \item[g\_OT]{ Possibly different false positive and false negative observation errors; false positives combine observation error and deviations from model, false negatives due only to observation error.}\\
  \item[g\_OBN]{ Observations can arise from a ``spontaneous activation model'' that represent deviations from specified model.}\\
  \item[g\_CBN]{ H-CBN: probability of error in  observation constant and independent for all sites. MC-CBN: model is a mixture of CBN and noise component (independent model).}\\
  \item[g\_HES]{ Same as H-CBN in CBN. Further details about error models in the Supplementary Material to \citet{diaz-uriarte2022a} ---section 2.3.6 of \url{https://rdiaz02.github.io/EvAM-Tools/pdfs/evamtools_methods_details_faq.pdf}.}
    }
  \end{tablenotes}

\caption{\label{tab:table-methods} \textbf{Cancer progression and monotonic accumulation models: main features.} Methods ordered approximately by increasing complexity of dependencies that can be reflected, as shown in the main text. See section \ref{sec:other-methods} for methods not shown.}
\end{threeparttable}
\end{minipage}
\end{sidewaystable}

\KOMAoption{fontsize}{11pt}

\subsection{Purpose, data, assumptions}\label{sec:assumptions}

These models are not limited to genetic alterations nor to cancer data, and it is arguably better to think of them as models for the ``monotonic accumulation of events'' \citep[][p.~2]{ramazzotti2019a}.  Our objective is to examine the existence of dependencies (deterministic or stochastic) between the occurrence of different irreversible events, events which are not limited to  mutations, but can include disease symptoms, tool use, student progress, etc. If we are dealing with cancer genomic data, we are typically  interested in understanding the dependencies between the different genetic alterations (e.g., mutations\footnote{When thinking about mutations, we are ignoring ploidy and dominance (e.g., for a genotype labelled ``A'' in any of the fitness landscapes, does it have both copies or only one of the single copies mutated in a diploid cell?). Some of the complications that could arise from different dominance patterns could be accommodated by a model where the double mutant necessarily depends on the single mutant.} in different genes) that are accumulated, non-reversibly, during tumour progression \citep{Hainke2012, Beerenwinkel2015,beerenwinkel_computational_2016}.

The models we will focus on accept as input binary data (or data that have been binarised), arranged as a matrix of subjects or samples by events. Thus, this is a cross-sectional data set, where each row of this matrix describes the state of a subject (or sample) with respect to the events, each subject (or sample) is represented only once in the data, and different subjects (or samples) are regarded as independent.

From the data, each method will try to infer the best fitting model. By best fitting, we mean that the predicted distribution of mutational profiles is as close as possible to the observed distribution of mutational profiles in the data. Thus, each method will find the parameters (trees/graphs, type of dependencies, and rates/conditional probabilities for models with deterministic dependencies ---section \ref{sec:deterministic-deps};  interaction and spontaneous rates or transition probabilities for models with stochastic dependencies ---section \ref{sec:stoch-deps}) that allow that model to provide the closest match between predicted and observed frequencies of mutational profiles\footnote{For simplicity, that wording is closest to maximum likelihood inference. Of course, nothing precludes using Bayesian inference and some of the methods discussed below use Bayesian inference}.

These models have some common assumptions. First, the events are gained one by one and irreversibly (thus the ``monotonic accumulation'', above)\footnote{This assumption might be sensible for some alterations but not others in the same domain. For example, \citet[p.~2457]{Misra2014}  indicate ``(...) this assumption may not be valid for large-scale karyotypic and copy number changes that are frequently observed in tumour samples.'', which leads them to restrict their analysis to point mutations and small indels.}.  Next, all the methods that do not consider phylogenetic information regard each of the different subjects or samples as replicate evolutionary experiments, or independent realisations of an evolutionary process. Here, all the subjects are under the same constraints or dependencies with respect to the accumulation of the events ---and the purpose of the analysis is, precisely, to understand these constraints \citep{Gerstung2011,Beerenwinkel2015,beerenwinkel_computational_2016,diaz-uriarte2018,diaz-uriarte2019a}. For example, for cancer data, we would regard all the subjects in a cross-sectional data set, corresponding to a homogeneous cancer type, as replicate evolutionary runs where all subjects share the same genetic constraints and dependencies, for example epistatic interactions; we discuss the consequences of not being able to assume a homogeneous type in \qpref{sec:heterogeneity}.

Data with more complex dependency structures can also be analysed by some of these methods, or by closely related methods. In particular, data with phylogenetic/genealogical relationships (where some samples share common ancestors that are not shared by other samples) are analysed by two methods discussed here (HyperTraPS, section \ref{sec:hypertraps}, and HyperHMM section \ref{sec:hyperhmm}) as well as by TreeMHN \citep{luo2023} and REVOLVER \citep{caravagna2018} (see section \ref{sec:other-methods}). HyperTraPS and HyperHMM can also analyse longitudinal data, where multiple observations of each subject have been taken over time\footnote{A single lineage can represent longitudinal data and is a subset of a phylogeny. Thus, it is possible that TreeMHN and REVOLVER might be used to analyse longitudinal data too.}. Much of the discussion in \qref{sec:violat}, in particular an explicit consideration of fitness landscapes and evolutionary models, can be relevant to phylogenetically related samples, but here we will focus on cross-sectional data.

Thus, cancer progression models or, more generally, monotonic accumulation models, use cross-sectional data, regarded as independent replicate evolutionary experiments, to infer the dependencies in the irreversible accumulation of discrete events. These dependencies can be formulated as deterministic restrictions (section \ref{sec:deterministic-deps}) or as inhibiting/facilitating stochastic dependencies (section \ref{sec:stoch-deps}). See Table \ref{tab:table-methods} for a summary of the methods discussed.

\subsubsection{Software}
\label{sec:software}
This manuscript is not concerned with the specifics of software, data handling, input, processing of output, and graphical representation. Details for software use should be sought in the original papers and associated repositories; EvAM-Tools \citep{diaz-uriarte2022a} is a web-based tool and R package, with ongoing development, that tries to provide a single, unified interface for analysing data with most of these methods, as well as for simulating random dependency structures and data under these models (as of now, notable missing methods are HyperTraPS and HyperHMM, which are planned to be added in the near future). A list of repositories is provided in the Appendix: \qref{sec:software-repos}.

\subsection{Deterministic dependencies}\label{sec:deterministic-deps}

The models in this section use directed acyclic graphs (DAGs) and trees\footnote{A tree is a Directed Acyclic Graph (DAG) in which a child can have only one parent; all trees are DAGs, but not all DAGs are trees.} to represent deterministic dependencies, or restrictions, in the accumulation of events. In these graphs, an edge from event $i$ (e.g., a mutation in gene $i$) to event $j$ (e.g., a mutation in gene $j$) means that event $i$ must occur before event $j$ can occur; in other words, an edge (or arrow) from $i$ to $j$ indicates a direct and necessary dependency of event $j$ on event $i$.

\subsubsection{Oncogenetic trees (OT)}
\label{sec:oncogenetic-trees}

\begin{figure}[t!]
 \centering \includegraphics[width=10.35cm,keepaspectratio]{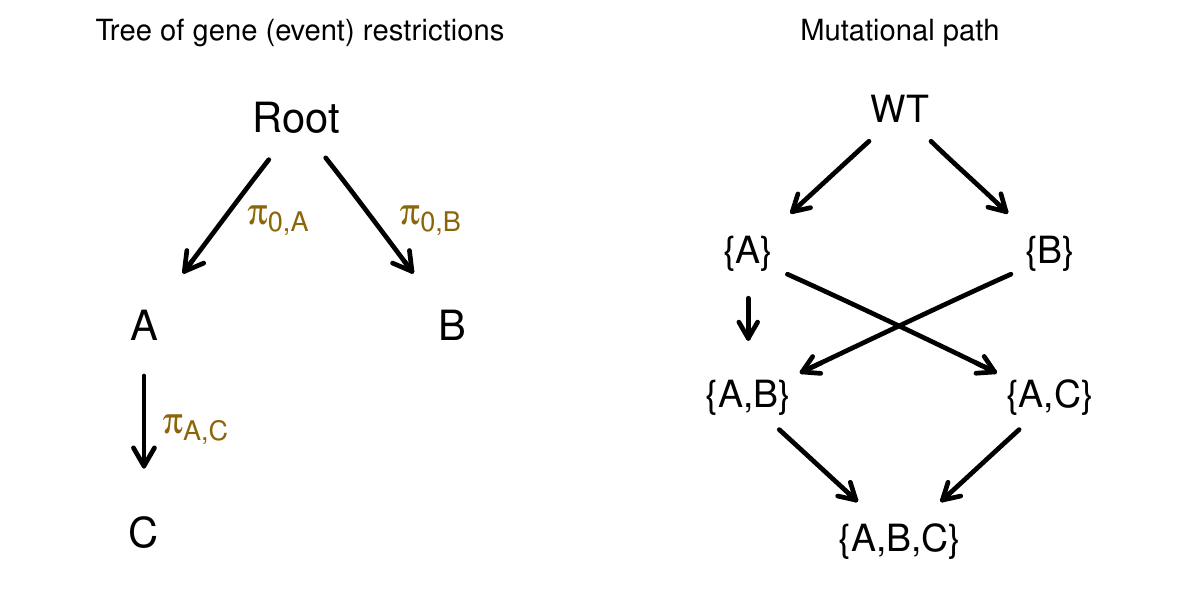}
 \caption{\textbf{Oncogenetic tree model}. Left: example dependencies in the acquisition of mutations; events A and B are independent of each other and depend on no previous event, but event C depends on event A. Right: possible
tumour states or mutational profiles (or genotypes, if the model applies to within-cell restrictions) and transitions between them. (For notation for genes, events, paths, and genotypes see \qpref{sec:notation-genots}). According to this OT model, and in the absence of errors, by the time of sampling we expect to see each mutational profile (cell type if the model pertains to within-cell restrictions) according to the parameters of the model, for example, $P(\{A\})  = \pi_{0,A}\ (1 - \pi_{0,B}) \ (1 - \pi_{A,C});  P(\{A,C\}) = \pi_{0,A}\ (1 - \pi_{0,B})\ \pi_{A,C}; \ldots$. The probabilities of transitions between mutational profiles/genotypes are not given by the OT model, as the OT model is an untimed model \citep{Desper1999JCB,Szabo2008} ---although timed versions have been used in \citet{Beerenwinkel2005b,Bogojeska2008}.
}
  \label{fig:ot-deps}
\end{figure}

OTs are among the earliest formal models of accumulation of mutations in cancer, described in  \citet{Desper1999JCB} (see also \citealp{SimonDesper2000,RadmacherSimon2001,Szabo2008, oncotree, Szabo2002}). OTs represent restrictions in the accumulation of events as a tree. Since OTs use trees, an event can only directly depend on one other event (its parent), but multiple events can directly depend on a previous event (as, in a tree, a parent can have multiple children). An example is shown in Figure \ref{fig:ot-deps}. OT has a model for errors that include possibly different false positive and false negative observation errors and deviations from the model (see \citealp{Szabo2002,Szabo2008}, and Supplementary Material to \citealp{diaz-uriarte2022a}).

\subsubsection{Disjunctive Bayesian Networks (DBN: OncoBN)}
\label{sec:oncobn}

\begin{figure}[t!]
 \centering \includegraphics[width=10.00cm,keepaspectratio]{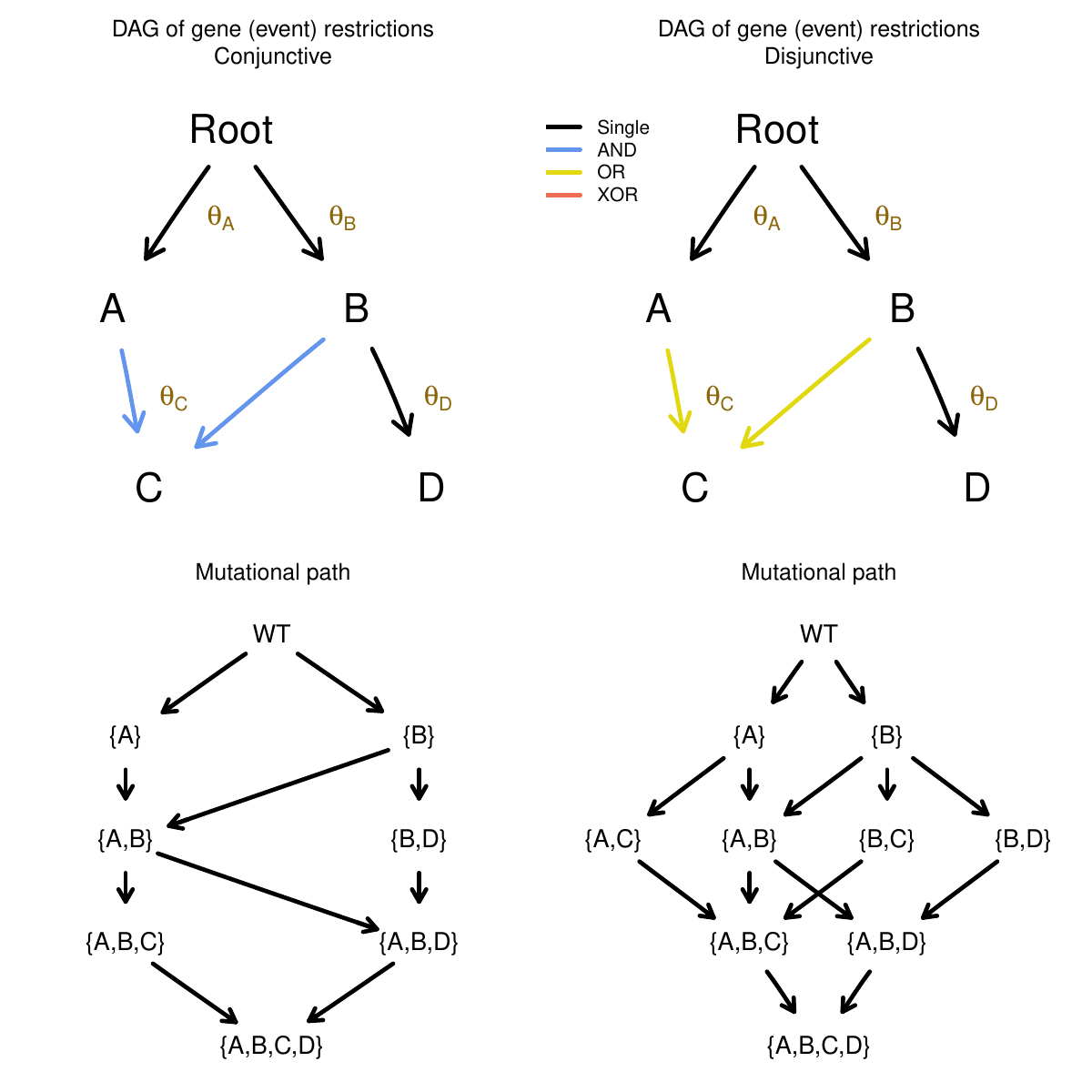}
 \caption{\textbf{OncoBN model}. Top: example dependencies in the acquisition of mutations with the conjunctive ---AND--- model (left) and the disjunctive ---OR--- model (right). Bottom: possible mutational profiles (or genotypes if the model applied to within-cell restrictions) according to the models on top, in the absence of errors. (For notation for genes, events, paths, and genotypes see \qpref{sec:notation-genots}). By the time of sampling, and in the absence of errors, the probabilities of some mutational profiles are $P(\{A,B\})  = \theta_A\ \theta_{B}\ (1 - \theta_C) \ (1 - \theta_D) $ for both the conjunctive and disjunctive model, whereas $P(\{A\})  = \theta_A\ (1 - \theta_B), P(\{A,C\})  = 0$ for the conjunctive model, and $\ P(\{A\})  = \theta_A\ (1 - \theta_B) \ (1 - \theta_C), P(\{A,C\})  = \theta_A\ (1 - \theta_B) \ \theta_C$ for the disjunctive model. As for OT, the probabilities of transitions between mutational profiles are not given by the OncoBN model, as OncoBN is an untimed model.
}
  \label{fig:oncobn-deps}
\end{figure}

OncoBN \citep{nicol2021} allows fitting an untimed oncogenetic model like OT, but it can model restrictions using a DAG, so an event can directly depend on multiple events. The OncoBN models in which events have multiple parents can be of two different types: a) Disjunctive Bayesian Network (DBN) models where the multiple dependencies are OR relationships (any one of the ancestor events is enough for the descendant to be possible), and this is the DBN \textit{sensu stricto}; b) Conjunctive Bayesian Network (CBN) models where the multiple dependencies are AND dependencies (all of the ancestor events must have occurred). The CBN models fitted by OncoBN are untimed: the parameters do not have the same interpretation as the parameters of the original CBN models (described in section \ref{sec:cbn}). In a given OncoBN all the multiple dependencies have the same meaning, so a model can be either a DBN or a CBN, but not both. An example is shown in Figure \ref{fig:oncobn-deps}. OncoBN also has a model for errors, that differs from the one of OT (see \citealp{nicol2021}, and Supplementary Material to \citealp{diaz-uriarte2022a}).

\subsubsection{Conjunctive Bayesian Networks (CBN)}
\label{sec:cbn}

\begin{figure}[t!]
 \centering \includegraphics[width=12.0cm,keepaspectratio]{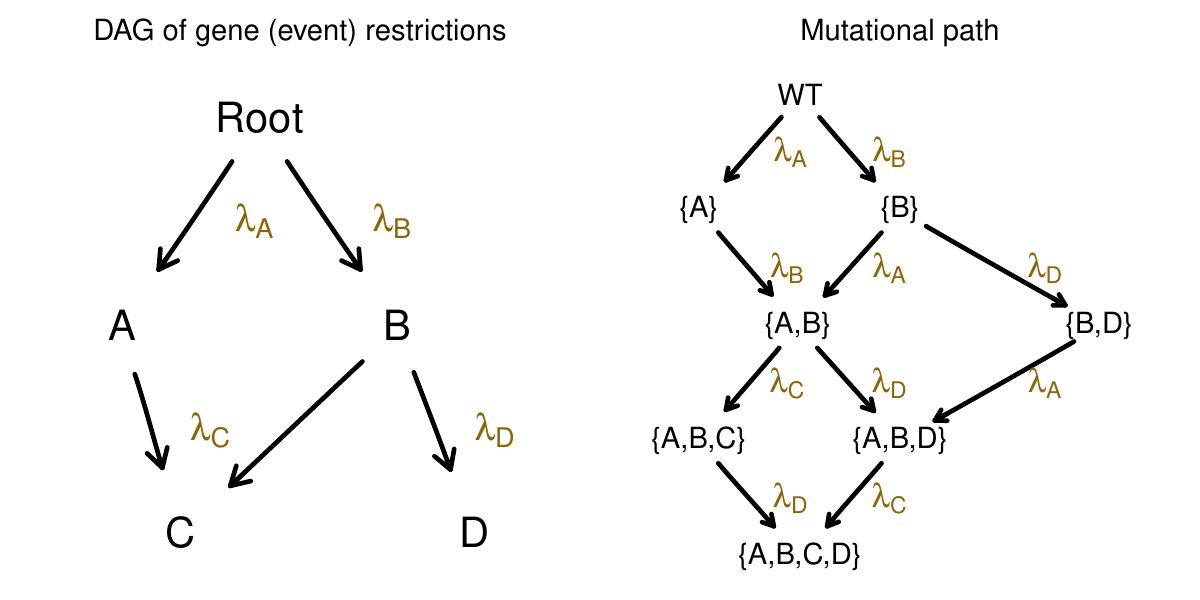}
 \caption{\textbf{CBN model}.
   Left: example dependencies in the acquisition of mutations; A and B are independent of each other and depend on no previous event, but C depends on both A and B; D depends only on B. Right: possible tumour states or mutational profiles (or genotypes, if the model applies to within-cell restrictions) and transitions between them; transitions are described by a continuous-time Markov Chain with transition rate matrix as shown in Table \ref{tab:cbn-trm}. In the figure, an edge is annotated with the rate that corresponds to that specific transition (and matches that in Table \ref{tab:cbn-trm}). (For notation for genes, events, paths, and genotypes see \qpref{sec:notation-genots}). Comparing this figure with the left of Fig.~\ref{fig:oncobn-deps},  we see that the same restrictions are captured in both cases using different modelling approaches.
}
\label{fig:cbn-deps}

\rowcolors{2}{gray!25}{white}
\vspace*{20pt}
   \centering
   {
     \fontsize{9}{10}\selectfont
\begin{tabular}{rcccccccc}

  \hline
  & \{WT\} & \{A\} & \{B\} & \{A,B\} & \{B,D\} & \{A,B,C\} & \{A,B,D\} & \{A,B,C,D\} \\
  \hline
  \{WT\}  &  & $\lambda_A$ & $\lambda_B$ &   &   &  &  &   \\
  \{A\}   &   &  &   & $\lambda_B$ &   &   &   &   \\
  \{B\}   &   &   &  & $\lambda_A$ & $\lambda_D$ &   &   &   \\
   \{A,B\}  &   &   &   &  &   &$\lambda_C$ & $\lambda_D$ &   \\
  \{B,D\} &   &   &   &   & &   & $\lambda_A$ &   \\
  \{A,B,C\} &   &   &   &   &   &  &   & $\lambda_D$ \\
  \{A,B,D\} &   &   &   &   &   &   & & $\lambda_C$ \\
  \{A,B,C,D\} &   &   &   &   &   &   &   &   \\
  \hline
\end{tabular}
\captionof{table}{\textbf{CBN transition rate matrix.} Transition rate matrix between mutational profiles (or genotypes, if the model refers to within-cell restrictions) corresponding to the CBN model in Figure \ref{fig:cbn-deps}. The diagonal, which is $-\sum$ of the row elements, is not shown for clarity; the remaining elements not shown are 0.}\label{tab:cbn-trm}
}
\end{figure}

A CBN  \citep{Gerstung2009,Gerstung2011,montazeri_large-scale_2016} represents restrictions using a DAG, and a node with multiple parents requires all parent events to have occurred. CBN differs from OT and OncoBN because the CBN model is a timed model: the parameters of the model ($\lambda$s), are the rates of the exponentially distributed times to fixation of an  event given that all parents of that event have been observed (i.e., given that the event restrictions, as specified in the DAG, are satisfied: \citealp[p.~i729]{montazeri_large-scale_2016}; \citealp[section 2.2]{Gerstung2009}). An example is shown in Figure \ref{fig:cbn-deps}. There are, in fact, two commonly used CBN models, H-CBN \citep{Gerstung2009,Gerstung2011} and MC-CBN \citep{montazeri_large-scale_2016}; they differ in the fitting algorithm and in the error model (see details  in \citealp{Gerstung2009,montazeri_large-scale_2016}, and Supplementary Material to \citealp{diaz-uriarte2022a}).

According to the model in Figure \ref{fig:cbn-deps}, the waiting times of, for example, events \textit{A, B, D, C} are:
$T_A \sim \mathit{Exp}(\lambda_A)$, $T_B \sim \mathit{Exp}(\lambda_B)$, $T_D \sim T_B + \mathit{Exp}(\lambda_D)$, $T_C \sim \max(T_A, T_B) + \mathit{Exp}(\lambda_C)$.  Figure \ref{fig:cbn-deps} shows the mutational path with edges annotated, which corresponds to the transition rate matrix in Table \ref{tab:cbn-trm}.

\subsubsection{Hidden Extended Suppes-Bayes Causal Networks (H-ESBCN, PMCE)} \label{sec:hesbcn}

\begin{figure}[t!]
 \centering \includegraphics[width=11.75cm,keepaspectratio]{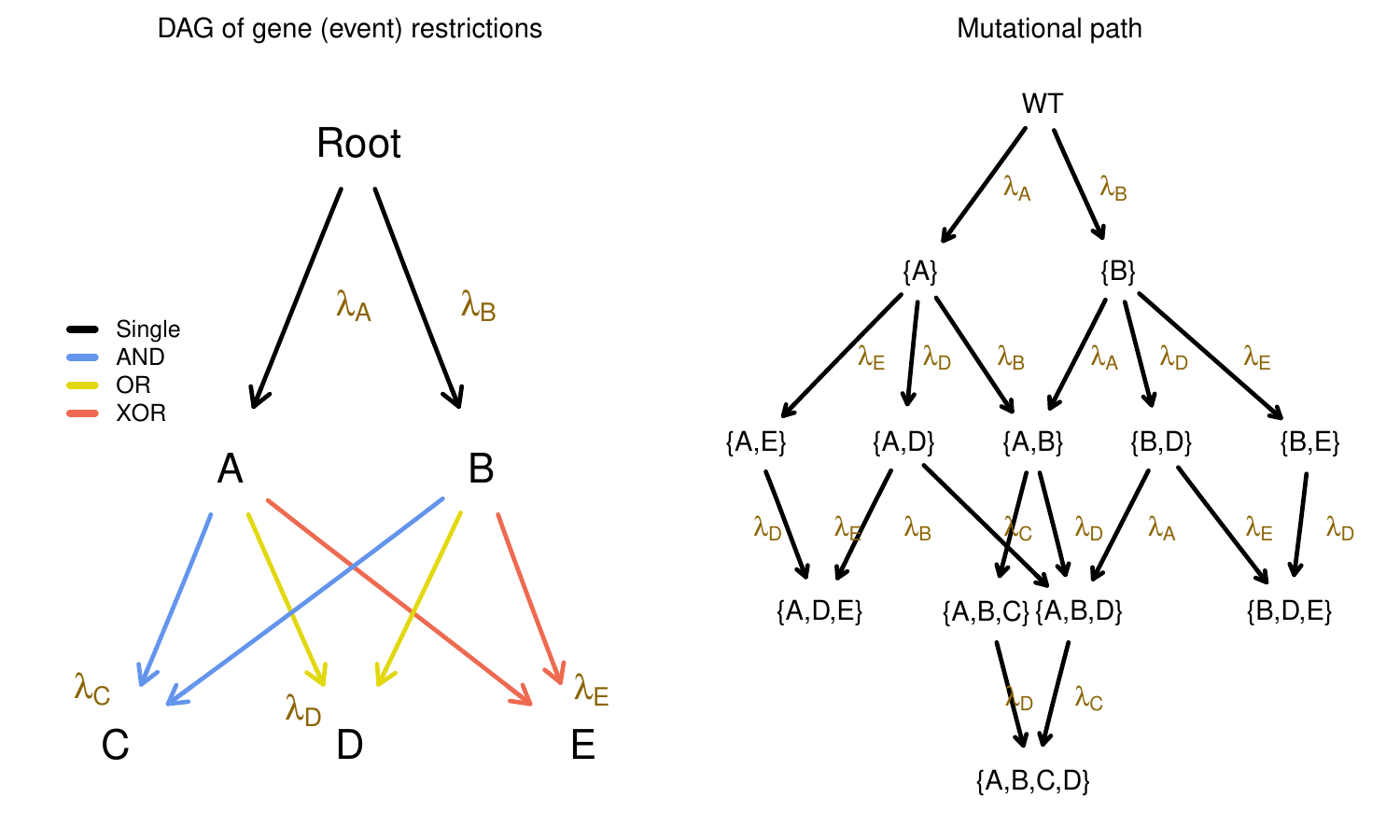}
 \caption{\textbf{H-ESBCN model}.
   Left: example dependencies in the acquisition of mutations;  A and B are independent of each other and depend on no previous event; C depends on both A and B;  D depends on A OR B, so either one of A or B (or both) is enough;  E depends on A XOR B: exactly one of A, B are needed for E.   Right: possible tumour states or mutational profiles (or genotypes, if the model refers to within-cell restrictions) and transitions between them; transitions are described by a continuous-time Markov Chain with transition rate matrix as shown in Table \ref{tab:hesbcn-trm}. In the figure, an edge is annotated with the rate that corresponds to that specific transition (and matches that in Table \ref{tab:hesbcn-trm}). (For notation for genes, events, paths, and genotypes see \qpref{sec:notation-genots}).
}
  \label{fig:hesbcn-deps}

\vspace*{20pt}
   \rowcolors{2}{gray!25}{white}
\centering
{\fontsize{6}{7}\selectfont
  \begin{tabular}{rccccccccccccc}
  \hline
  &     \{WT\} & \{A\}           & \{B\} &        \{A,B\} & \{A,D\} & \{A,E\} & \{B,D\} & \{B,E\} & \{A,B,C\} & \{A,B,D\} & \{A,D,E\} & \{B,D,E\} & \{A,B,C,D\} \\
                                                                                                                                                    \hline
  \{WT\} &  & $\lambda_A$ & $\lambda_B$ &  &  &  &  &  &  &  & & & \\
  \{A\} &  &  &  &  $\lambda_B$ & $\lambda_D$ & $\lambda_E$ &  &  & &  &  &  & \\
  \{B\} &  &  &  &  & $\lambda_A$ &  $\lambda_D$ & & $\lambda_E$ &  & &  &  &\\
  \{A,B\} &  &  &  &  &  &  &  &  &  $\lambda_C$ & $\lambda_D$ &  &  & \\
  \{A,D\} &  &  &  &  &  &  &  &  &  & $\lambda_B$ &  $\lambda_E$ &  &  \\
  \{A,E\} &  &  &  &  &  &  &  &  &  &  & $\lambda_D$ & & \\
  \{B,D\} &  &  &  &  &  &  &  &  &  &  $\lambda_A$ &  & $\lambda_E$ & \\
  \{B,E\} &  &  &  &  &  &  &  &  &  &  &  & $\lambda_D$  &  \\
  \{A,B,C\} &  &  &  &  &  &  &  &  &  &  &  &  & $\lambda_D$ \\
  \{A,B,D\} &  &  &  &  &  &  &  &  &  &  &  &  & $\lambda_C$ \\
  \{A,D,E\} &  &  &  &  &  &  &  &  &  &  &  &  &  \\
  \{B,D,E\} &  &  &  &  &  &  &  &  &  &  &  &  &  \\
  \{A,B,C,D\} &  &  &  &  &  &  &  &  &  &  &  &  &  \\
  \hline

  \end{tabular}
}
\captionof{table}{\textbf{H-ESBCN transition rate matrix.} Transition rate matrix corresponding to the H-ESBCN model in Fig.~\ref{fig:hesbcn-deps}. Diagonal ($= -\sum$ of the row elements) not shown for clarity; the remaining elements not shown are 0.
}\label{tab:hesbcn-trm}

\end{figure}

H-ESBCN (Hidden Extended Suppes-Bayes Causal Networks) is part of a suite called Progression Models of Cancer Evolution, PMCE \citep{angaroni2021}. Like CBN, it is a timed model, where the parameters of the model ($\lambda$s),  are  the rates of the exponentially distributed times to fixation of an  event given that the parents of that event have been observed. The main difference with CBN is that a node with multiple parents can have three different types of dependencies: AND (like CBN), OR (any one of the parent events is enough, like in OncoBN), XOR (exclusive OR: exactly one parent is necessary). An H-ESBCN model can include all three types of dependencies for different child nodes (whereas OncoBN could have AND or OR, but not both, in the same model) but a child node has the same type of dependency on all of its parents (so, for example, it would not be possible for an event to depend with an OR relationship on some parents and with an AND relationship on some other parents). The error model of H-ESBCN is the same as the one in H-CBN \citep{Gerstung2009}.

It is important to notice that the  XOR (as well as OR and AND) relationships refer to the dependency of a child event on its parents. This is why under the model in Figure \ref{fig:hesbcn-deps}, genotype $\{A,B\}$ is a possible genotype  but $\{A,B,E\}$ is not. It is not possible with the current implementation of the XOR to model exclusive relationships between sets of mutations that do not involve a descendant\footnote{It might be possible to hack around this by creating a ``NULL'' event that descends from parents we want to make non-viable when combined, but this does not seem possible unless we know, beforehand, which are the lethal non-viable combinations.}.
That the XOR is defined with respect to descendants also affects the AND and OR operators but, since we are modelling accumulation, the AND and OR can only be defined as they currently are. However, the XOR operator could be defined independently of the accumulation of subsequent mutations to allow modelling that combinations of mutations are not possible, regardless of descendant mutations. The latter is what models with stochastic dependencies do, and this issue will be relevant in our discussion in \qref{sec:heterogeneity}.

\subsection{Stochastic dependencies}\label{sec:stoch-deps}

\begin{figure}[tbhp]
 \centering \includegraphics[width=12.0cm,keepaspectratio]{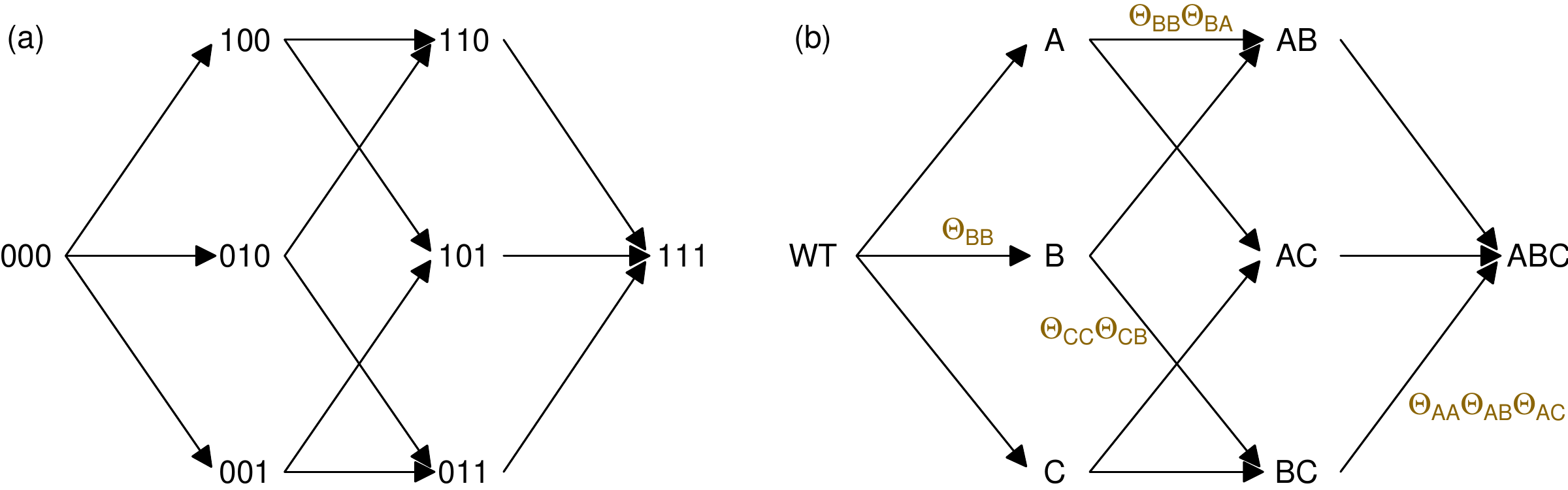}
 \caption{\textbf{Models with stochastic dependencies}. HyperTraPS, MHN, and HyperHMM model a hypercubic transition graph between genotypes. Both (a) and (b) show all possible one-step transitions for a three event case, but they differ in the coding. In (a), a 1 denotes that a particular event or mutation is present and a 0 it is not. (b) is identical to (a), but using letters for the mutations or events present. MHN and HyperTraPS model those transitions as a function of a reduced number of parameters. For example, with MHN the transition rate from 010 or the genotype with only \textit{B} mutated, to 011, or the genotype with both \textit{B} and \textit{C} mutated, is, as shown in eq.~\ref{eq:mhn}, $\Theta_{CC} \ \Theta_{CB}$. And the transition rate from genotype \{B, C\} to \{A, B, C\} is $\Theta_{AA} \ \Theta_{AB} \Theta_{AC}$; these and two other transition rates are shown next to the corresponding edges.}
  \label{fig:hypercube}
\end{figure}

In models with stochastic dependencies, events can alter, increasing or decreasing, the probability of acquiring other events. These methods model the transitions in a hypercubic transition graph linking the genotypes (Fig.~\ref{fig:hypercube}); two of the models, MHN and HyperTraPS, model those transitions as a function of a (potentially) reduced number of parameters, whereas HyperHMM  directly models the transitions between genotypes without expressing them as a function of a smaller number of parameters.

\subsubsection{Mutual Hazard Networks (MHN)}\label{sec:mhn}

MHN is described in \citet{schill2020}. This is a continuous time Markov model (like CBN ---section \ref{sec:cbn}--- and H-ESBCN ---section\ref{sec:hesbcn}) in which the rate of occurrence of an event is modelled by a spontaneous rate of acquisition ($\Theta_{ii}$ in eq.~\ref{eq:mhn}) and a multiplicative effect that each of these events can have on the rate of other events via pairwise interactions ($\Theta_{ij}$ in eq.~\ref{eq:mhn}). These pairwise interactions model promoting ($\Theta_{ij} > 1$) and inhibiting ($\Theta_{ij} < 1$) dependencies. The transition rate from a genotype \textbf{x} to a genotype with mutation \textit{i} added to genotype \textbf{x} is specified by eq.~\ref{eq:mhn} \citep[eq.~2]{schill2020}:

\begin{equation}\label{eq:mhn}
Q_{\mathbf{x},\mathbf{x}_{+i}} = \exp (\theta_{ii} + \sum_{j=1}^{n} \theta_{ij} x_j) = \Theta_{ii} \prod\limits_{x_j = 1} \Theta_{ij}
\end{equation}
where   $x_j$ is 1 if event $j$ has already occurred (e.g., gene $j$ is already mutated) in genotype \textbf{x}, and $Q_{x,y}$ is the transition rate from $x$ to $y$ (we are using the transpose of the notation in \citealp{schill2020}, for consistency with the transition rates shown for CBN and H-ESBCN in Tables \ref{tab:cbn-trm} and \ref{tab:hesbcn-trm}). $\Theta_{ii}$ is the baseline hazard or the rate of event $i$ and  $\Theta_{ij}$ is the multiplicative effect of event $j$ on the rate of event $i$; see also example in Fig.~\ref{fig:hypercube}.  Because the $\Theta_{ij}$ reflect interactions between pairs of genes, MHN cannot incorporate higher-order interactions (such as three-way epistasis).

The MHN model has recently been extended \citep{schill2024a} to account for selection bias (also know as collider bias or Berkson's bias: \citealp{hernan2004, pearl_causality_2009})\footnote{It should be possible to extend other models, in particular HyperTraPS and HyperHMM, to account for observation bias following a procedure similar to the one explained in section 2.2 of \citet{schill2024a}: introducing additional parameters to model the effects of events on the rate of observation.}.

\subsubsection{HyperTraPS}
\label{sec:hypertraps}

Prior to MHN, the concept of mutual hazards (though not the name) was developed in HyperTraPS (hypercubic transition path sampling). Originally used to study mitochondrial genome evolution \citep{johnston2016}, HyperTraPS has also been applied to cancer \citep{greenbury2020} (as well as, among others, learner behaviour in online courses, \citealp{peach2021}; tool use in animals, \citealp{johnston2020a}; malaria progression in children,  \citealp{johnston2019}). In contrast to MHN, HyperTraPS in its original description was a discrete time model, although a continuous time version of the model is now also available \citep{aga2024}. The transition probability between a state $s$ and a state with all the events in $s$ plus event $i$ is given by eq.~\ref{eq:hypertraps} (see \citealp[Equation 2 in Supplementary Material of][]{greenbury2020}; \citealp[p.~6 of Supp.\ Mat., ``Inferring transition matrices'' in][]{johnston2016}; \citealp[p.~10 of Suppl.\ Mat.~in][]{johnston2020a}):

\begin{equation}\label{eq:hypertraps}
P_{\mathrm{gain}\ i | \mathrm{state}\ s} \propto \exp (\theta_{ii} + \sum_{j=1}^{n} \theta_{ij} s_j)
\end{equation}

The transition probability is the contribution of $\theta_{ii}$, the log of the basal rate of occurrence of event $i$, and the $\theta_{ij}$, the influence of the other $j$ events present in $s$, on the occurrence of $i$.

Although the model is  similar to MHN, HyperTraPS is developed considering a Hidden Markov Model for the transitions in the hypercube defined by the $2^L$ genotypes or possible combinations of events, where $L$ is the number of events. The original implementation of HyperTraPS models the hypercubic transitions in terms of pairwise interactions between events  for practical computational reasons, but the model is not inherently limited to two-way interactions, and it can incorporate three-way or higher-order interactions between events \citep{aga2024}. For example, the non-additive influence of pairs of acquired features can be captured in an $\mathcal{O}(L^3)$ model:

\begin{equation}\label{eq:hypertraps-3}
P_{\mathrm{gain}\ i | \mathrm{state}\ s} \propto \exp (\theta_{iii} + \sum_{j=1}^{n} \sum_{k=j}^{n} \theta_{ijk} s_j s_k)
\end{equation}
where $\theta_{iii}$ is the base rate for feature $i$, $\theta_{ijj}$ is the influence of feature $j$ on the base rate of $i$, and $\theta_{ijk}$ is the addition contribution from the feature pair $j,k$. Influences of triplets and larger sets of features can similarly be captured through progressively higher-dimensional parameter sets.

A key difference between HyperTraPS and MHN, is HyperTraPS's ``(...) ability to account for samples linked by temporal or phylogenetic relationships, rather than only independent cross-sectional samples'' \citep[][p.~2]{moen2023}. Instead of single states, it treats transitions between states as fundamental observations. This format captures longitudinal and reconstructed phylogenetic data, and cross-sectional data is represented as a set of individual transitions from the precursor state to each individual observation.

\subsubsection{HyperHMM}
\label{sec:hyperhmm}

HyperHMM was developed by \citet{moen2023} using, like HyperTraPS, a Hidden Markov Model to represent the hypercubic transitions. One major difference with respect to HyperTraPS is the fitting algorithm (an EM algorithm instead of MCMC). The other key difference is that HyperHMM is not limited to two-way interactions between events: transitions are modelled directly, without restrictions in the order (two-way, three-way, \ldots) of event interactions, and this allows it to capture arbitrarily complex interactions between events (see, e.g., \citealp[Fig.\ S6 in Supplementary Material of][]{moen2023}).

In the same way as HyperTraPS, HyperHMM can be used not only for cross-sectional data, but also for non-independent data with longitudinal or phylogenetic relationships \citep{moen2023}.

\subsection{Other methods}\label{sec:other-methods}

In Table \ref{tab:table-methods}, we have not listed TreeMHN \citep{luo2023},  Hintra \citep{khakabimamaghani2019}, or REVOLVER \citep{caravagna2018}, as they focus on inferring recurrent evolutionary trajectories from multiple within-patient (or intra-tumour) phylogenetic trees; these data often arise from multi-region, single-cell sequencing (i.e., the data come from multiple patients, where each patient contributes multiple samples, from different tissues or sampling times).
ToMExO is focused on a somewhat different task \citep[][p.~3]{neyshabouri2022a}: ``(...) simultaneously identify critical driver genes, group them as sets of mutually exclusive genes (driver pathways), and arrange them in a tree structure representing the order in which they get mutated'' (see also discussion in section \ref{sec:heterogeneity}). MASTRO focuses on ``discover[ing] all conserved trajectories in a collection of phylogenetic trees describing the evolution of a cohort of tumours, allowing the discovery of conserved complex relations between alterations'' \citep[][p.~ii49]{pellegrina2022}. CAPRESE \citep{caprese_2014} and CAPRI \citep{capri_bioinformatics, capri_pnas} are previous methods from the authors of H-ESBCN and seem to have been superseded by H-ESBCN, and thus are not discussed further (see also Supporting Information, file S4 Text of \citealp{diaz-uriarte2019a} for discussion on difficulties obtaining probabilities of paths of tumour progression from CAPRESE and CAPRI).

Bayesian Mutation Landscape (BML), developed by \citet{Misra2014}, is focused on identifying ``evolutionary progression paths'' and  reconstructing ancestral genotypes, and it does this taking into account epistatic interactions between pairs of genes. Thus, its objectives are somewhat similar, though not identical, to other CPMs, and the leveraging on epistatis is explicit.
However, the method does not seem to have been used or developed further, and it is unclear that the predicted probabilities of genotypes and paths under the model can be computed from the output. Its main output are the likely paths as a figure (where the detail and, thus, number of paths shown, depends on a threshold that the user gives), the ``evolutionary probabilities'' for the top ordered genes, and the evolutionary probabilities and departures from independence for pairs of genes. See Appendix for software (section \ref{sec:software-repos}).

Finally, we have only mentioned models and methods with an existing, working, free software implementation and without dependencies on proprietary software. Specifically, some methods either have ``a rough and very preliminary implementation'' \citep{Tofigh_2011}  or are too slow for routine use (e.g., \citealp{Sakoparnig2012}). TimedHN \citep{chen2023} and DiProg \citep{FarahaniLagergren2013} depend on proprietary software (Matlab and IBM's ILOG CPLEX, respectively); and other methods  \citep[e.g.,][]{Hjelm2006, Cheng2012} have no existing, publicly available working implementations.

\subsection{Deterministic vs.\ stochastic dependencies: possible genotypes and error models}\label{sec:observable_genots}

Under the models with deterministic dependencies (section \ref{sec:deterministic-deps}), an event can only occur if its dependencies are satisfied; as a consequence, under these models, some genotypes (or combinations of events) are possible whereas others, those that do not satisfy the dependencies, are not. In Figures \ref{fig:linear-deps} to \ref{fig:hesbcn-deps}, genotypes not shown under ``Mutational path'' should not be observable (except for observational error or deviations from the model). In models with stochastic dependencies all genotypes are, in principle, possible, so all transitions than involve gaining a mutation have a probability $> 0$, albeit arbitrarily small. Note, though, that as discussed in the Supplementary Material of \citet{schill2020}, it is possible to consider MHN (and, therefore, HyperTraPS and HyperHMM) as stochastic approximations to the deterministic dependencies of CBN (see also section \ref{sec:assumpt_stoch}).

Error models thus become a necessity for deterministic models, because under the deterministic models some genotypes are not possible; even minimal rates of observational error would then render any dependency model (except for a star topology, with all events descending directly from the Root event) incompatible with observed data. The lack of explicit error models for MHN, HyperTraPS and HyperHMM does not have this effect: with stochastic dependencies, observational error and error from deviations from the model can be fully aliased without leading the fitting algorithms  astray.  Note that regardless of the existence of an explicit error model in a given method, simulations to assess the performance of these methods can incorporate observational error in the simulation process \citep[e.g.,][p.~244]{schill2020}.

\section{Uses and entities}\label{sec:uses_entities}

\subsection{Uses -- inference}\label{sec:uses_all}
The majority of CPM publications use a particular method to infer dependencies in existing cancer datasets, and then discuss how these dependencies match what is already known and also expand and suggest new possible dependencies. Some papers  further use the output from a given method to examine the predictive value of progression (as given by the model) and patient survival \citep[e.g.,][]{Gerstung2009,angaroni2021,Bogojeska2008appnote}. Some other work has focused on the possible uses of CPMs to estimate cancer evolutionary predictability \citep{hosseini2019a}, predict the paths of tumour evolution \citep{diaz-uriarte2019a}, or the next genotype conditional on the currently observed genotype \citep{diaz-colunga2021}.

These models have also been used to model mutation acquisition in HIV \citep{beerenwinkel_evolution_2006, Beerenwinkel2007}, and possible differences between HIV  subtypes \citep{montazeri_estimating_2015,posada-cespedes2021}; most of that work has used CBN but also  mixtures of mutagenetic (oncogenetic) trees \citep{Beerenwinkel2005a, Beerenwinkel2005b, Yin2006}.

Accumulation models have also found use in a diverse range of disciplines outside cancer progression. HyperTraPS was originally developed to understand the selective pressures behind mitochondrial genome reduction \citep{johnston2016}. It has  also been used to model malaria progression in children \citep{johnston2019}, where differences in symptom ordering as reflected by the model could be used for triage and other interventions, and to examine drug resistance acquisition in tuberculosis  \citep{greenbury2020}, where results help identify what drug has the highest probability of future resistance, conditional on the current state. In \citet{johnston2020a}, HyperTraPS was applied to model tool acquisition in animals, including predicting future and unobserved behaved; \citet{peach2021} used HyperTraPS to study learner behaved in online courses which allowed them to identify learning tasks correlated with higher performance. HyperHMM \citep{moen2023} has been applied to cancer, but also to some of the above problems (tuberculosis drug resistance).  A precursor to HyperHMM was used to study the evolution of \(C_4\) photosynthesis \citep{williams2013}. Note that in the tuberculosis, mitochondrial gene loss, animal tool use, and learner behaviour examples, HyperTraPS and HyperHMM are used precisely because of their capability to deal with phylogenetically and longitudinally related data, and the analysis of phylogenetically related data  involved inferring ancestral character states. In most of the cases in which HyperTraPS and HyperHMM have been used, the empirical entities to which the models are applied,  and the traits modelled, are of a  different kind from those in studies of genetic alteration acquisition in  cancer and, to a large extent, HIV.

All of these models and other CPM approaches can be viewed as instances of the Mk model of discrete character evolution in phylogenetic comparative methods (\citealp{pagel1994}; \citealp{lewis2001}; see also \citealp[ch.~7]{harmon2019}, and \citealp[ch.~6]{revell2022a}); this is a Markov model with \textit{k} states on a phylogeny, and independent data (as from cross-sectional data) can be represented as a star phylogeny. Using the Mk model directly for these problems, however, is not practical because the monotonic accumulation assumption is not leveraged.

\subsection{Uses -- intervention/stratification}\label{sec:uses_intervention}
The examples above also highlight different objectives for predictions and conditional interventions (see also section \ref{sec:model-choice}). A model of accumulation of symptoms/markers can be used to divide subjects into those with better and worse prognosis to then, in principle, triage or apply different medical treatment to each group \citep{johnston2019}. Here the medical treatment, the intervention, involves altering variables that are not part of the modelled system: what variables to alter, and how, is not knowledge gained from the model of the sequential accumulation of symptoms. Thus, the model need not have a direct causal interpretation. Alternatively, in other cases we might want to intervene in variables that are part of the model, guided by the model itself. In this case, it is crucial to  endow the relationships in the model with a causal, counterfactual interpretation \citep{pearl_causality_2009,hernan2020}. For example, we might try to guide the choice of therapeutic target based upon the results of the model:  what is the gene that, if altered, will guide evolution to a particular destination \citep{zhao2016,diaz-uriarte-rios-arroyo-interv}.

\subsection{What are the ``entities'' under study}\label{sec:entities}

These methods have been applied to a variety of problems, but the entities to which the inferences pertain are not always explicit.

Most of the uses of HyperTraPS and HyperHMM use clearly defined entities: 2904 children coded with respect to a list of clinical features (coughing, jaundice, fever, diarrhoea, etc.) in the study of malaria progression \citep{johnston2019}; animal species, coded with respect to tool use \citep{johnston2020a} for tool use emergence across animal taxa\footnote{As the authors recognise, ``there is substantial debate and flexibility in what constitutes tool use, or a specific mode within that class'' and ``data also omit the finer-grained dynamics of tool use acquisition in individual animals, instead reporting the capacity for a species to develop a particular mode of tool use.'' \citep[][pp.~9 and 10]{johnston2020a}. There is, it might be argued, room for some fuzziness, but the terms of the possible debate are clear.}; mitochondrial genomes when examining mitochondrial gene loss \citep{johnston2016}; 395 drug-resistant tuberculosis isolates in studies of antimicrobial resistance \citep{moen2023,greenbury2020}.

In the case of models for cancer there seem to be different interpretations about the modelled entities, for example cell, tumour, patient, clonal lineage, sample. In the Appendix, \qref{sec:quotes-entitites}, we give full quotations, and here we summarise what we understand most authors mean. \citet{Desper1999JCB}, one of the earliest references (OT model), refer to cancer cells and understanding the cause-effect relationships between genetic (CGH) events, but this paper is previous to the wide awareness about intra-tumour heterogeneity. \citet[][pp.~220-221]{Szabo2002} write ``Some alterations present in the tumour might be missed because of the spatial heterogeneity'', which indicates that for these authors the models are primarily about tumours, possibly without stronger commitments (cells, evolutionary regime, epistasis) and \citet{Szabo2008} refer exclusively to tumours (not cells), but mention a ``mechanistic interpretation''. Other papers, especially those that use and develop CBN, frame restrictions in terms of epistatic interactions and show an explicit commitment to a fitness landscape interpretation \citep{Beerenwinkel2015}. The fitness landscape interpretation under a strong selection, weak mutation (SSWM ---\citealp{sniegowski_beneficial_2010,Desai2007,krug2021}) regime is unambiguous in \citet{hosseini2019a}, so this ought to imply that the model is about restrictions that apply within cells. This interpretation is consistent with  the wording in \citet{Gerstung2009,Gerstung2011} (and also in the BML paper by \citealp{Misra2014}), that refers to restrictions in the fixation of events in the population, presumably in a model of successive clonal sweeps, not unlike SSWM; finally, in other cases it is said that the model describes transitions between (HIV) genotypes \citep{montazeri_large-scale_2016}. In the first MHN publication, \citet{schill2020} mention individual tumour cells and their fitness, though their latest publication \citep{schill2024a} explicitly refers to bulk and tumour genotypes.

\citet{angaroni2021}, authors of PMCE (H-ESBCN), make it unequivocal that their model is to be interpreted as a model of the restrictions and probabilities of the bulk samples, not necessarily of the restrictions in individual cells within tumours.  The OR and XOR in their model are presented as more complex dependencies that should allow us to deal with heterogeneity, but this is heterogeneity between patients within a tumour (sub)type; moreover, these OR/XOR say nothing, by themselves, about within-tumour selection, competition, or cooperation between cells.

\citet[][p.~3]{nicol2021}, who developed the DBN model in OncoBN, refer to a tumour being composed of subclones. Their model's parameters and predictions of probabilities of different states refer to the states of individual tumours (or mutational profiles), not clones or cells. But the restrictions identified by their model apply also within cells, and thus for each tumour profile. Of note, they implicitly exclude the ``until fixated'' assumption (in contrast to CBN) because the mutational profile can show mutations that are only present in some of the subclones (which also applies to PMCE/H-ESBCN).

Thus, it is not always clear if each point in the data (the actual samples used to fit the model) is supposed to reflect a single type of cell/clone/genotype (with respect to the mutations modelled), or if each data point could be a mutational profile that results from bulk sequencing (see \qpref{sec:scenarios_no_rse}, item \ref{scenario_collapse}). If the latter, it is not obvious how the method is able to make correct (within-cell? epistatic?) inferences with a sample that contains a mixture of different cell types.

\subsection{Why worry about the entities?}\label{sec:why-entities}

Ambiguities about the entities can lead to mistakes in the interpretation of the meaning of the parameters. For example, and in line with the discussion in section \ref{sec:entities}, does an inferred model describe the behaviour of single cells or of groups of cells aggregated over a whole tumour? There are clear implications for the variability, but also the meaning of inferred behaviours.

This lack of precision about the entities can also blur the distinction between qualitatively different types of evolutionary models, in particular frequency-dependent fitness (see \qpref{sec:freq-dep}). And, methodologically, what the entities are affects how we assess the performance of these methods.  If we expect these methods to correctly infer within-cell restrictions, we will simulate processes happening within cells and evaluate if, after sampling, processing, and analysing, we recover the within-cell restrictions.  But this assessment might be irrelevant to other tasks, such as indicating how to intervene to drive the overall state of a tumour to a certain destination, or for using the sequence of events to stratify subjects for further treatment.

The results of \citet{Sprouffske2011} and \citet{Diaz-Uriarte2015,diaz-uriarte2018}  showed poor performance of some of the above methods to infer within-cell restrictions. In the above sense, this might say little of relevance if we are interested in using these models to separate patients into prognostic groups. Indeed, this is a problem that other papers that use tumour samples suggest could be approached fruitfully \citep{Gerstung2009,angaroni2021,fontana2023}; outside cancer, this problem  has also been addressed in malaria \citep{johnston2019} and tool use in animals \citep{johnston2020a}, by looking at differences in the dynamics of event accumulation between different subsets of data (patients that lived or died, terrestrial vs.\ aquatic animals).

On the other hand, careful consideration of the entities underscores that the common practice of assessing the performance of methods using data simulated from the generative models encoded by the DAGs (or the generative models encoded by the stochastic dependencies) can convey an overoptimistic impression of their performance. This is a serious problem  highlighted by \citet{Sprouffske2011} and \citet{Diaz-Uriarte2015}. For example, that a method can detect XOR patterns or unbiasedly estimate rates in XOR patterns from data simulated from a generative model, does not mean that the method can recover within-cell XOR patterns or provide unbiased estimates of within-cell XOR rates under bulk sampling.

\section{Modelling within-cell restrictions: Interpretations and violations of assumptions}\label{sec:violat}

\begin{figure}[tbph]
  \centering
  \begin{adjustbox}{width=1.00\linewidth}

\renewcommand{\arraystretch}{0.01}
\begin{tabularx}{\textwidth}{m | q | w | s | t | u | v | e}
  \toprule
  Issue \& Figure number & True DAG & Fitness landscape & Genotypes, mutational profile & Inferred DAGs  & RSE diagram & LOD, POM, paths & \quad Example figure layout (for reference) \\
         &
           \includegraphics[width=0.4\linewidth,valign=t]{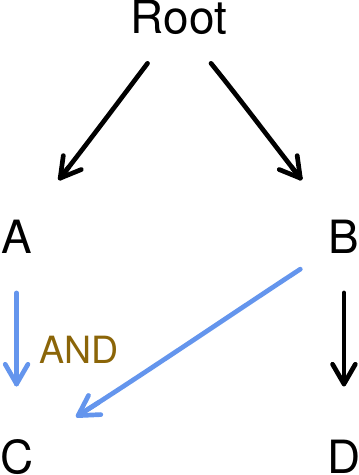}
                           &  \includegraphics[width=1.1\linewidth,valign=t]{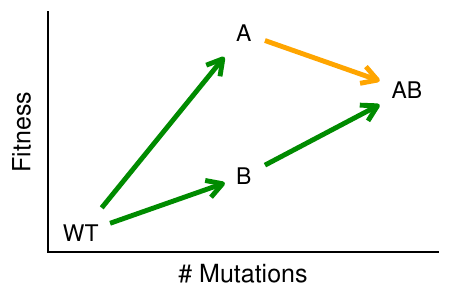} &
\vspace*{-5pt}\resizebox{1.15\linewidth}{!}{
    \begin{tabular}[t]{ll}
      \hline
      \rowcolor{white}
      Cells & Tumor\\
      (clones) & (mutational profile) \\
      \hline
      \{A\}, a few others &         \{A\}     \\
       \rowcolor{gray!55}
      \{B\}, a few others &         \{B\}     \\
      \rowcolor{white}
      \{A\}, \{A,C\}, a few others &         \{A,C\}   \\
      \rowcolor{gray!55}
      \{B\}, \{B,C\}, a few others &         \{B,C\}   \\
      \rowcolor{gray!25}
      \{A\}, \{B\}, \{A,B\}, a few others &         \{A,B\}   \\
      \rowcolor{white}
      \{A\}, \{B\}, \{A,B\}, \{A,C\}, a few others &         \{A,B,C\}   \\
      \{A\}, \{A,B,C\}, a few others &         \{A,B,C\}   \\
      \{B\}, \{A,B,C\}, a few others &         \{A,B,C\}   \\
      \ldots & \{A,B,C\}\\
      \hline
    \end{tabular}
 }
  &
 \includegraphics[width=0.610 \linewidth,valign=t]{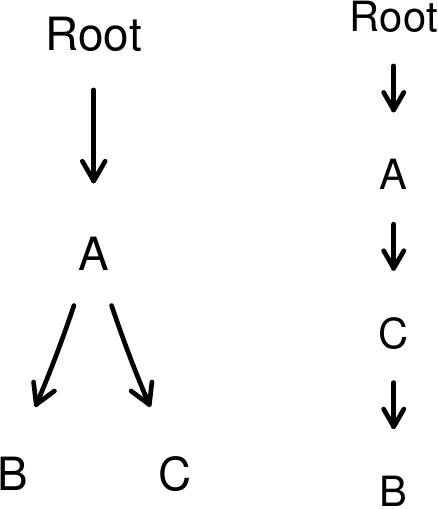} &                                                         \vspace*{2pt}\includegraphics[width=0.37\linewidth]{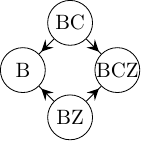} &
 \vspace*{1pt}\includegraphics[width=0.62\linewidth,valign=t]{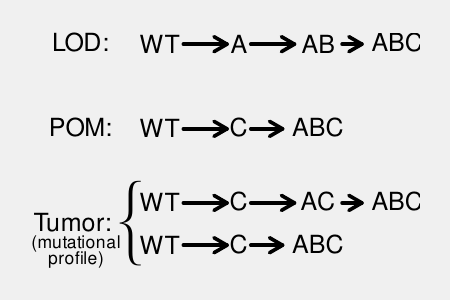} &
  \\
  \midrule
  \midrule
  Represent-able landscapes: \ref{fig:representable} & X & X & & & & & \vspace*{-1pt}\includegraphics[width=3.2\linewidth,valign=t]{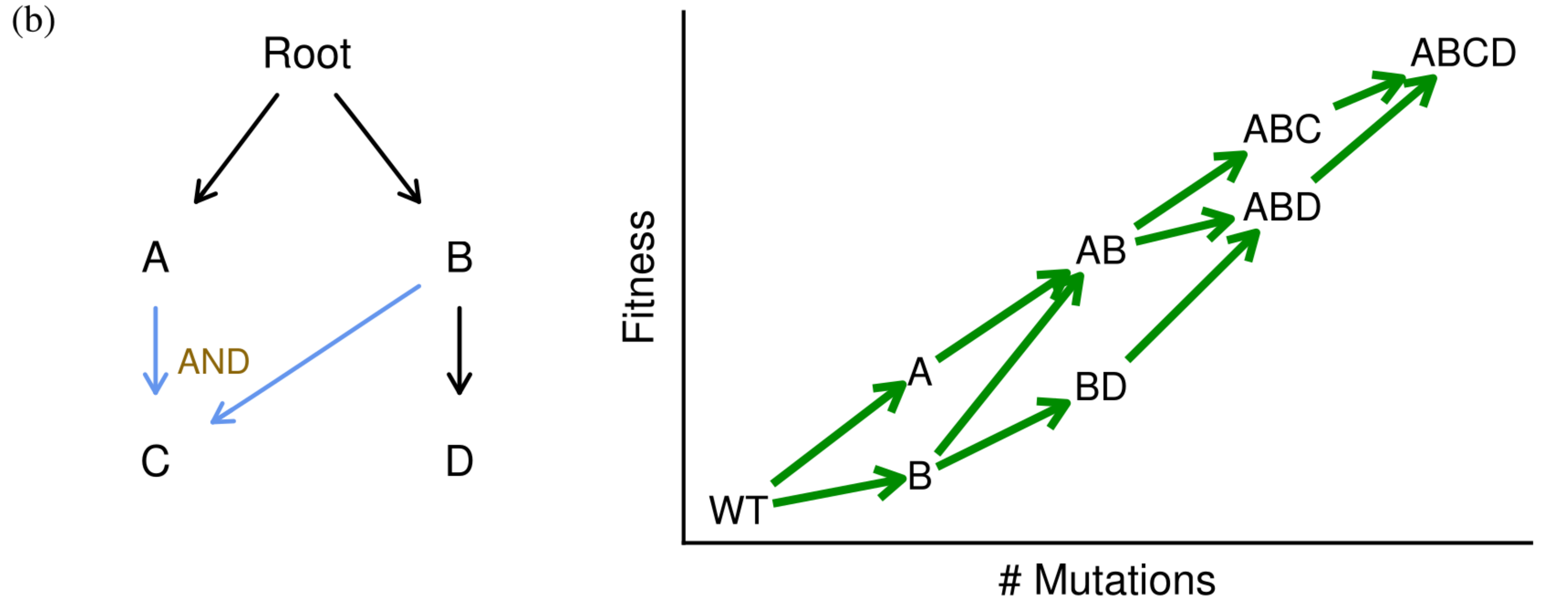}\\
  \arrayrulecolor{gray}\hline
  \\[3pt]

  Scenarios: \ref{fig:scenarios} &&&&&&& \vspace*{-1pt}\includegraphics[width=2.2\linewidth,valign=t]{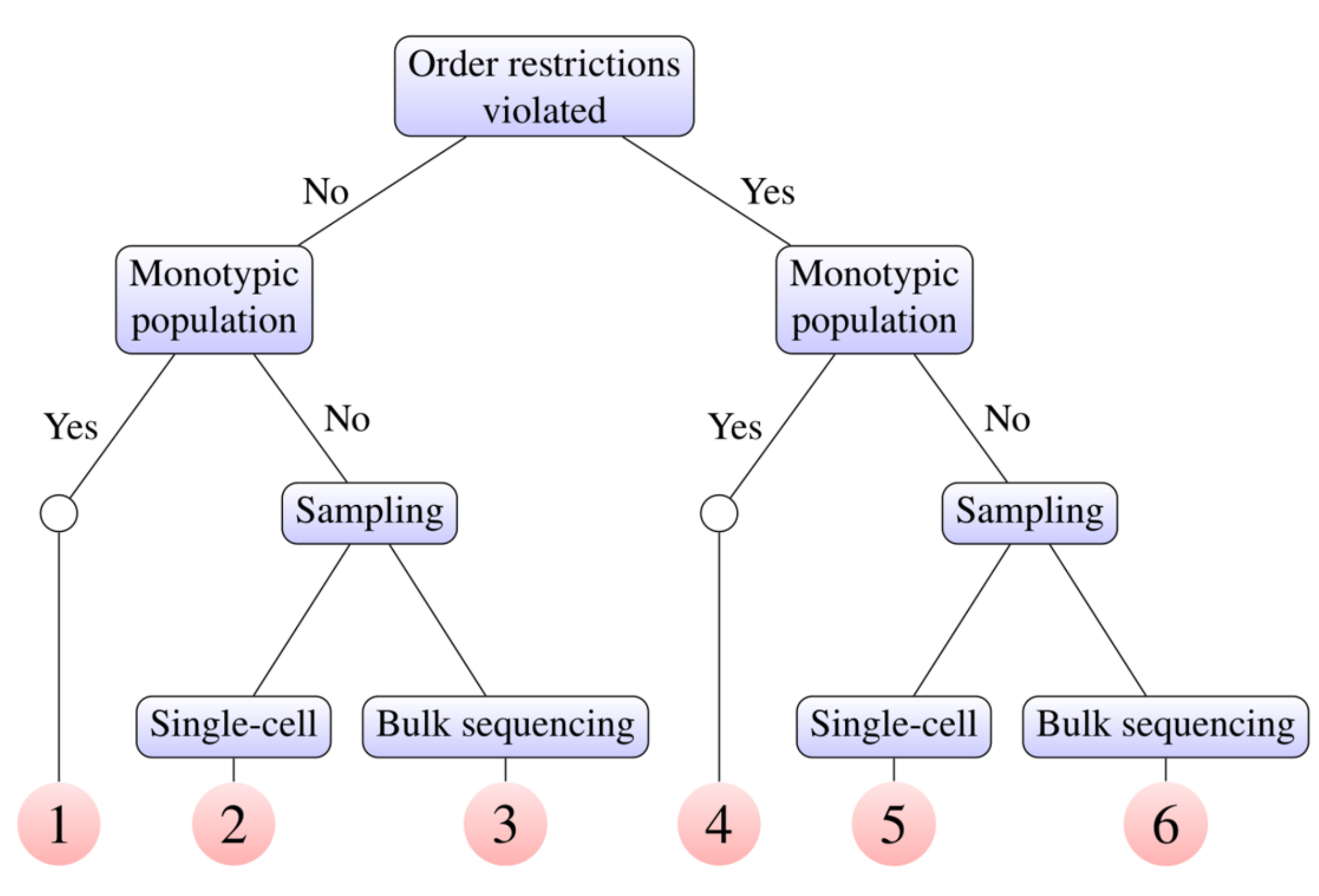}\\
  \arrayrulecolor{gray}\hline
  \\[3pt]

  Bulk sequencing: \qquad \ref{fig:collapse}
                & X & & X & X & & & \vspace*{-1pt}\includegraphics[width=3.2\linewidth,valign=t]{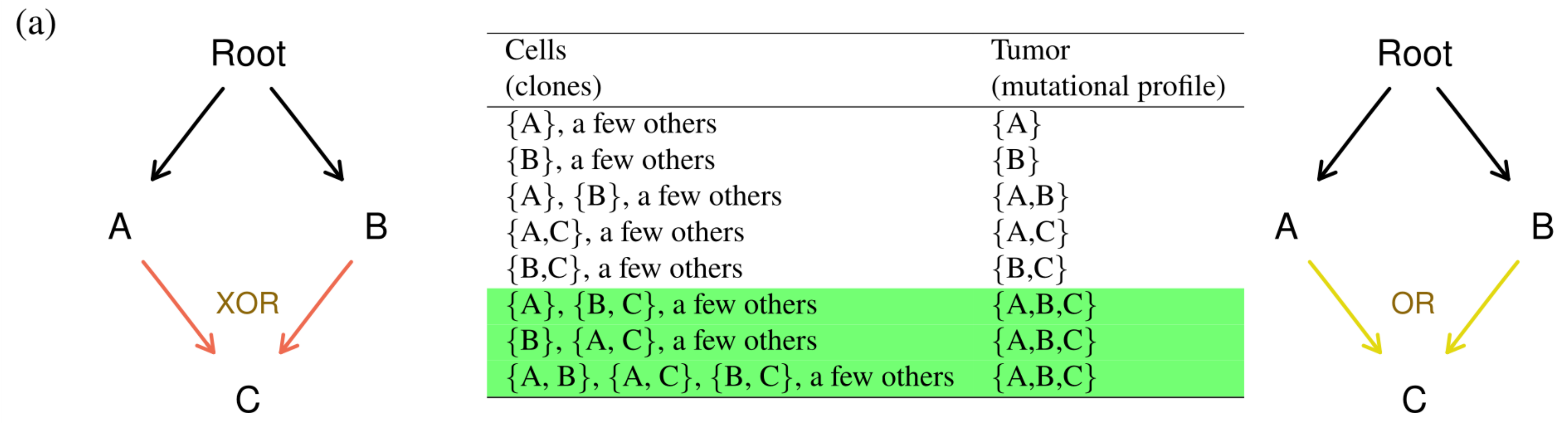}\\
  \arrayrulecolor{gray}\hline
  \\[3pt]

  Order restrictions violations: \ref{fig:order_restrictions_violated} &&&&&&& \vspace*{-1pt}\includegraphics[width=2.2\linewidth,valign=t]{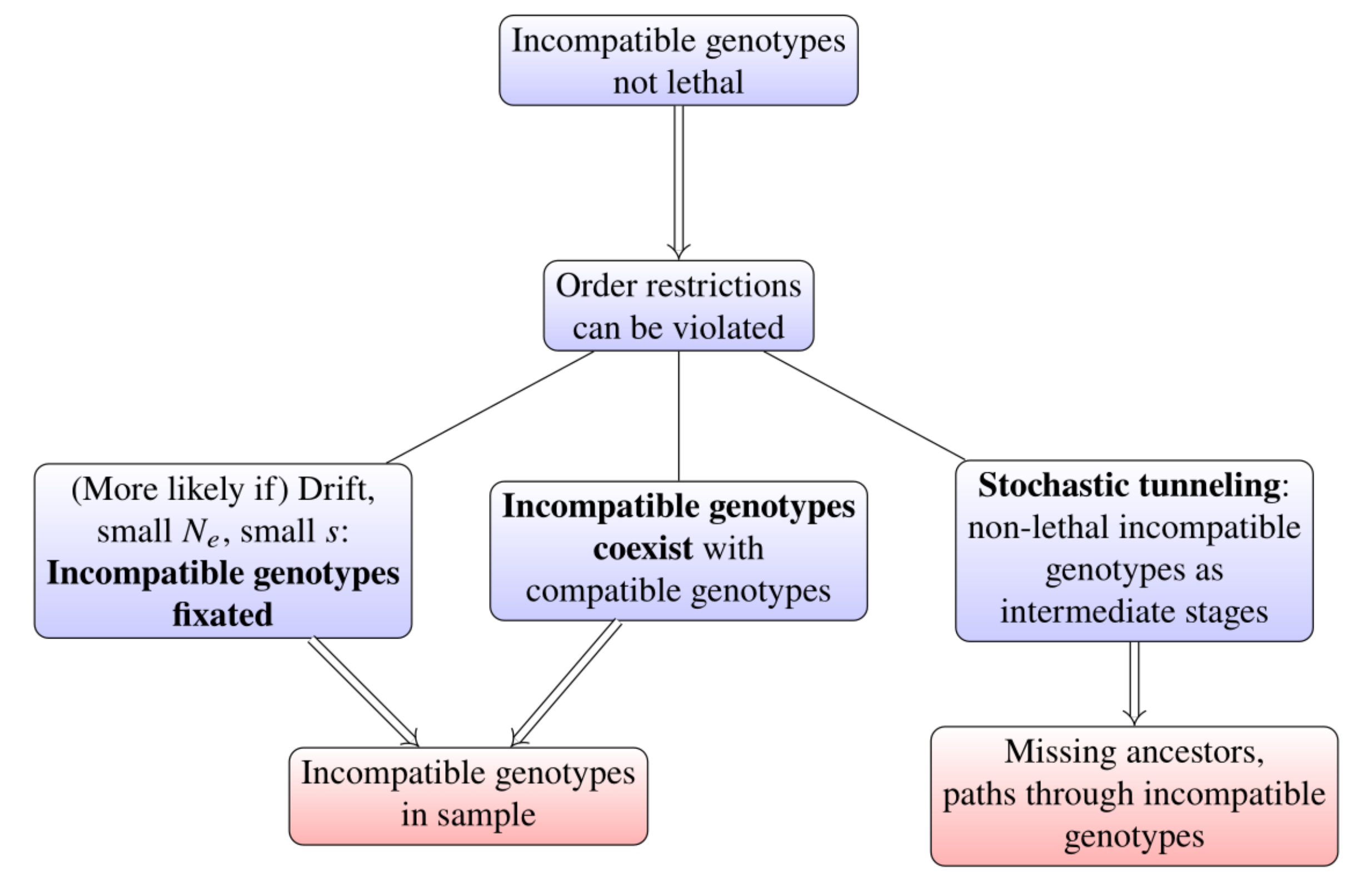}\\
  \arrayrulecolor{gray}\hline
  \\[3pt]

  Tunneling: \ref{fig:tunnel} 
                & X & X & & & & &  \vspace*{-1pt}\includegraphics[width=3.2\linewidth,valign=t]{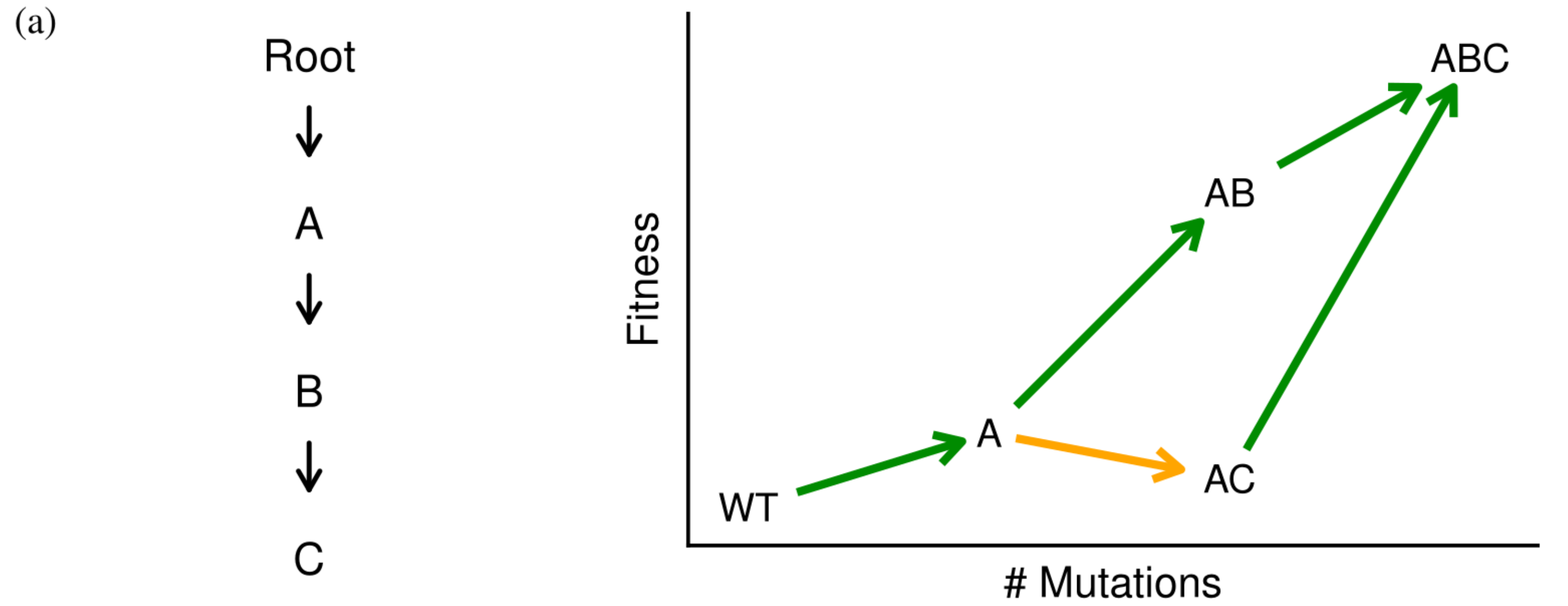}\\
  \arrayrulecolor{gray}\hline
  \\[3pt]
  Reciprocal sign epistasis (RSE): \qquad \ref{fig:rse}, \ref{fig:rse-complications} \vspace*{3.2pt}  & X & X & & & X & & \vspace*{-1pt}\includegraphics[width=3.2\linewidth,valign=t]{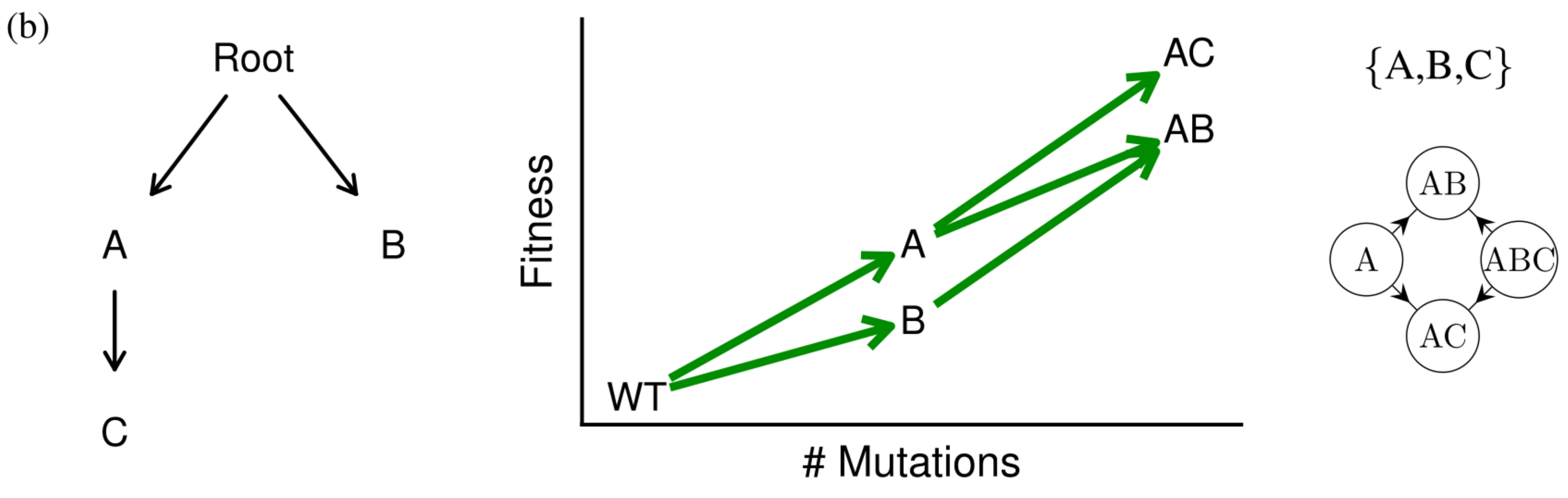}\\
  \arrayrulecolor{gray}\hline
  \\[4pt]
  RSE \& reverting restrictions: \qquad \ref{fig:undo} & X & X & & & & &  \vspace*{-1pt}\includegraphics[width=3.2\linewidth,valign=t]{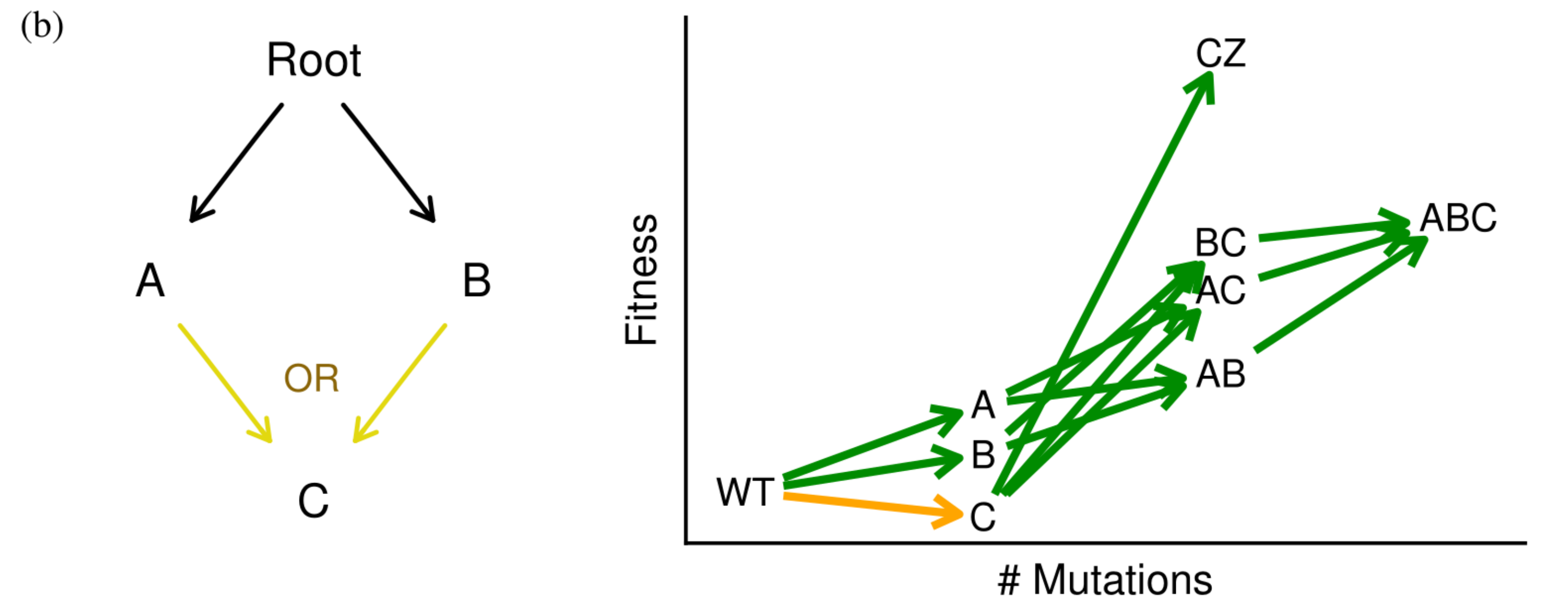}\\
  \arrayrulecolor{gray}\hline
  \\[3pt]
  Lines of descent (LOD), path of the maximum (POM): \ref{fig:lod_pom} \vspace*{2.01pt} & & X & & X & & X & \vspace*{-1pt}\includegraphics[width=3.2\linewidth,valign=t]{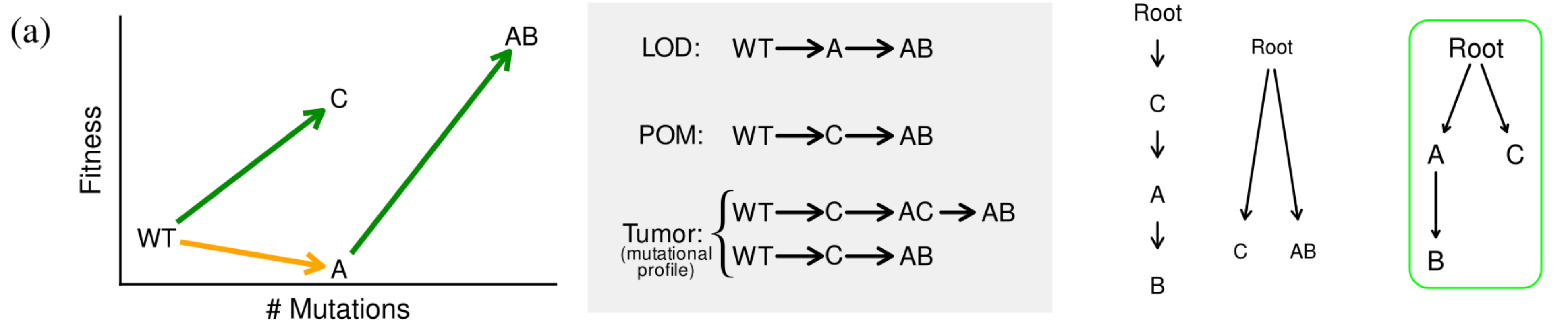}\\
  \arrayrulecolor{gray}\hline
  \\[3pt]
  Hetero-geneity, mixtures of DAGs: \ref{fig:heterog} \vspace*{2.01pt} & X & X & & & & & \vspace*{-1pt}\includegraphics[width=3.2\linewidth,valign=t]{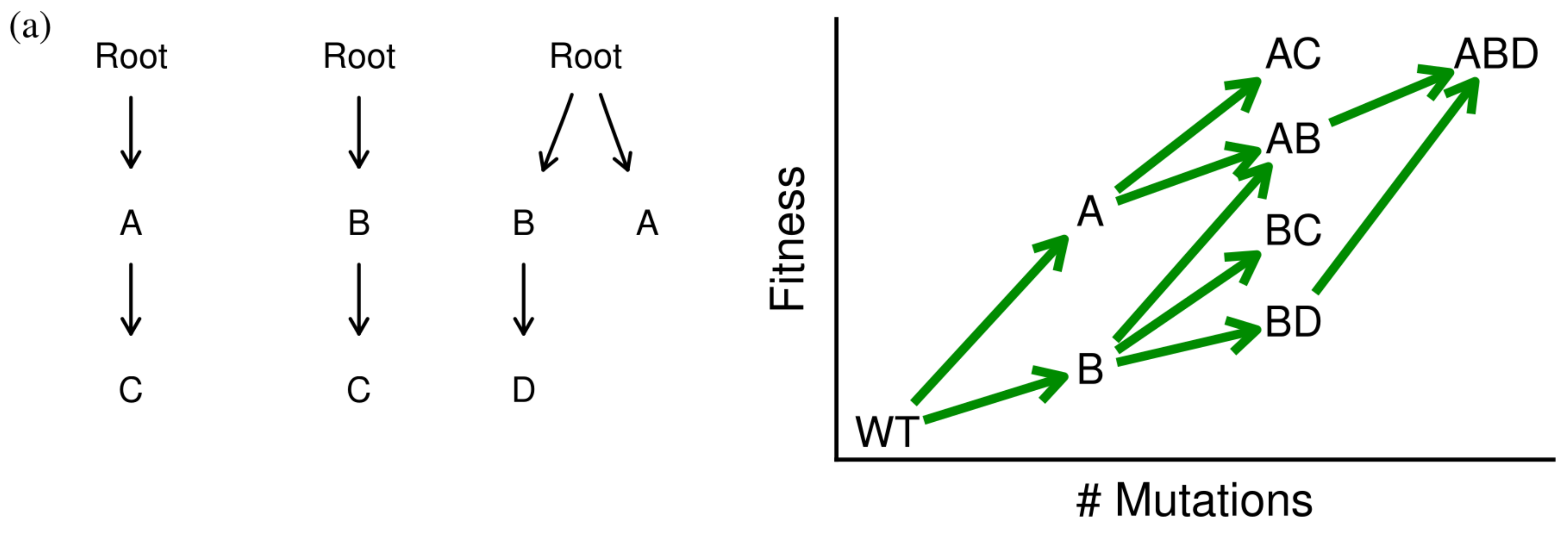}\\
  \arrayrulecolor{gray}\hline
  \\[3pt]
  Frequency-dependent fitness: \ref{fig:freq_dep} & X & & X & X & & & \vspace*{-1pt}\includegraphics[width=3.2\linewidth,valign=t]{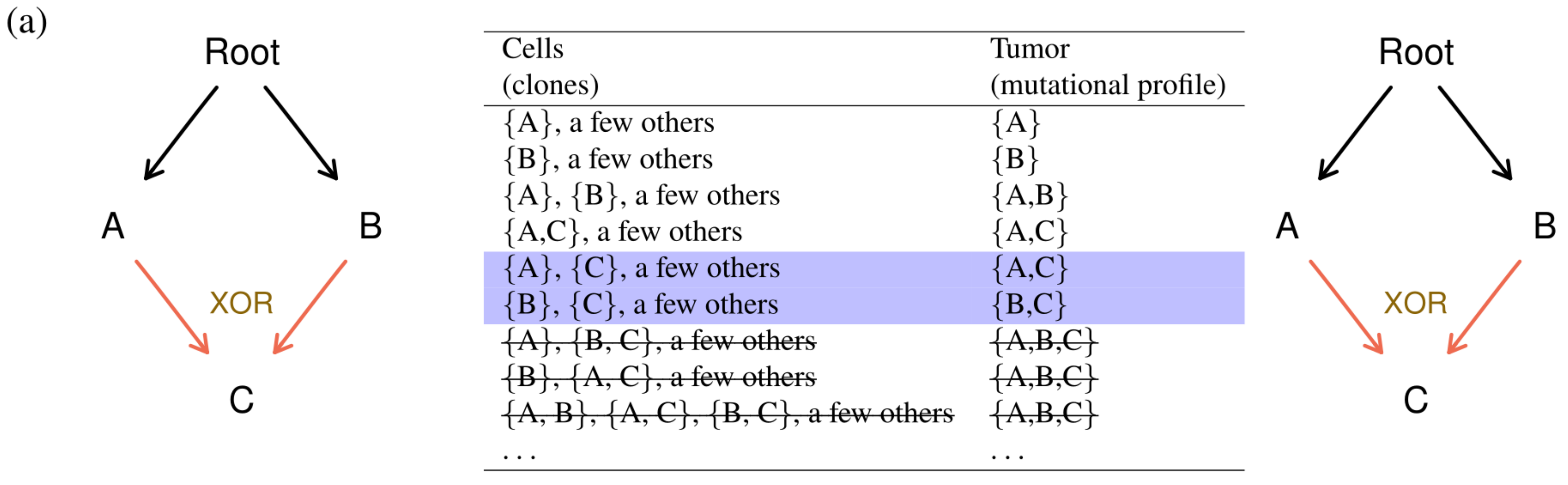}\\
  \bottomrule
\end{tabularx}


\end{adjustbox}

 \caption{\textbf{Issues illustrated in the figures that follow: navigational guide.} Figure insets in the table header show example pieces of the individual figures. An ``X'' denotes that element is present.
 }
  \label{fig:metafigure}
\end{figure}

We have already mentioned above (section \ref{sec:introduction}) how sign epistasis leads to restrictions in the order of accumulation of mutations (see \citealp{crona_peaks_2013,weinreich_perspective_2005-1, gong2013}, and references therein). Thinking about epistasis, as argued by \citet{crona_peaks_2013}, allows us to use the qualitative perspective offered by fitness landscapes\footnote{In this ms.\ ``(...) \textit{fitness landscape} refers to the assignment of fitness values to genotypes that are connected by mutations'' \citep[][p.~2]{krug2021}; this is sometimes referred to as the ``genotype interpretation'' \citep[][p~19]{skipper2012} of Sewall Wright's adaptive landscape concept} \citep[see][and references therein for adaptive and fitness landscapes]{svensson2012a,krug2021}.  Briefly, fitness landscapes offer a useful metaphor: we are attaching a scalar (fitness) to each state (genotype) and, in the simplest scenarios, evolution will move uphill in fitness (and very rarely downhill).

In fact, \citet[][p.~2]{crona_peaks_2013} connect sign epistasis directly to cancer progression models: ``A related field is the study of constraints for orders in which mutations accumulate (see e.g., Desper et al., 1999; Beerenwinkel et al., 2007a). It is well known that a drug resistance mutation is sometimes selected for, only if a different mutation has already occurred. Such a phenomenon requires sign epistasis. Indeed, if a particular mutation is beneficial regardless of background, then it can occur before or after other mutations.''. \citet{Misra2014} independently framed the problem addressed by cancer progression models in the context of sign epistasis and fitness landscapes, and their BML method ``(...) takes into account (...) unknown epistatic gene interactions, before inferring a probabilistic model for the accumulation of somatic mutations in a population of cancer cells.'' \citep[][p.~2456]{Misra2014}.

\begin{figure}[tbhp]
 \centering \includegraphics[width=14.0cm,keepaspectratio]{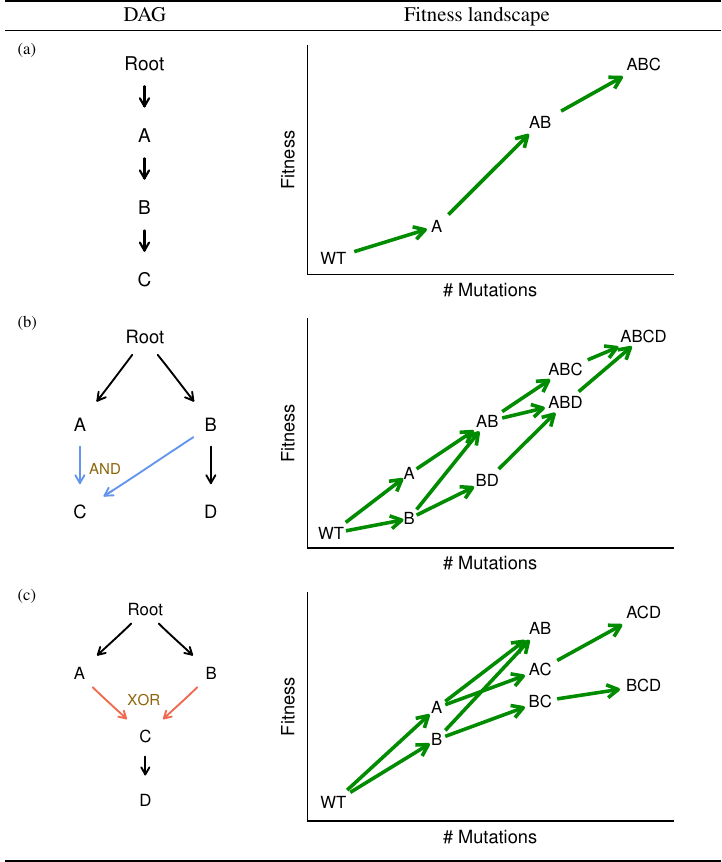}
 \caption{\textbf{DAGs of restrictions and representable fitness landscapes}. Three DAGs of restrictions and corresponding representable fitness landscapes; genotypes not shown are strongly deleterious.  All the genotypes that should exist according to the DAG are accessible genotypes when evolution moves uphill in fitness; in \citet{diaz-uriarte2018} these were called representable fitness landscapes. The DAGs encode sign epistasis relationships; for example, in (a) mutation in \textit{B} alone leads to a non-viable genotype, whereas a mutation in \textit{B} when \textit{A} is also mutated leads to an increase in fitness.}  \label{fig:representable}
\end{figure}

In this section, we will use fitness landscapes as a heuristic for evaluating how and why different methods can fail, using figures to support the reasoning (to navigate these figures, Figure \ref{fig:metafigure} shows a summary of the elements of each figure with an illustrative example of it). We will start by considering required links between fitness and accumulation (section \ref{sec:scenarios_no_rse}).
We then introduce reciprocal sign epistasis (section \ref{sec:rse}) and expand on the concepts of ``lines of descent'' (LOD) and ``path of the maximum'' (POM) (section \ref{sec:lod-pom-sampling}) and how they relate to  monotonic accumulation models. Next, we examine how models with stochastic dependencies (section \ref{sec:assumpt_stoch}) are affected by the previous issues, and finish discussing the problems that arise when there are two or more underlying fitness landscapes (section \ref{sec:heterogeneity}).

\subsection{Deterministic models as within-cell restrictions; no reciprocal sign epistasis}\label{sec:scenarios_no_rse}

\begin{figure}[!t]
  \centering
  \includegraphics[width=12.0cm,keepaspectratio]{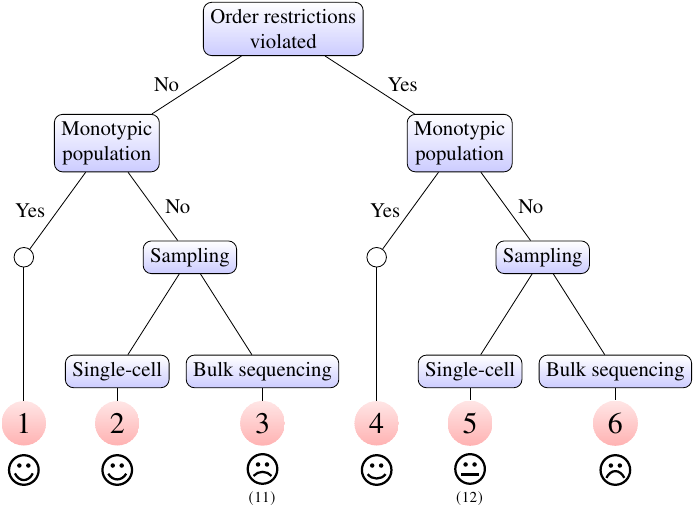}
 \caption{\textbf{Deterministic models as within-cell restrictions: scenarios with respect to order restriction violations, evolutionary dynamics, and sampling}. The numbers correspond to the scenario identified in the main text, section \ref{sec:scenarios_no_rse}. Order restrictions violated means that genotypes that are incompatible with the DAG (i.e., genotypes that should not exist if the model of the DAG were faithfully respected) are viable and, thus, have a non-zero probability of existing (see also Figure \ref{fig:order_restrictions_violated}). Branches that lead to nodes 1 and 4, which are monotypic, have no splits corresponding to sampling, since single-cell and bulk sequencing result in identical samples. ``Monotypic population'' means that for virtually all the time (except for very brief periods of fast clonal sweeps) there is a single genotype, defined with respect to the mutations being modelled. For example, if the driver mutation is \textit{A}, and from a clone with \textit{A} mutated arise two different clones, with passenger mutations \textit{P1, P2}, so that \{A,P1\} and \{A,P2\} have identical fitness and behaviour, we regard the population as monotypic even if it is composed of \{A,P1\} and \{A,P2\}. Even if \{A,P1\} arises from a clone that only had \textit{P1} and \{A,P2\} arises from a clone that only had \textit{P2}, we regard the population as monotypic, despite the clones acquiring \textit{A} independently from two different ancestors. Heterogeneity of the tumour types plays no role in this figure, as we regard it as a different problem (see \qpref{sec:heterogeneity}). Emojis provide a rough indication of how bad/good each scenario is; numbers in parentheses below some emojis indicate figures with additional details of the scenario.}
  \label{fig:scenarios}
\end{figure}

Consider that the \textbf{within-cell} order of accumulation of mutations follows deterministic restrictions rules that can be described by a single DAG. This allows us to differentiate between genotypes that should and should not exist, according to the DAG. All the genotypes that the DAG predicts will exist (e.g., sets of possible genotypes in Figs.~\ref{fig:ot-deps} to \ref{fig:hesbcn-deps}) should be accessible genotypes in a fitness landscape, in the sense that evolution, moving uphill in fitness, can go through those genotypes, as shown in Fig.~\ref{fig:representable}. This also means that all the genotypes that the DAG predicts will exist are genotypes that can appear in our samples and be part of the ``lines of descent'', which are the ``lineages that arrive at the most populated genotype at the final time'' \citep[][p.~572]{szendro_predictability_2013}. And we define as genotypes incompatible with the DAG those genotypes that do not respect the restrictions encoded by the DAG (for example, in Fig.~\ref{fig:representable} (a), genotypes \(\{B\}, \{C\}, \{B, C\}, \{A, C\}\); in (b) genotypes \(\{A, C\}, \{B, C\}, \{A, D\}\), among others; and so on); these genotypes would not exist if the DAG restrictions were faithfully respected.

Understanding that we are dealing with within-cell restrictions, Figure \ref{fig:scenarios} shows the six different scenarios that can result from the combination of sampling, evolutionary dynamics, and violation of order restrictions. By violation of order restrictions we mean that genotypes that are incompatible with the DAG (the genotypes that should not exist) are, however, present at low frequency. 
Reciprocal sign epistasis is a  different phenomenon discussed later (section \ref{sec:rse}). The six scenarios have the following consequences:

\begin{enumerate}
\item Ideal case: for large enough sample sizes, the correct DAG and rates/conditional probabilities should be recoverable.\label{scenario_best}

\item Second ideal case, though we can expect larger variability in the estimation of rates/conditional probabilities/DAG structure (arising from the sampling variability of the genotypes in the sample).\label{scenario_2nd}

\item \label{scenario_collapse}See Fig.~\ref{fig:collapse}.  Bulk sequencing can create mutational profiles that do not correspond to any existing genotype \citep{alves_multiregional_2017,luo2023}.  Under within-cell XOR restrictions we can observe mutational profiles (``tumour genotypes'') that violate the XOR and, thus, that are not possible under the DAG; this would lead to inferring incorrect restrictions. For other dependencies (AND, OR, forking single dependencies, linear dependencies), bulk sequencing can exacerbate the low representation of some cells in the population, increasing the probability of inferring the incorrect DAG. In all cases, the rates/conditional probabilities inferred will be biased: they will not match the true ones that correspond to transitions between clones because we underestimate the frequency of clones with fewer mutations and overestimate the frequency of clones with more mutations. These problems are discussed in more detail in section \ref{sec:collapse}.

\begin{figure}[tbph]
 \centering \includegraphics[width=1.00\linewidth,keepaspectratio]{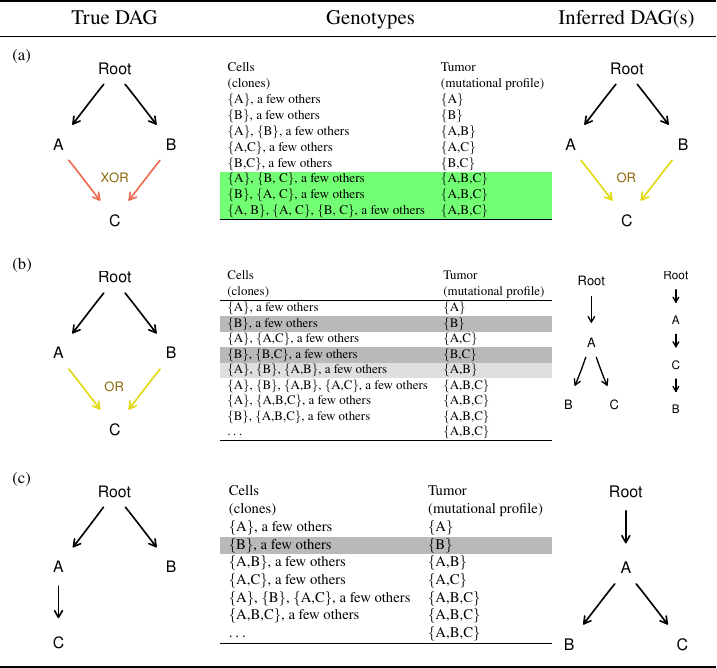}
 \caption{\textbf{Bulk sequencing and mutational profiles that do not correspond to any existing genotype: scenario \textbf{3} in Figure \ref{fig:scenarios}.}
   ``True DAG'': the true DAG within-cell that generates genotypes.   ``Genotypes'': the true  ``Cell'''s genotypes in the population possible under the DAG, and the ``Tumour'' genotype or mutational profile obtained from bulk sequencing; rows whose presence or absence leads to problems highlighted; ``a few others'': other possible genotypes, with other mutated genes, but with a frequency too small to be reported in the mutational profile.   ``Inferred DAG(s)'': one or more inferred DAGs from the mutational profile.
   In (a), bulk sequencing leads to observing genotype \{A,B,C\} (shown in green) which is not possible under an XOR, and if sufficiently common would lead to inferring the DAG under ``Inferred DAG(s)''.
   Only under (a) do we observe mutational profiles that are not compatible with the true DAG. In (b) and (c) (and similarly for AND and linear dependencies), bulk sequencing will not lead to observing genotypes incompatible with the DAGs.
   However, bulk sequencing can exacerbate the low representation of some cells in the population.
   In (b), if \{B\} and \{B,C\} (shown in dark grey) are infrequent enough, we will always see \textit{B} with \textit{A}, leading to the inference of the forking DAG; if there is also too small a frequency of \{A,B\} (shown in light grey), so that we tend to observe only \{A\}, \{A,C\}, \{A,B,C\},  the linear DAG will be inferred.
   In (c), too small a frequency of \{B\} (dark cray) will lead to inferring a DAG where \textit{B} depends on \textit{A}.
   For all of (a), (b), (c), as well as linear DAGs and DAGs with only AND, even if the true DAG is inferred, the transition rates inferred will not match those between genotypes: when we observe \{A,B\} we can have \{A\} and \{B\} and \{A, B\}, so we are underestimating the frequency of \{A\} and \{B\} and overestimating that of \{A,B\}. }
  \label{fig:collapse}
\end{figure}

\item For practical purposes, very similar to \ref{scenario_best}: violations of assumptions are largely inconsequential because the incompatible genotypes are  unlikely to appear in samples with large frequency (we are in a monotypic population ---see definition of monotypic in legend of Fig.~\ref{fig:scenarios}). If they do, error models should accommodate them (see next).\label{scenario_inconsequential}

\item \label{scenario_5} This scenario can lead to three different cases, shown in Fig.~\ref{fig:order_restrictions_violated}.

\begin{figure}[!t]
  \centering
\includegraphics[width=0.80\linewidth,keepaspectratio]{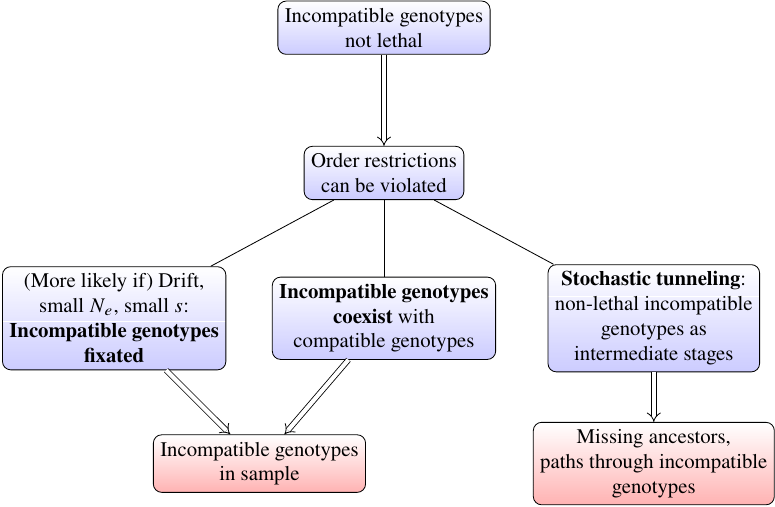}
 \caption{\textbf{Order restrictions violated for deterministic models: cause and consequences}. A simple scheme of the cause and two main consequences (and mechanisms) of violation of order restrictions. These consequences are relevant for the three scenarios on the right of Figure \ref{fig:scenarios}. The presence of small fractions of incompatible genotypes should be handled by the error models. Missing ancestors and paths through incompatible genotypes lead to incorrect estimation of rates, but should not affect structure (see Figure \ref{fig:tunnel}). More complicated scenarios related to tunnelling can appear when combined with reciprocal sign epistasis (Figure \ref{fig:undo}). See text for details. }
  \label{fig:order_restrictions_violated}
\end{figure}

  The first and second (occasional fixation of incompatible genotypes, occasional sampling of incompatible genotypes) should be handled by the error models (OT, OncoBN, and MC-CBN have error models, different among themselves, that explicitly accommodate observing genotypes that are not compatible with the model; H-CBN does not, but the observational error model can accommodate these departures too).

  The third (stochastic tunnelling: moving through low fitness incompatible genotypes ---\citealp{iwasa2004,Weissman2009,Weinreich2005}) is illustrated with examples in Fig.~\ref{fig:tunnel}. This case is also unlikely to be problematic: it should not lead to inferring the incorrect DAG, and should at most result in incorrect estimates of some rates/conditional probabilities (with error increasing with the frequency of incompatible genotypes in samples).\label{scenario_sample_incomp_tunnel}

\item \label{scenario_worst}This is arguably the worst scenario, as it combines \ref{scenario_collapse} with \ref{scenario_inconsequential} and \ref{scenario_sample_incomp_tunnel}. A possible workaround might be to try a deconvolution approach \citep{qin2020,chu2022,wang2020a} so that scenario \ref{scenario_worst} is turned into scenario \ref{scenario_5} (see also section \ref{sec:collapse}).

\end{enumerate}

\begin{figure}[!t]
 \centering \includegraphics[width=0.85\linewidth,keepaspectratio]{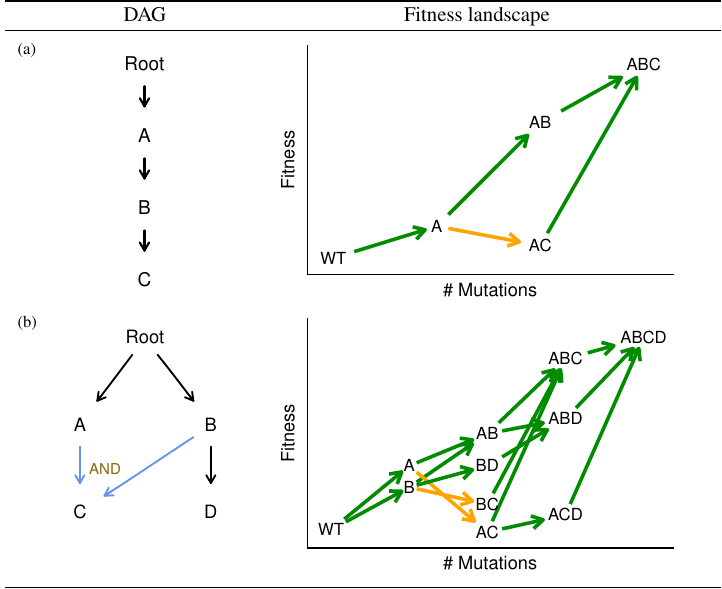}
 \caption{\textbf{Tunnelling and moving through incompatible genotypes.} Two DAGs of restrictions where some incompatible genotypes are only mildly deleterious, and evolution can move through fitness valleys, as in stochastic tunnelling \citep{iwasa2004,Weissman2009,Weinreich2005};  transitions of decreasing fitness shown in orange in the fitness landscape. In both (a) and (b), if the deleterious genotypes are infrequent, they will not lead to incorrect estimation of the DAG, but can result in underestimation of some rates (those for the orange arrows, and from the valley genotypes to other genotypes). Genotypes not shown are strongly deleterious.}
  \label{fig:tunnel}
\end{figure}

Standard evolutionary dynamics models are often described as strong selection, weak mutation (SSWM), strong selection, strong mutation (SSSM), weak selection, strong mutation (WSSM) \citep{Desai2007,sniegowski_beneficial_2010,krug2021}. The six scenarios above show why SSWM is so convenient: in the absence of violations of restrictions, we are in scenario \ref{scenario_best} (no tunnelling is possible since violations of restrictions are lethal). Under SSSM or WSSM we would be in scenarios \ref{scenario_2nd} or \ref{scenario_collapse}. SSWM plus violation of restrictions would lead to  \ref{scenario_inconsequential}, but SSSM or WSSM plus violation of restrictions would lead to \ref{scenario_sample_incomp_tunnel} or \ref{scenario_worst}. As emphasised by \citet[][p.~1181]{Weinreich2005}, polymorphism can play a key role in evolutionary dynamics even under SSWM (see also \citealp{Weissman2009}, who show that valley crossing can be  common). Theoretical work could help get a more nuanced understanding of how likely each of these six scenarios is and the precise extent of their effects on the inferences from monotonic accumulation models\footnote{\citet{Misra2014} present in their Supplementary Material, for three scenarios involving two genes, a detailed examination of their BML approach for estimating ``evolutionary probabilities'' under the stochastic model of \citet{Attolini2010a}.}.

On the other hand, and if we insist on an interpretation of bulk-inferred restrictions as within-cell restrictions, most cancer data still fall under \ref{scenario_collapse} and \ref{scenario_worst}. A key difference between scenario \ref{scenario_best} on the one hand and scenarios  \ref{scenario_collapse} and \ref{scenario_worst} on the other, is that in these last two an increase in sample size will not improve our inferences much, since there is a fundamental mismatch between what we are trying to infer and what the methods are giving us, for the data at hand. That  bulk sequencing can create serious problems for cancer progression models was already discussed by \citet{Sprouffske2011}; the cases they examine, however, are not identical to our scenarios \ref{scenario_collapse} or \ref{scenario_worst}, since they allowed different clones to evolve along possibly different paths, so no single model is necessarily enough to describe the process (see also \qpref{sec:heterogeneity}).

\subsection{Bulk sequencing and the difficulties of inferring XOR relationships}\label{sec:collapse}

The difficulty of correctly inferring XOR relationships has already been mentioned by \citet[][]{angaroni2021}. However, the problem they discuss and the suggested cause (``This result is likely related to the presence of spurious dependencies among events due to the properties of the XOR logical formula''\footnote{Incidentally, we do not understand the meaning of ``spurious dependencies'': is this intended to mean an observed association that does not reflect a causal dependency between events? we subscribe here to the idea in \citet[][section 7.1]{hernan2020} that there is nothing spurious about non-causal associations: they are real associations even if they cannot be given a causal interpretation. }, \citealp[][p.~758]{angaroni2021}) is different from the phenomenon we are discussing in Fig.~\ref{fig:collapse} and scenario  \ref{scenario_collapse}: they observe the difficulty of inferring XOR using simulations that use a generative model where XORs exist at the level of whichever entity their simulations are simulating (see also section \ref{sec:why-entities}). In contrast, what we are discussing here is that bulk sequencing can result in the within-cell XOR being undetectable by bulk sequencing. This is the same phenomenon discussed by \citet[p.~3]{kuipers2021}: ``Intra-tumour co-occurrence and exclusivity patterns may differ substantially from the patient level ones. For example, if two genes are strictly mutually exclusive at the patient level, they never both occur in the same or different clones and cannot have subclonal interactions. On the other hand, if genes are exclusive at the clone level and only appear in different subclones, they can still both occur and produce co-occurrence at the patient level.''

A possible workaround for bulk sequencing might be to use a deconvolution approach \citep{qin2020,chu2022,wang2020a} so that scenario 3 is turned into scenario 2\footnote{Or, alternatively, but that is outside the scope of this ms., so that each subject provides multiple observations that are then analysed with methods such as TreeMHN \citep{luo2023}, Hintra \citep{khakabimamaghani2019}, or REVOLVER \citep{caravagna2018}}. We are not aware of anyone having done this with CPMs and monotonic accumulation models; this procedure raises some additional methodological questions (e.g., how to properly account for the  uncertainty in the deconvolution and the estimation of the cell fractions), but this might be a fruitful research program.

It should be noted, though, that the problems in Fig.~\ref{fig:collapse}  can appear  even under scenario \ref{scenario_best}: for example if $\lambda_A \gg \lambda_B, \lambda_C \gg \lambda_B$ it is easy to reproduce the patterns in (b) and if $\lambda_A \gg \lambda_B$ those in (c), as it will be rare to sample \{B\}, \{B,C\}, and \{A,B\} or \{B\} cells, respectively.

\subsection{Reciprocal sign epistasis (RSE)}
\label{sec:rse}

\begin{figure}[tbhp]
 \centering \includegraphics[width=0.85\linewidth,keepaspectratio]{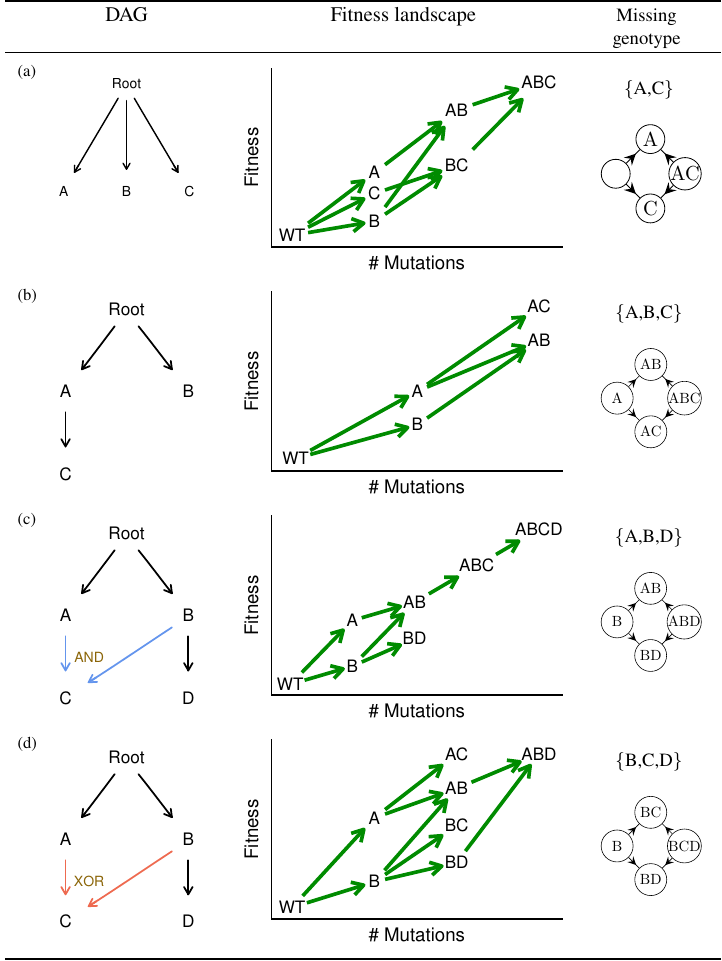}
 \caption{\textbf{Reciprocal sign epistasis.} In each row we show a best-approximation DAG of restrictions to a fitness landscape that cannot be fully represented by a DAG because of reciprocal sign epistasis (these are non-representable fitness landscapes in the sense of \citealp{diaz-uriarte2018}). In the examples, reciprocal sign epistasis (RSE) turns some genotypes, which should exist according to the DAG, into non-viable genotypes, which are therefore missing from the landscape and the set of observable genotypes. These are indicated in the right-most column, with the notation \{A,B,C\} to mean the genotype with genes A, B, and C mutated, as explained in section \ref{sec:notation-genots}. If the missing genotypes were accessible, the fitness landscapes would be representable by the DAG.  Figure insets under the missing genotypes show the reciprocal sign epistasis relationships using figures as in \citet[][p.~3]{crona_peaks_2013} where an arrow denotes increasing fitness ---in contrast to \citet{crona_peaks_2013}, to facilitate comparisons with the DAGs and fitness landscapes, the non-mutated gene is not shown---; note that in (b), (c), (d), the epistasis is between B and C with A as background, A and D with B as background, and C and D with B as background, respectively. In (a), RSE does not lead to local maxima.
}
  \label{fig:rse}
\end{figure}

The violations of restrictions presented in section \ref{sec:scenarios_no_rse} result in the observation of genotypes that are incompatible with the model (Fig.~\ref{fig:order_restrictions_violated}). In contrast, under reciprocal sign epistasis (RSE), genotypes (and paths) that should exist in the fitness landscape under the model might not exist. RSE \citep{crona_peaks_2013, poelwijk_empirical_2007, poelwijk_reciprocal_2011} is a genetic interaction where two mutations that individually increase fitness, reduce fitness when combined (or two mutations that individually decrease fitness, increase fitness when combined). In fitness landscapes where back mutation is possible, RSE is a necessary condition for multipeaked fitness landscapes \citep{poelwijk_reciprocal_2011}. Simple examples are presented in Fig.~\ref{fig:rse}.

DAGs of restrictions can encode sign epistasis, as discussed above (section \ref{sec:scenarios_no_rse}; see also \citealp[][p.~2]{crona_peaks_2013}; \citealp[][pp.~2457, 2461]{Misra2014}). But RSE leads to fitness landscapes that cannot be represented by any DAG because there will always be genotypes that should be present under a given DAG, possibly in large frequencies, that are not present because they have low fitness. For this reason, these fitness landscapes were called non-representable fitness landscapes in \citet{diaz-uriarte2018}. The problems originally described concerned DAGs with AND restrictions, but the problems also affect DAGs with OR and XOR restrictions (e.g., Fig.~\ref{fig:rse}(d)) and models with stochastic dependencies that only consider low-order (e.g., pairwise) interactions.

If we add RSE to Fig.~\ref{fig:scenarios} we have three new scenarios to consider, all with RSE: (a) a monotypic population; (b) a non-monotypic population from which we  take single-cell samples; (c) a non-monotypic population from which we take bulk sequencing samples. The last one is the case examined in \citet{diaz-uriarte2018}, where it was shown that RSE can have very detrimental effects on the performance of CPMs. These issues are due to the RSE + bulk sequencing, without the additional factor of tunnelling\footnote{Less than 1 in a 1000 of the simulations had a non-accessible genotype in the ``lineages that arrive at the most populated genotype at the final time'' (line of descent, or LOD) or in the
``time ordered sets of genotypes that at some time contain the largest subpopulation'' (path of the maximum or POM); definitions of LOD and POM from \cite{szendro_predictability_2013}.}. We mentioned above (\qpref{sec:scenarios_no_rse}, item \ref{scenario_sample_incomp_tunnel}) that the presence in a sample of a few genotypes not predicted by a model should not be a major problem, as those can easily be accommodated by the error models. With RSE the main problem are the missing genotypes: trying to choose the underlying model is difficult because using DAGs there is no such thing as a correct model.

\begin{figure}[tbhp]
 \centering \includegraphics[width=0.85\linewidth,keepaspectratio]{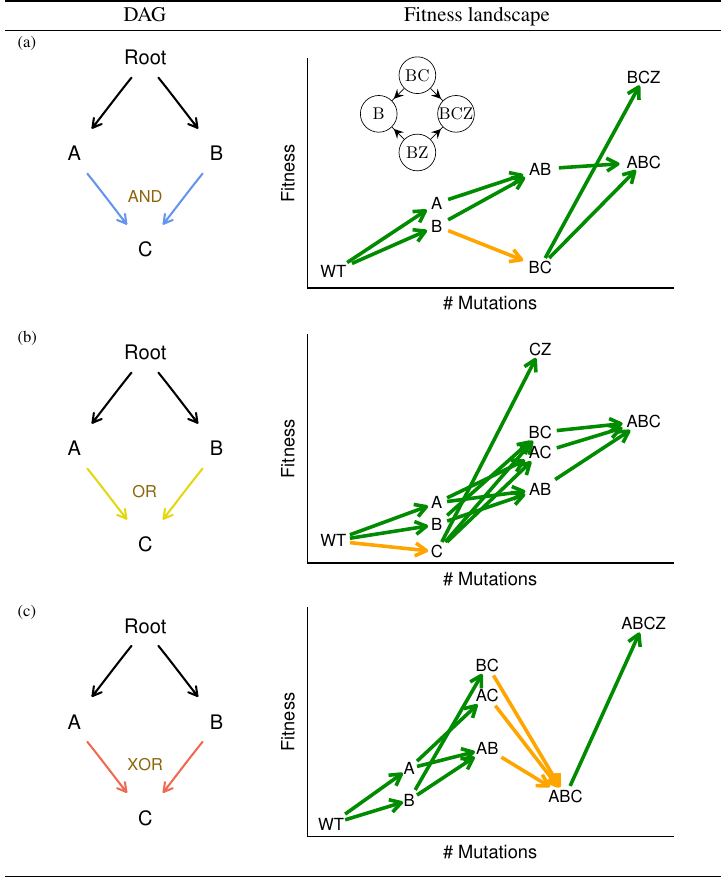}
 \caption{\textbf{Reverting or ``undoing'' restrictions.} In each row we show a DAG of restrictions and a fitness landscape that cannot be represented by a DAG because a mutation in gene Z reverts some of the restrictions.
      (For a particular data set generated under these fitness landscapes, where Z is placed in the DAG will depend on the frequencies of the different genotypes, but none of the landscapes are representable.)
   These examples are also cases of reciprocal sign epistasis (this can easily be seen applying the criterion in \citealp[p.~3 of][]{crona_peaks_2013}, as in Fig.~\ref{fig:rse}; this is shown as an inset in (a)). Some paths in the fitness landscapes require going through genotypes in shallow fitness valleys, by moving downwards in fitness, as indicated by the orange arrows; this might only be possible in reasonably large populations ---see details and references in \citet{Weissman2009}. Genotypes not shown are strongly deleterious.
 }
  \label{fig:undo}
\end{figure}

RSE is a biologically relevant phenomenon: it is  a necessary condition for multipeaked fitness landscapes under reversible mutations \citep{poelwijk_reciprocal_2011}, and it is the cause of the so-called ``synthetic lethality'', where several mutations that individually are tolerable (or even beneficial) are lethal when they co-occur. Synthetic lethality is common in both cancer cells and the human genome and is often sought for targeted molecular therapy in cancer \citep{blomen_gene_2015, beijersbergen2017,leung_synthetic_2016}. RSE is also an interesting phenomenon because, when thinking about DAGs, it can ``undo'' or revert restrictions, as shown in Fig.~\ref{fig:undo}; panels (a) and (b), for example, show how \textit{Z} reverts the mildly deleterious effects of \textit{C}, with \textit{Z} itself being strongly deleterious. Together, what these figures show is that DAGs (with AND, OR, XOR) might suffer from limited expressive power to represent relationships that could be of high relevance, whereas models such as HyperHMM and HyperTraPS-CT could capture these. Note, however, that we are not suggesting that only models capable of reflecting arbitrary-order dependence will have sufficient expressive power: strong-effect high-order epistatic relationships, involving more than three genes, are probably uncommon \citep{weinreich_should_2013}, and high-order interactions probably will show diminishing effects as seen with  global epistasis \citep{diaz-colunga2023}.

\begin{figure}[tb!]
 \centering \includegraphics[width=1.0\linewidth,keepaspectratio]{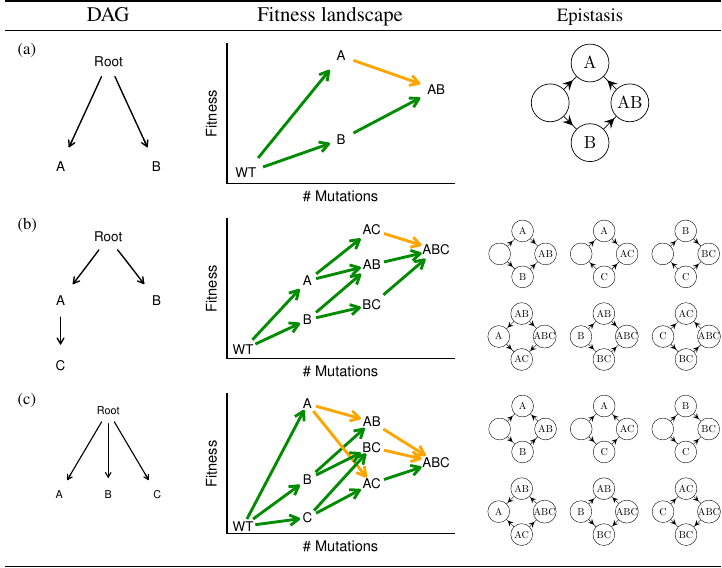}
 \caption{\textbf{Local maxima without reciprocal sign epistasis, and reciprocal sign epistasis and representable landscapes.}
   In each row we show the DAG, the fitness landscape, and epistasis relationships using figures as in \citet[][p.~3]{crona_peaks_2013} where an arrow denotes increasing fitness ---but, for clarity, the non-mutated gene is not shown.
   (a): two local fitness maxima (\{A\}, \{A,B\}) without reciprocal sign epistasis (as seen on the right, there is sign epistasis between A and B). (b): two local fitness maxima (\{A,C\}, \{A,B,C\}) without reciprocal sign epistasis (there are three instances of sign epistasis).
   (c): reciprocal sign epistasis but all genotypes predicted by the DAG are accessible (there is one instance of reciprocal sign epistasis and four of sign epistasis; there are four local fitness maxima).
   In all cases, though, paths that should be available according to the DAG are not available; transitions of decreasing fitness shown in orange in the fitness landscape.
   Genotypes not shown are strongly deleterious.
}
  \label{fig:rse-complications}
\end{figure}

Before leaving the discussion about RSE, it is important to point out that in models such as those of monotonic accumulation of events, we can have fitness landscapes with local maxima without RSE, as shown in Fig.~\ref{fig:rse-complications}(a), (b) (see additional discussion in \citealp[][S2 Text]{diaz-uriarte2019a}). In addition, the consideration of whether a fitness landscape is representable or not in \citet{diaz-uriarte2018} focuses only on accessible genotypes. However, it is possible to have RSE and a representable landscape, as shown in Fig.~\ref{fig:rse-complications}(c). In cases like this, RSE gives rise to local fitness maxima and eliminates some paths (that should exist according to the DAG) between genotypes, but does not turn any genotype predicted by the DAG into inaccessible (see additional discussion in \citealp[][Supplementary Material]{diaz-uriarte2018}).

\subsection{Lines of descent (LOD), path of the maximum (POM), bulk sequencing}
\label{sec:lod-pom-sampling}

\begin{figure}[t!]
 \centering \includegraphics[width=\linewidth,keepaspectratio]{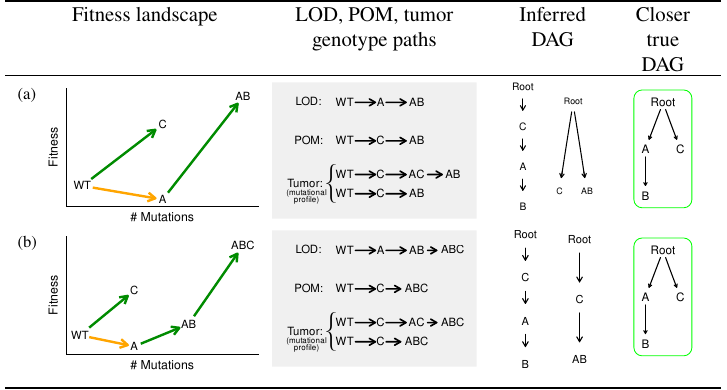}
 \caption{\textbf{Fitness landscapes, lines of descent (LOD), path of the maximum (POM), and bulk sequencing.} In each row we show a non-representable fitness landscape. Next, the LOD, the sequence of unique states of the POM, and paths observable in the tumour with bulk sequencing. LOD: line of descent, ``lineages that arrive at the most populated genotype at the final time''; POM: path of the maximum: ``time ordered sets of genotypes that at some time contain the largest subpopulation''; LOD and POM definitions from \citet[][p.~572]{szendro_predictability_2013}. See Fig.~\ref{fig:lod_pom_histogram} for example genotype frequencies through time that would yield the LOD and POM shown in panel (a).  The tumour paths correspond to the mutational profiles or the observed  ``Tumour genotypes'' and their sequence, under bulk sequencing using  different thresholds for detection of rare mutations. The third column shows the DAGs that would be inferred from the bulk sequencing samples (the leftmost DAG corresponds to the first path under ``Tumor''); we are here inferring a single DAG, but which one is inferred will depend on the actual frequencies of the different genotypes (in contrast to the situations depicted in Fig.~\ref{fig:heterog}, where multiple DAGs are inferred). The last column is the DAG closer to the truth that could be inferred, in the sense that it reflects the true dependencies between mutations; but this DAG cannot fully capture all the genotypes because of reciprocal sign epistasis (e.g., in the second row, genotypes \{A,C\} and \{A,B,C\} are not viable), so these fitness landscapes are non-representable, a problem introduced in Figure \ref{fig:rse}. The DAG has been enclosed in a green rectangle to make it easier to differentiate it from the inferred DAGs. In these figures we assume stochastic tunnelling \citep{iwasa2004,Weissman2009,Weinreich2005}, so evolution arrives at the highest fitness peak in the landscape crossing the fitness valley, but the genotypes in the valley are rare and never become part of the POM.
   Genotypes not shown are strongly deleterious. }
  \label{fig:lod_pom}
\end{figure}

\begin{figure}[t!]
  \centering
  \includegraphics[height=6cm,keepaspectratio]{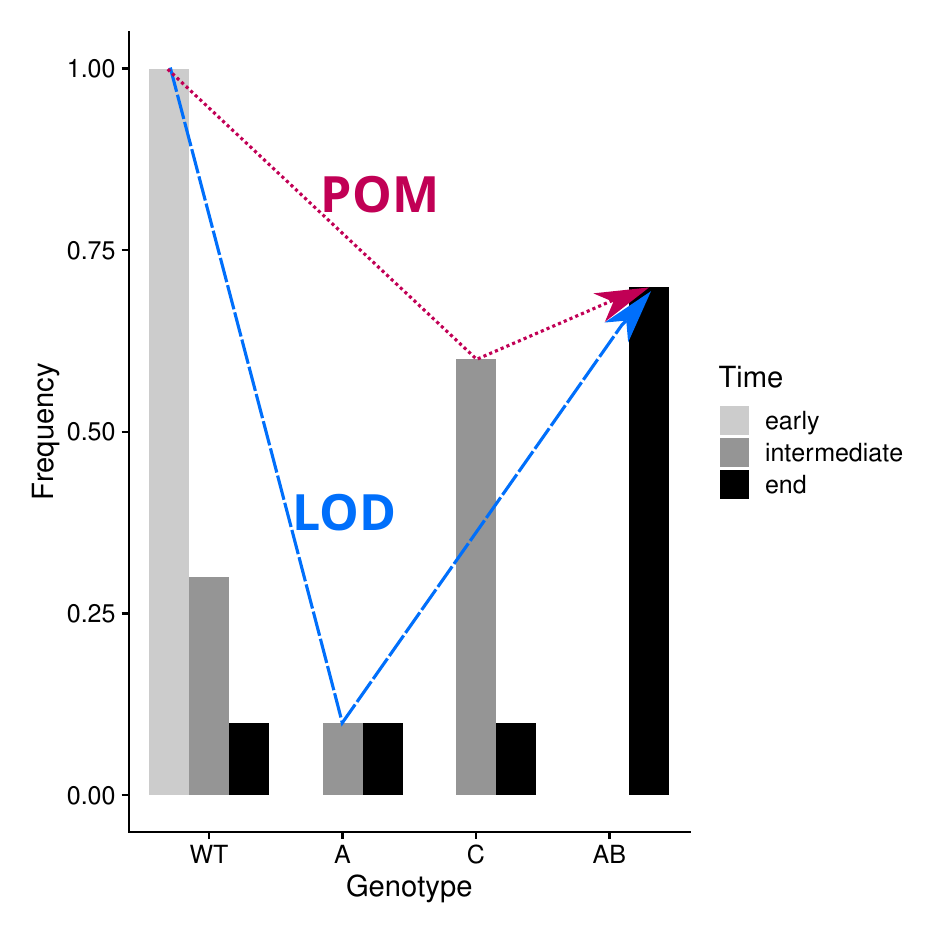}
  \caption{\textbf{Lines of descent (LOD), Path of the maximum (POM): example genotype frequencies}.  Possible genotype frequencies corresponding to the early, intermediate, and end time points that would yield the LOD and POM shown in the centre panel of  Fig.~\ref{fig:lod_pom}(a). We start with a population where \{WT\} has a frequency of 1; at the intermediate time, frequencies of genotypes are 0.6, 0.3, and 0.1 for genotypes \{C\}, \{WT\}, and \{A\}, respectively; at the end time shown, genotype frequencies are 0.7 for  \{A,B\}, and 0.1 for each of \{WT\}, \{A\}, and \{C\}.
LOD: line of descent, “lineages that arrive at the most populated genotype at the final time”. Here, this lineage is \(WT \rightarrow A \rightarrow AB\), as  \{A,B\} is the final most populated genotype, and we arrive at \{A,B\} through \{A\}. POM: path of the maximum: “time ordered sets of genotypes that at some time contain the largest subpopulation”. Here, these sets are \{WT\}, \{C\}, \{A,B\}, as each of these genotypes forms the largest subpopulation at some observed timepoint. LOD and POM definitions from \citet{szendro_predictability_2013}.}\label{fig:lod_pom_histogram}
\end{figure}

The discussion of challenges and difficulties in the sections above (\ref{sec:scenarios_no_rse}, \ref{sec:collapse}, \ref{sec:rse}) have implicitly assumed that we are interested in inferring what are called ``lines of descent'' (LOD): ``lineages that arrive at the most populated genotype at the final time'' \citep[][p.~572]{szendro_predictability_2013}; see Figures \ref{fig:lod_pom} and \ref{fig:lod_pom_histogram}. The restrictions/dependencies we care about are those that pertain to the genotypes in the LOD. And for designing interventions, the transitions that matter are those estimated from the LOD.

A seemingly possible alternative would be to focus on the ``path of the maximum'' (POM): ``time ordered sets of genotypes that at some time contain the largest subpopulation'' \citep[][p.~572]{szendro_predictability_2013} (see Fig.~\ref{fig:lod_pom_histogram}). This, however, is unlikely to be what we want: (a)  substitutions along the POM can involve multiple mutations (even if the underlying true substitutions in cells happen one-by-one); (b) what is changing along the POM are not clones or genotypes with a parent-descendant relationship, so modelling dependencies that pertain to the POM is not equivalent to modelling within-cell restrictions; (c) as a consequence, the POM does not seem to be what is most relevant for guiding interventions.

Therefore, LODs will be the focus of our models when we think about restrictions \textbf{within cells}. Contrasting LODs and POMs, which are both made of clones, with mutational profiles (which collapse over clones), can be useful to clarify different scenarios, including those involving bulk sequencing and tunnelling. The models we build for LODs and POMs are different. For example, in Fig.~\ref{fig:lod_pom}(b), when we focus on the LOD we want to estimate $P(A|WT)$ and $P(AB|A)$, even if \{A\} rarely, or never, appears in the sample.  It is what happens in the future (what eventually becomes fixated or becomes the largest population) that determines what estimates of what happened in the past are relevant or not; \citet{Misra2014} use a very similar idea when they estimate \(P(g)\), the ``\textit{evolutionary probability} of genotype \(g\)''\footnote{\(P(g)\), the \textit{evolutionary probability} of genotype \(g\), is defined as ``the sum of path probabilities for every mutation path from the normal genotype that passes through g and ends as a tumor genotype.'' (p.~2456).  In p.~2457 they  explain that \(P(g)\) ``(...) depends on the fitness landscape over all ancestral cell states traversed during somatic evolution, as well as the details of evolutionary dynamics of cell populations.''; see also their Supplementary Material.}. In contrast, if we focus on the POM, both of $P(A|WT)$ and $P(AB|A)$ are 0. The criterion of the LOD, when we estimate $P(A|WT)$, is not necessarily ``how likely is it that A will become fixated'' but, rather, ``how much does the transition  from WT to A contribute to what eventually becomes fixated''. Under SSWM and very rare tunnelling, LOD and POM coincide and, thus, ``how likely is it that A will become fixated'' and ``how much does the transition  from WT to A contribute to what eventually becomes fixated'' are identical questions.

Under bulk sequencing, the mutational profiles we observe depend on detection thresholds, and these mutational profiles  can be neither part of the LOD nor the POM. For example, in Figure \ref{fig:lod_pom}, under large detection thresholds, the acquisition of A and B will appear as a single event and, thus, many of the inferred DAGs will not tell the two apart. In this figure, the closest DAGs to the truth either ignore C or, as shown in the figure, add C, correctly reflecting that A and B do not depend on C.  But these DAGs cannot represent the fitness landscape because genotypes predicted under the DAG are not possible under the fitness landscape (\{A,C\} and \{A, B, C\} in (b), and \{A, C\} in (c)). Therefore, these DAGs would be of questionable use for designing interventions. Moreover, the estimates of $P(X|Y)$ need not match those under the LOD or the POM (as we mention later,
section \ref{sec:what-if-bulk-seq}, this problem is similar to the ecological inference problem).

Finally, it is also possible to understand some of the problems with RSE in this context. With non-representable fitness landscapes, the DAG says there are, in the landscape, paths to the LOD that go through certain genotypes. But in the true fitness landscape there are no such paths because those genotypes (the ones that cannot exist due to RSE) cannot be part of the LOD.

\subsection{How do the above scenarios affect models with stochastic dependencies: HyperTraPS, MHN, HyperHMM}\label{sec:assumpt_stoch}

The starting points in section \qref{sec:scenarios_no_rse} were fitness landscapes and sign epistasis. With MHN, HyperTraPS, and HyperHMM, there is no guarantee that there is a well-defined mapping from the dependencies learned by these methods to a single fitness landscape\footnote{Note that none of the original papers of those methods have framed their model in the context of fitness landscapes.}, as explained also in \citet[][section 1.7 of the Supporting
Information]{diaz-colunga2021}. If we tried to interpret these models in terms of fitness landscapes:

\begin{enumerate}
\item \textbf{The implicit fitness of a genotype could depend on the order in which mutations were acquired.}\label{fl_stoch_path} This is a consequence of the possibly asymmetrical effects of genes on each other: $\theta_{ij} \ne \theta_{ji}$ in eq.~\ref{eq:mhn} and \ref{eq:hypertraps}.  This is at odds with how the different kinds of epistasis are defined \citep{crona_peaks_2013, weinreich_perspective_2005-1}: comparing the fitness of different genotypes, regardless of how the mutations were acquired. In a ``standard'' fitness landscape, there is a fitness value for each genotype, but for MHN/HyperTraPS there can be multiple implicit fitness values for a genotype, fitnesses that differ depending on the order in which mutations were acquired.
  Therefore, for a fitness landscape for $L$ loci there are not $2^L$ possibly different implicit fitness values but up to $L!$ different values.

  Can we interpret these possibly multiple implicit fitness values per genotype as  fitness variability due to, for example, variation in the environment, or variation in the fitness landscape over time, or phenotypic plasticity \citep{frank_nonheritable_2012,cooper2012}? It does not seem so: the key issue here is not that there is fitness variability for a genotype, but rather that the fact that there are different implicit fitnesses is a consequence of the different orders in which mutations were acquired.

  Fitness values that depend on the path could be useful, though, to represent cases where the fitness of a genotype can genuinely be affected by the order in which mutations were acquired; this seems to be an uncommon phenomenon, but has been reported in cancer \citep{Ortmann2015}.

  For a direct interpretation of MHN and HyperTraPS in terms of ``standard'' fitness landscapes and epistasis, $\theta_{ij}$ should equal $\theta_{ji}$\footnote{We also need to decide how to relate the $\Theta$s (using now the capitalised parameter, as it simplifies the expressions below ---see eq.~\ref{eq:mhn}) to fitness.\\
    A simple choice is to make $W_{i}$ (the fitness of genotype \textit{i}) be (proportional to) the product of the $\Theta$'s. For example, $W_{AB} = W_{WT}\ \Theta_{AA} \Theta_{BB} \ \Theta_{BA}$ if the path was $WT \rightarrow A \rightarrow AB$. When $\Theta_{AB} = \Theta_{BA}$ fitness does not depend on path. However, transition probabilities, $P(i \rightarrow j)$ have to be computed as $\frac{W_j}{\sum W_g}$ instead of the more common  (\citealp[e.g.,][p.~14]{nichol2015}; \citealp{weinreich_darwinian_2006}) $\frac{W_j-W_i}{\sum W_g - W_i}$. The last expression corresponds to $\frac{s_{i \rightarrow j}}{\sum s_{i \rightarrow g}}$, where $s_{i \rightarrow j}$ is the relative increase in fitness, or the  ``selection coefficient for the mutation that carries allele i to allele h'' \citep[][Supplementary Material, p.~4]{weinreich_darwinian_2006}. Using $P(i \rightarrow j) = \frac{W_j}{\sum W_g}$ with the above fitness definition also implies that a transition from a genotype to a genotype of lower fitness is possible (see also main text) and, thus, all one-step transitions are possible.\\
    Other intuitive mappings of functions of $\Theta_{ij}$ to fitness will not work. In particular, setting  $W_{AB} = W_{WT}\ (1 + \Theta_{AA})\ (1 + \Theta_{BB} \ \Theta_{BA})$ if the path was $WT \rightarrow A \rightarrow AB$ has a nice interpretation as ``$s_{A \rightarrow AB} = \Theta_{BB} \ \Theta_{BA}$'': the addition of \textit{B} to a genotype with \textit{A} results in  a (relative) increase in fitness of $\prod\limits_{x_j = 1} \Theta_{ij}$ in  eq.~\ref{eq:mhn}. But it does not lead to fitness of $W_{AB}$ being independent of path when $\Theta_{AB} = \Theta_{BA}$ (unless $\Theta_{AB} = \Theta_{BA} = 1$). \\
    Setting  $W_{AB} = W_{WT}\ (1 + \Theta_{AA})\  (1 + \Theta_{BB}) \ (1 +\ \Theta_{BA})$ if the path was $WT \rightarrow A \rightarrow AB$ yields path independence when $\Theta_{AB} = \Theta_{BA}$, but no simple functions of $W_i$ for $P(i \rightarrow j)$, neither $\frac{W_j}{\sum W_g}$ nor $\frac{W_j-W_i}{\sum W_g - W_i}$, yield $\frac{\prod \Theta}{\sum \prod \Theta}$.}. This would require modifying the methods and their code (see also \qpref{sec:simul_based_inf}). Symmetrising the coefficients, however, might not be what we want to do if we drop the requirement of endowing the model with a direct interpretation in terms of standard fitness landscapes, as argued in \qref{sec:what-if-bulk-seq}.

\item \textbf{All genotypes would be accessible, since all one-step transitions are possible.}\label{fl_stoch_access}
  This precludes modelling inaccessible/lethal genotypes regardless of evolutionary regime, and can make it much harder to reason about different cases. It could be argued that:
  \begin{enumerate}[label=\alph*)]
  \item This is of little practical consequence, as it is possible to make a transition probability arbitrarily small (make the $\theta \ll 0$). In fact, this is how \citet[][Supplementary Material]{schill2020} use MHN as a stochastic approximation of CBN.
  \item Tunnelling and moving through fitness valleys \citep{iwasa2004,Weissman2009,Weinreich2005} are almost always a possibility, even if small. However, having all genotypes be implicitly accessible can still make reasoning harder, and can conflate the fitness of a genotype with what happens under different evolutionary regimes.

  \item If one really insists, it might be possible to modify the models to make some genotypes really inaccessible, but this requires modifying the code (see also \qpref{sec:simul_based_inf}), and could have undesirable consequences for the current fitting algorithms. The new versions of HyperTraPS will make this much easier by imposing priors on individual parameters.
  \end{enumerate}

\end{enumerate}

Now, if we go to Fig.~\ref{fig:scenarios}, only the left branch from ``Order restrictions violated'', ``No'', is possible since, as mentioned right above, all genotypes are accessible. Thus, we are left only with scenarios \ref{scenario_best}, \ref{scenario_2nd}, and \ref{scenario_collapse}, depending on evolutionary regime \citep{Desai2007,sniegowski_beneficial_2010,krug2021}.

Let us walk through the remaining figures that covered possible problems for deterministic models:

\begin{itemize}

\item Fig.~\ref{fig:collapse}: bulk sequencing issues. This can affect stochastic models. Interestingly, models with higher order dependencies, such as HyerTraPS or MHN extended to use $L^3$, $L^4$, \ldots, or the model in HyperHMM (which models transitions directly, without restrictions of order of interaction) would be impacted more severely. For example, think about the value of  $\theta_{C,\{A,B\}}$, the effect on \textit{C} of having both \textit{\{A,B\}}. Are the \{A,B,C\} we see (the mutational profile) mostly true \{A,B,C\} clones, or are they \{A,B\}, \{A,C\}, \{B,C\} clones that give a mutational profile of \{A,B,C\}?

  If we want to enhance the interpretation of these models in terms of interactions occurring within cells, it will  be worth examining what is the effect of bulk sequencing in the estimates of the $\theta$s with different sampling schemes and different parameterisations ($L^2, L^3, L^4, \ldots, 2^{L-1}$); modifying the likelihood function to account for the fact that we are observing a union of individuals should also be possible.

\item Fig.~\ref{fig:order_restrictions_violated} and Fig.~\ref{fig:tunnel}, order restrictions and tunnelling, are not relevant: we said above that these models do not preclude any transition and, in addition, it is difficult to interpret them as ``standard'' fitness landscapes.

\item Fig.~\ref{fig:rse}, Fig.~\ref{fig:undo} and Fig.~\ref{fig:rse-complications}, sign epistasis and landscape structures,  are examples of insufficient expressive power, which also affect stochastic dependencies models, arguably even the $2^{L-1}$.

  \begin{itemize}
  \item Because of the possible conflation of fitness with evolutionary regime (see above), it can be hard to tell if a genotype is a local maximum. For example, consider genotypes \{A\} and \{B\} and $\Theta_{A,A} > 1,\  \Theta_{B,B} > 1,\ \Theta_{A,B} \ll 1,\  \Theta_{B,A} \ll 1$. Are \{A\} and \{B\} local maxima from which evolution can occasionally escape via a downwards in fitness move through \{A,B\}, or is \{A,B\} a genotype upwards in fitness to which we rarely move?  To try to differentiate both options it seems we would need to know also what are other possible movements from \{A\} and \{B\} and their associated $\Theta_{C,A}, \Theta_{C,B}$, etc. But the usual epistasis criteria do not require us to do this.

    A way of solving this problem by making it explicit we are modelling a local maximum would be to add an ``END'' state, to which we can transition from each genotype with a certain probability (this was done in \citealp[][p.~7]{diaz-colunga2021}, as a way to try to accommodate local maxima in MHN, H-CBN and MC-CBN). This requires modifying the code of the fitting algorithms (see also \qpref{sec:simul_based_inf}).

  \item Some models that are easily expressed with DAGs can not be represented by stochastic dependencies. For example, a model where acquiring a mutation in gene \textit{D} depends on having all of \textit{A, B, C}. \citet[][Supplementary Material]{schill2020} show how to approximate a dependency of one event on three other events; but notice that, as explained there, ``the event becomes more and more likely the more parent events have occurred'', which is not exactly what we are trying to model. And this is not easier with OR or XOR operators.

    This problem, however, does not affect HyperHMM or HyperTraPS-CT, since these methods can reflect that all transitions to a state \textit{D} are zero except those from a state that has all of \textit{A, B, C} (see also \qpref{sec:heterogeneity}).

  \item Effects that depend on specific combinations of three or more genes cannot be accommodated by $L^2$ models. This includes reciprocal sign epistasis (and negative effects of RSE on the performance of MHN have been reported in \citealp{diaz-colunga2021}). However, this does not necessarily mean we will always want to use models of order $2^{L-1}$ (see above for comments above about the conflation of fitness with evolutionary regime, as well as strong-effect high-order epistatic relationships involving more than three genes being probably uncommon ---\citealp{weinreich_should_2013}--- and with diminishing effects as shown by global epistasis ---\citealp{diaz-colunga2023}). HyperTrapS-CT allows implementation and comparison of different model structures.

  \end{itemize}

\item Fig.~\ref{fig:lod_pom}, lines of descent vs.\ path of the maximum, is also relevant for stochastic dependency models, as the differences between LOD, POM, and mutational profile, exist regardless of the dependency model, but are hidden by bulk sequencing. A model of stochastic dependencies of the mutational profile is unlikely to correspond to the stochastic dependencies in the LOD (see section \ref{sec:lod-pom-sampling}).

\end{itemize}

\subsection{Mixtures of models, heterogeneity, and ``mutual exclusivity''}\label{sec:heterogeneity}

\begin{figure}[t!]
 \centering \includegraphics[width=\linewidth,keepaspectratio]{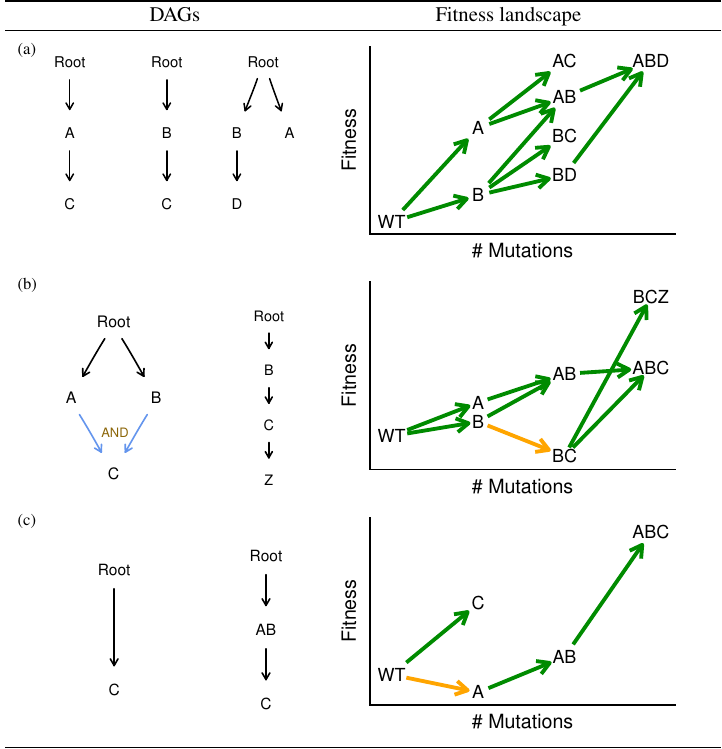}
 \caption{\textbf{Fitness landscapes with RSE, heterogeneity, and mixtures of DAGs.} In each row we show a collection of DAGs that, together, can account for the genotypes in the non-representable DAG shown on the right (where genotypes not shown are strongly deleterious). (a) This is the same fitness landscape shown in Fig.~\ref{fig:rse}(d);  none of the three DAGs here use any XOR relationship (to capture the true XOR dependency of C on A XOR B), in contrast to Fig.~\ref{fig:rse}(d). (b) This is the same fitness landscape shown in Fig.~\ref{fig:undo}(a). (c) This is the same fitness landscapes shown in Fig.~\ref{fig:lod_pom}(b);  the model might not be able to differentiate events \textit{A} and \textit{B} (if the frequency of single \{A\} is small enough), thus detecting a single event \textit{AB}. Similar multiple DAGs could be used to represent all the fitness landscapes shown in Figures \ref{fig:rse}, \ref{fig:undo}, and \ref{fig:lod_pom}.
 }
  \label{fig:heterog}
\end{figure}

In the sections above, a single model is fitted to a dataset. But some work has focused on modelling cancer progression as a mixture of models, specifically as a mixture of mutagenic (oncogenetic) trees \citep{Beerenwinkel2005a, Beerenwinkel2005b, Bogojeska2008appnote, Bogojeska2008}\footnote{Some of these, such as \citet{Beerenwinkel2005b,Bogojeska2008}, use a timed OT model.}.
Using a mixture of models allows us to deal with two conceptually different situations:

\begin{enumerate}
\item \label{true_different_models}  The data include subjects under different true models.
\item \label{mixture_as_approximation} We use mixtures to accommodate a single true model that our specific modelling approach cannot represent directly.
\end{enumerate}

In the first case, the collection of subjects is really heterogeneous, in the sense that the dependencies between events correspond to two or more different true models. For example, the data correspond to evolutionary processes happening under two or more different fitness landscapes, such as from  different tumour types (where evolutionary pressures and constraints are different), or from a single tumour type but with patients of different ages if the fitness landscape changes significantly with age, as posited by the adaptive oncogenesis hypothesis \citep{degregori2018}.
A common way of trying to deal with this problem that does not involve using mixtures is fitting different models to different subsets of data, based on an initial stratification of subjects \citep[e.g.,][]{capri_pnas,nicol2021,angaroni2021}.  An alternative, possibly more challenging approach, is to use a single model that is expanded in such a way that some of its parameters depend on other variable(s); for example, under adaptive oncogenesis \citep{degregori2018} some model parameters might depend on the age of the subjects (not unlike the expansion of fitness landscapes using contextual analysis to incorporate social interactions and frequency and density dependence: \citealp{goodnight2012}). We are not aware of this expanded model procedure having been applied.

In the second case, the collection of subjects is homogeneous in the sense that there is a single true underlying model, but the model we are using is too limited to reflect that situation. Here, a mixture of models of a simple type might be able to accommodate data generated from a single, more complex model\footnote{MC-CBN \citep{montazeri_large-scale_2016} uses a mixture of a CBN model and an independent model ---a star topology,  with all events descending directly from the Root event, so that events are independent of each other.  This is not what we are  referring to in this case; in MC-CBN the star model is used to account for observations that cannot be accommodated by the CBN model in a way similar to what the error models of OT, H-CBN, or OncoBN do.}. For example, the accessible and inaccessible genotypes under RSE and DAGs with XOR can be accommodated using a mixture of DAGs that include only DAGs with AND conjunctions, as shown in Fig.~\ref{fig:heterog}. This  approach can be sensible for many modelling objectives (e.g., prediction of genotype frequencies) but does not allow for mechanistic, causal interpretations, and thus can be limited for other objectives (e.g., identification of therapeutic targets) ---see also section \ref{sec:model-choice}.

In this sense, models with OR and XOR allow us to represent, with one single model, data that would require a mixture of models with only ANDs. For example, data that show  ``mutual exclusivity'' patterns, which other methods also try to identify (e.g., \citealp{neyshabouri2022a}; \citealp{cristea_pathtimex:_2017};  \citealp{leiserson_comet_2015}; \citealp{raphael_simultaneous_2015}).  \citet[][p.~4]{nicol2021} have argued that OncoBN's DBN model (OR relationships) ``can naturally accommodate distinct progression paths for subtypes and is expressive enough to capture the mutual exclusivity of alterations present in the data. Therefore, one can skip two preprocessing steps necessary for the state-of-the-art models: stratifying samples by subtype and mutual exclusivity detection.'' \citet[][p.~756]{angaroni2021} also refer to mutual exclusivity, distinguishing between ``soft exclusivity'' (OR) and ``hard exclusivity'' (XOR), and argue (p.~754) that ``available methods can not infer logical formulas connecting events to represent alternative evolutionary routes or convergent evolution'' which, presumably, can be inferred by PMCE/H-ESBCN. For models with stochastic dependencies, \citet{schill2020} have argued that mutual exclusivity is directly incorporated by MHN, and the \(L^2\) models in HyperTraPS \citep{johnston2016,greenbury2020} and HyperTraPS-CT \citep{aga2024} have been shown to definitely capture ``repression'' between features\footnote{Methods such as PathTimex \citep{cristea_pathtimex:_2017} or  PLPM \citep{raphael_simultaneous_2015} return explicit groups of ``mutually exclusive'' mutations; from MHN, HyperTraPS, OncoBN, or PMCE/H-ESBCN the ``mutual exclusivity'' should be indirectly inferred from the value of the coefficients (MHN, HyperTraPS) or the OR/XOR relationship (OncoBN, PMCE/H-ESBCN). But see main text.}.

However, ``mutual exclusivity'' is a possibly ambiguous term with different definitions and underlying biological causes \citep[e.g.,][and references therein]{luo2023,kuipers2021,dao2017,cristea_pathtimex:_2017,leiserson_weighted_2016,pena-llopis_cooperation_2013} that ``(...) is frequently misinterpreted owing to insufficient statistical power'' \citep[][p.~4174]{pena-llopis_cooperation_2013} and not necessarily due to epistatic interactions \citep[][p.~2459]{Misra2014}. \citet[p.~11]{kuipers2021} discuss biological explanations of different ``exclusivity'' patterns, including: (a) mutation pairs that result in complementary phenotypes that cooperate; (b) synthetically lethal pairs of mutations; (c) different mutations with similar phenotypic effects that lead to parallel and convergent evolution.

Now, the ``mutual exclusivity'' that OncoBN's OR  captures is not one where, within cells, mutations are exclusive in the sense of being synthetic lethal, but rather a relationship where anyone of several mutations, \textit{A, B, \ldots}, are enough for the acquisition of other, downstream mutations. The XOR relationship that can be modelled by PMCE (H-ESBCN) is also a disjunctive relationship with respect to the acquisition of further downstream mutations, not directly with respect to the acquisition of sets of mutations \textit{per se}; as was explained in section \ref{sec:hesbcn}, that is why under the H-ESBCN model in Figure \ref{fig:hesbcn-deps}, genotype $\{A,B\}$ is a possible, viable genotype, but $\{A,B,C\}$ is not. We also saw in Figure \ref{fig:rse} other examples of reciprocal sign epistasis or synthetic lethality that are non-representable by the XOR implementation of H-ESBCN.

Remember also we are modelling dependencies, not directly fitness: these OR and XOR relationships need not mean that the parent mutations are redundant in terms of fitness, when considered in terms of some additional, downstream mutations. For example, if \textit{C} depends on \textit{A} OR \textit{B} we are only saying that gaining \textit{C} requires at least one of \textit{A} OR \textit{B}, and that is what makes them equivalent requirements for the acquisition of \textit{C}. This does not entail that, without \textit{C}, \textit{A} and \textit{B} have the same selection coefficients, $s_A = s_B$, i.e., it does not entail that both have the same fitness benefit, nor that the fitness of the \textit{A,C} and \textit{A,B} combinations are the same. Identical comments apply to XOR. This also means that the use of expression  ``fitness-equivalent'' alterations, as in \citet{capri_pnas}, cannot be justified based solely on OR/XORs inferred from bulk sequencing data.

In contrast, HyperTraPS, MHN, and HyperHMM can model inhibiting relationships between events. In MHN (and the original version of HyperTraPS) inhibiting relationships between pairs of genes are directly incorporated with a $\theta_{ij} < 0$ (i.e., a $\Theta_{ij} < 1$). HyperTraPS-CT and HyperHMM can model exclusivities of arbitrary type involving any complex combination of genes. None of these methods, however, allow us to model synthetic lethality or mutations whose joint presence is strongly deleterious, since all  probabilities of transition are $>0$.

We have also seen (Fig.\ref{fig:collapse}, scenario \ref{scenario_collapse} in section \ref{sec:scenarios_no_rse}) that truly exclusive mutations for the acquisition of other downstream mutations (XOR of H-ESBCN), can be missed, since within-cell XORs might not be detectable with bulk sequencing. The cases depicted in Fig.~\ref{fig:collapse} would also be misleading for models with stochastic dependencies.  This raises the question of what is the underlying biological cause of the OR/XOR patterns reported in \citet{angaroni2021} and \citet{nicol2021}: these could represent true within-cell XOR/OR but can also be instances of redundancy of effects or intra-tumour cell cooperation (\citealp{kuipers2021}; see also \qpref{sec:freq-dep}). Similar questions apply to the inhibiting relationships reported in \citet{schill2020}.

\subsection{Frequency-dependent fitness} \label{sec:freq-dep}

\begin{figure}[tbph]
 \centering \includegraphics[width=0.975\linewidth,keepaspectratio]{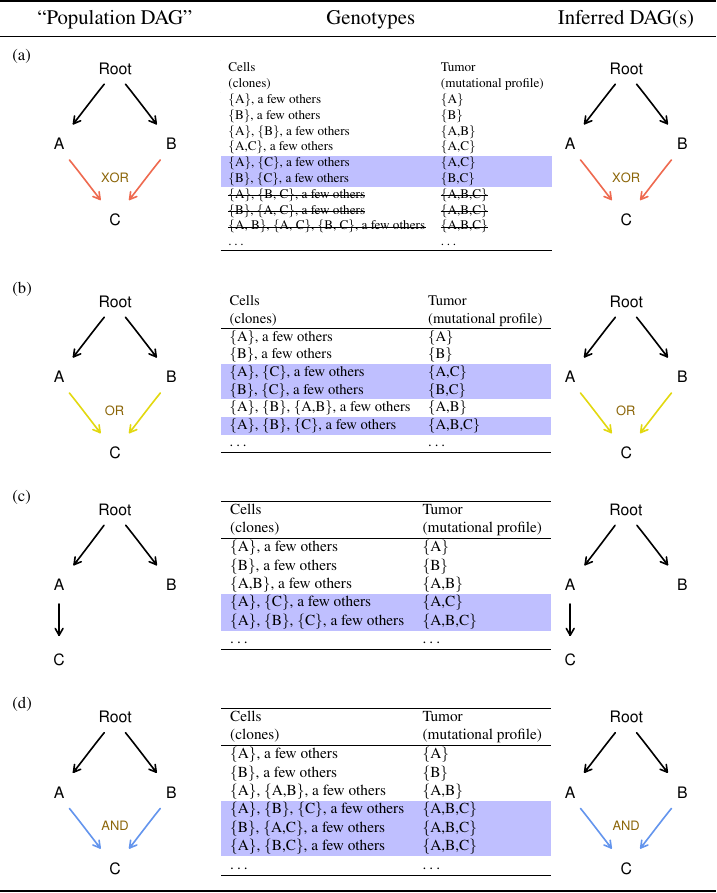}
 \caption{\textbf{Frequency-dependent fitness and bulk sequencing.}
   ``Population DAG'', the true DAG that describes the restrictions in the accumulation of mutations in the complete population. Following \cite{axelrod_evolution_2006}, we consider dependency restrictions to be satisfied if diffusible factors are available in the population; e.g., a dependency of C on A is satisfied if A is available in the population, which can happen either because the same cell has both A and C mutated or because there are already some cells in the population with A mutated at the time a cell without A mutated acquires a mutation in C. For an XOR, we interpret it as meaning that the simultaneous presence of the terms involved in the XOR leads to a non-viable tumour.
   ``Genotypes'' and ``Inferred DAG(s)'' as in Fig.~\ref{fig:collapse}.
   In (a), and in contrast to what happened in Fig.~\ref{fig:collapse}, no genotypes incompatible with the DAG can be observed; the ones that lead to problems in Fig.~\ref{fig:collapse}(a) haven been strikethroughed.
   We show in blue combinations of  genotypes that, under that DAG, are possible because  the restrictions operate at the population level  via diffusible factors, but would not be possible if the restrictions operated  within cells.
}
  \label{fig:freq_dep}
\end{figure}

Under frequency-dependent fitness, the fitness of a genotype depends on the frequency of other genotypes\footnote{These problems are often studied using evolutionary game theory and adaptive dynamics \citep{mcnamara2020,waxman2005,brannstrom2013}, and they have been applied to different problems in cancer \citep[][]{kuipers2021,stankova2019a,kaznatcheev2019a,kaznatcheev2017,archetti2019}. If we wanted to use fitness/adaptive landscapes, it would be necessary to modify these concepts to incorporate the changes in the mapping between genotype and fitness as a function of the frequency of other genotypes \citep{goodnight2012}.}. \citet[][p.~13475]{axelrod_evolution_2006} suggested that tumour growth can be supported by commensalism via diffusible products; the tumour is composed of a mixture of cell types, that enhance each other type's growth via these diffusible products, until one of the clones acquires the ``full deck'' of mutations. The paper ended asking, among others, the questions ``What are the implications for the expected order of mutations, given that some can be in parallel?'', and ``How can cooperation among partially transformed tumor cells be interrupted to stop, or at least slow, the progression to malignancy?'' \citep[][p.~13478]{axelrod_evolution_2006}. The first is precisely what CPMs try to do, and the second is  an intervention that might benefit from answering the previous question. As far as we know, using CPMs to understand (restrictions/dependencies in) the order of accumulation of mutations under frequency-dependent fitness has not been addressed.

In  Fig.~\ref{fig:freq_dep}, we show the consequences of bulk sequencing when there is frequency dependence and the restrictions of accumulation of mutations  apply not within cells, but over the whole population. For simplicity, we use DAGs of restrictions, but similar scenarios could be constructed with stochastic inhibiting/facilitating relationships between genes coding for diffusible factors. A key idea here is that frequency-dependent scenarios, of the type considered by \citet{axelrod_evolution_2006}, could be profitably examined using CPMs, focusing on the mutational profile from bulk sequencing. And if the dependencies captured by the models represent population-level dependencies, these would be relevant to design therapeutic interventions. A second key idea is that dependencies inferred from bulk sequencing data can reflect dependencies at the population level, not the within-cell level, and this could explain some OR and XOR patterns at the bulk sequence (tumour profile) level \citep[see also][]{kuipers2021}.

\subsection{Selection bias, sampling, and bulk sequencing}\label{sec:obs_bias_bulk}

Recent work by \citet{schill2024a} addresses bias arising from selection bias (also know as collider bias or Berkson's bias: \citealp{hernan2004, pearl_causality_2009}). In the case of CPMs, this would be caused by the effect that the events we are modelling have in the probability of observation itself. In fact, this problem underscores the importance of clarifying the entities under study and the role of bulk sequencing. Effects that arise from probabilities of detection changing with tumour size have already been considered in, for example, the simulations in \citet{diaz-uriarte2018} and \citet{diaz-uriarte2019a}, though those papers never considered collider bias as a possible explanation of the results. In these two papers, it is tumour size that directly affects the probability of detection, and genotypes affect probability of detection indirectly via the (very complex) relationship between fitness, growth, and tumour composition and size. In \citet{schill2024a}, in contrast, rate of detection is directly a function of the progression events. This seems to posit a different sampling model.  Moreover, as mentioned before (section \ref{sec:why-entities}), whether our models are about cells or about tumours will affect how we conduct simulation studies  to assess the performance of methods, including observation bias-correcting approaches. In this regard, a possibly relevant question is the role of thresholds for mutation detection/calling on the performance and interpretation of observation-bias corrections under different evolutionary regimes and tumour heterogeneity. Thus, considerations of entities and sampling should also play a role in modelling efforts to reduce additional biases.


\subsection{Can we use these models to tell drivers from passengers?}\label{driver-passenger}

In the cancer literature in particular, a distinction is often made between ``driver'' mutations, those that drive cancer progression, and ``passenger'' mutations, those that are not responsible for cancer progression but can sometimes show large frequencies if they ``hitchhike'' on drivers (e.g., \citealp{Greaves2012, Beerenwinkel2015}). Can CPMs be used to tell them apart?

The short answer is that CPMs are not designed for this purpose, a point that has been made before, implicitly (e.g., \citealp{Beerenwinkel2015}) and explicitly (e.g., \citealp{Misra2014}); in fact, the applied use of these methods with cancer data generally involves a pre-processing step with the purpose of eliminating passengers, so that the input data includes only drivers (see comparison and discussion of approaches in \citealp{Diaz-Uriarte2015}). Without this pre-filtering, and if our data includes both passengers and drivers, we might expect drivers to appear as events with few edges in the DAGs or small coefficients in stochastic methods, and drivers to show more edges/larger influences. However, there could be strong effects for different reasons, and there could be drivers (in the sense of ``lead to large growth'') that show no strong effects.

Suppose event A has only a moderate effect on growth rate (i.e., has a small positive fitness contribution) and event B is highly deleterious, but the combination of A and B has a very large growth rate (i.e., fitness of genotype AB is much larger than that of A). What most CPM methods will tell us is that A has a strong effect on the downstream acquisition of B. But it is debatable that A is really a ``driver'' if, by driver, we mean a genetic alteration that, on its own, provides a strong fitness advantage and leads to tumour growth.

Suppose, in contrast, that C is a driver in the usual sense, and it leads to a very large increase in population size. To make the scenarios simple, further suppose that once C is gained, additional mutations are either deleterious or neutral, so additional mutations cannot increase fitness. Now, after C is mutated, the population will grow in size. As the population is now much larger, over the whole population, neutral mutations will appear sooner (given a constant mutation rate, we have more cells to have that mutation appear), but none of these additional mutations confer any benefit and their fate follows that of neutral mutations. So C is not making any difference for acquiring other mutations beyond the generic effect of "the population is larger".

In the above scenario, consider now what happens with passengers. Neutral mutations already present in cells that later acquire C will increase in frequency. However, over multiple patients, we expect there to be no consistent pattern to these "hitchhiking" events, so we would not find that C makes any difference for acquiring specific neutral mutations nor that specific neutral mutations make any difference for acquiring C.

As a third scenario, consider D as a driver in the usual sense that its mutation, \textit{per se}, provides a strong fitness advantage and leads to tumour growth. E, on its own, is deleterious, but the combination of D and E provides a further fitness advantage. Now, we will see a large effect of D on E. Acquiring D leads, quickly, to further adding E.

Thus, the relation between driver and passengers and the rates estimated by these methods is not simple, and can depend strongly on the details of the evolutionary model and the precise definition of ``driver''. In terms of fitness, and assuming we are modelling within-cell phenomena (i.e., our entities are cells), event \(X\) increases the probability of acquiring event \(Z\) if the \{X,Z\} genotype has larger fitness than \{X\} and much larger fitness than \{Z\} (genotype  \{Z\} on its own might even be deleterious). But, as already discussed, this interpretation need not hold with bulk sequencing and phenomena such as frequency-dependent fitness. This is why it is best to think of these methods as modelling "difference making for acquiring other events".

\section{Modelling just mutational profiles}
\label{sec:what-if-bulk-seq}

If our samples come from bulk sequencing, our models can only reflect within-cell restrictions if either (a) we are in a SSWM evolutionary regime (so the sample is close to monotypic) and, additionally, tunnelling is rare enough to be irrelevant, or (b) we use detection thresholds for mutations of 100\%, so that we ensure all detected mutations are simultaneously present in a cell.  Condition (b) has a consequence similar to that of condition (a): a clone is fully replaced by another clone, where clone is defined with respect to the mutations of relevance; in other words, for the mutations of relevance, the population is monotypic (see legend of Fig.~\ref{fig:scenarios} for the definition of monotypic population). If (a) or (b)\footnote{Condition (b) requires assuming that such a phenomenon (some mutations are present in 100\% of the cells of an otherwise possible non-monotypic populations) occurs; if it never happens, we will not find mutational profiles beyond the ``WT'' genotype, the genotype without any additional mutations.} apply, we would be in scenarios \ref{scenario_best} or \ref{scenario_inconsequential}\footnote{In these cases, difficulties caused by reciprocal sign epistasis, \qpref{sec:rse}, can severely affect the interpretations of models, but this problem is different from that of bulk sequencing \textit{per se}.} of section \ref{sec:violat}. Otherwise, the model is a model about other entities.

Thus, unless we make very strong assumptions about evolutionary regimes (and too-rare-to-be-relevant movements through fitness valleys), we cannot endow our bulk-inferred models with an interpretation in terms of what happens in cells or clones: the rates and conditional probabilities, and the DAGs of relationships, pertain to  mutational profiles, not cell types or genotypes. This has three consequences:
\begin{enumerate}
\item The dependency relationships between events need not reflect epistatic, within-cell interactions between mutations.
  \item  That some models can accommodate parallel and divergent trajectories does not mean that these trajectories really exist at the cell level (see \qpref{sec:freq-dep}); nor does it mean that all divergent trajectories at the cell level would be captured even with infinite sample size (see Fig.~\ref{fig:collapse}(a)).
  \item If different bulk samples contain different proportions of cell types, we might be in a situation similar to the one explained in section \ref{sec:heterogeneity}, where we discussed the problem of collections of heterogeneous subjects.
\end{enumerate}
In fact, these issues are  similar to the ecological inference problem \citep[][and references therein]{kim2019,wakefield2010,jamesgreiner2009}, where relationships observed (inferred) using aggregate or grouped data do not necessarily hold at the individual level.

The suggestion of this section is to build models about the data at hand, which is what we are doing already, without endowing them with an interpretation in terms of what happens in cells or clones\footnote{In this sense,  we find that the expression ``selective advantage'' (e.g., \citealp{fontana2023,capri_pnas}) can misleadingly evoke the idea of selection coefficients of specific mutations that happen in individual cells.}. That these models apply to bulk sequencing data is, of course, not a new idea \citep[e.g.,][]{nicol2021,angaroni2021}, but the suggestion here is to be explicit about the previous three consequences.
Those consequences limit the transportability\footnote{Transportability is often used in the causal inference literature to refer to how extrapolable between populations are causal inferences \citep[][ch.~4]{hernan2020}.} of our inferences, since the rates and DAGs we estimate can be affected by the specifics of our data (e.g., thresholds for detection, proportions of cell types in the samples), and we might also capture effects that are irrelevant, and fail to capture with precision others that are key for intervention (\qpref{sec:lod-pom-sampling}).

This also forces us to be explicit about additional nuisance factors, such as thresholds or limits of detection of events. In fact, this raises the question of what thresholds to use: what is the minimum frequency of a mutation before we call it present in the sample. For many scenarios,  larger thresholds are probably better if we  want to  model fixation of mutations (e.g., see \citealp{Gerstung2009,Gerstung2011}, and section \ref{sec:entities}): if there is no frequency-dependent fitness, fixation of mutations might be easier to understand than mere presence of a mutation. Large thresholds might also make it easier that other assumptions are also sufficiently satisfied to allow inferences that go beyond the mutational profile. But in other scenarios smaller thresholds will be more appropriate: see \qref{sec:lod-pom-sampling} and \qref{sec:freq-dep}. These are underexamined matters, and it is worth exploring these conditions, taking into account the intended uses of the inferences in future studies. It bears repeating the difference between, for example, on the one hand \citet{Gerstung2009} and \citet{Misra2014}, who seem to emphasise fixation of mutations, with that of, on the other hand,   \citet{nicol2021} and \citet{angaroni2021} who, implicitly, exclude ``until fixated'', as the mutational profile can show mutations that are present only in some subclones.

A consequence of explicitly giving up a within-cell interpretation is that problems related to limited expressive power (including questions motivated by reciprocal sign epistasis), or whether specific scenarios are better served by AND/OR/XOR or stochastic dependencies of different $L^x$  order would be judged, exclusively, by how well we can make relevant predictions \citep{billheimer2019,candes2020a,yu2020} (see also section  \ref{sec:model-choice}).

Indeed, there is no indication that  models interpreted at face value as models that apply ``just'' to mutational profiles  would be less useful for many prediction tasks. For example, predictions of patient survival based on CPM's output, as in \citet[][]{Gerstung2009,angaroni2021,Bogojeska2008appnote}, do not depend on a specific fitness landscape interpretation. In contrast, predicting the next genotype in the LOD has been shown to be problematic \citep{diaz-colunga2021}, but it might turn out to be much easier to predict the \textit{next observable mutational profile}; this is a question worth exploring, even if the next observable mutational profile is less useful for some intervention objectives.

Moreover, by uncoupling these models from cells/clones/genotypes we make it more clear that these models are  useful in a much wider range of problems, as demonstrated for questions that range from animal tool use to malaria progression or antimicrobial resistance (see \qpref{sec:uses_all}).
Finally,  using these models for scenarios with frequency-dependent fitness (see \qpref{sec:freq-dep}) requires giving up a direct fitness landscape interpretation based on an ideal monotypic sample.

Let us return to the scenarios in \qref{sec:scenarios_no_rse}, but without trying to endow the inferences with interpretations in terms of what happens in cells or clones:

\begin{itemize}
\item In Fig.~\ref{fig:scenarios}, where we considered six possible scenarios determined by order restriction violations, evolutionary dynamics, and sampling, we are now only left with the first split, as considerations about monotypic populations and sampling disappear.

\item In Fig.~\ref{fig:collapse}, that focused on the effect of bulk sequencing and mutational profiles that do not correspond to any existing genotype,  the answer to the question ``What XORs can we detect?'' is ``Those that are manifested in the data, in the mutational profile''. By not attempting any mapping to underlying evolutionary process in cells or clones, the XOR pertains to the XOR in the mutational profile, and likewise for any other pattern. As we mentioned above, this does not answer the question of what XORs really represent: are they within-cell XORs? or are they different evolutionary trajectories due to frequency-dependent fitness?

\item Fig.~\ref{fig:order_restrictions_violated} (cause and consequences of violations of order restrictions) and Fig.~\ref{fig:tunnel} (tunnelling) are no longer a concern either.

\item Fig.~\ref{fig:rse}, \ref{fig:undo}, \ref{fig:rse-complications}, that explore complications arising from reciprocal sign epistasis (including their role in ``undoing'' restrictions and local maxima without reciprocal sign epistasis), now become a problem about the appropriate expressive power, a direct question of how complex we want our models to be to predict correctly (see also \qpref{sec:model-choice}).

\item Fig.~\ref{fig:lod_pom}, that discusses LOD and POMs, and contrasts them with mutational profiles,  is no longer a concern as there is no notion of LOD or POM of relevance here.  Focusing, for example, in  Fig.~\ref{fig:lod_pom}(a), if \textit{C} is really irrelevant for \{A, B\}, and we infer the linear DAG on the left,  our model will not reflect any real causal relationship. If we end up inferring the DAG on the right, our model will not be misleading about the causal effect of \textit{C}, but will not give any information about the causal relationship between \textit{A} and \textit{B}. This difficulty cannot be solved unless we have finer resolution data.

\end{itemize}

A research question that arises from this discussion is understanding what is the ``bulk sequencing view'' of evolution under different fitness landscapes, different evolutionary dynamics (population size, mutation rate), and different mutation detection thresholds, and how it affects these models, including possible observation biases. A conjecture is that the ``landscapes'' will be smoother, with fewer local maxima (e.g., \citealp[][pp.~80-82]{goodnight2012}), observed data patterns will look more like stochastic dependencies with possibly asymmetric coefficients as in HyperTraPS and MHN (because of the averaging over different cell populations), and that most states predicted by the models will be observable (especially if mutation detection thresholds are small, because then mutations with even small frequencies are detected, and the observed data cannot mask transitions via genotypes of small frequency). In other words, the problems of predicting future mutational profiles or tumour states will be easier.


\section{Model evaluation, experimental evolution, and model fitting}

\subsection{Model evaluation and model performance on experimental evolution data and single-cell data}\label{sec:model-choice}

For CPMs applied to a given dataset, one option for model evaluation and model choice can focus on the ability of different models to make relevant predictions \citep{billheimer2019,candes2020a, yu2020}. For example,  and in the absence of additional information, this could be assessed using the cross-validated likelihood, which should be straightforward, given that all methods discussed return the probability of the mutational profiles. This is more time-consuming that using, say, AIC, but avoids the difficulty of comparing models of different kinds (e.g., DAGs with only ANDs, vs.\ DAGs with all of OR, XOR, AND, vs.\ models with stochastic dependencies) where it is unclear how to quantify the number of parameters comparably \citep[e.g.,][]{schill2020}. Note, though, that a key idea is the ability of making relevant predictions, and for a particular scenario predicting the frequencies of mutational profiles might not be the relevant question.  The relevant question might be one that we can not evaluate with the cross-validated likelihood, such as predicting the next genotype or response to therapy, for which there is no information  in the available data set. These types of questions require simulations, and  we have already underscored that assessing performance via generative models can lead to an overoptimistic impression of method performance ---see  \qref{sec:why-entities}.

What about using data from experimental evolution to compare and assess the overall performance of different methods? Experimental evolution studies have shown that, because of epistasis, the order of acquisition of mutations has a strong effect in the adaptive process, leading to highly contingent histories that can depend on which mutation(s) become fixed first \citep[][and references therein]{elena2003, cooper2012,losos2018}.
Already existing data from these experiments could be used to examine the performance of these approaches, and this would be particularly important for all of those questions that cannot be settled by comparing methods via the cross-validated likelihood. This includes parameter estimates for genes with known epistatic interactions, as well as comparing model-derived with true mutational paths. Experimental data would also allow us to examine the effects of aggregating samples ---like in bulk sequencing---, as well as the ability to estimate and make predictions of the future evolutionary trajectories \citep{hosseini2019a,diaz-uriarte2019a,diaz-colunga2021}, or the detection of mutual exclusivity \citep{neyshabouri2022a,cristea_pathtimex:_2017}.

But evaluation of inferential procedures using experimental evolution data is not necessarily better than using simulation studies. Addressing the value of experimental phylogenetics, \citet{oakley2009} writes: ``(...) experimental phylogenetics is subject to the same constraints as simulations: in either situation, it is necessary to assume the evolutionary processes present in the tests apply universally (Sober 1993; Grant 2002). This assumption is especially true when trying to establish the efficacy of methods, as opposed to the shortcomings. Any one replicate history can call into question the reliability of a method, but because any single replicate could be nongeneral, establishing reliability of methods requires generating replicates under many different assumptions or parameter values'' \citep[][p.~663]{oakley2009}.
For example, what would successful application of these accumulation models to bacterial experimental evolution tell us about their applicability to cancer evolution? The possibility of examining the performance of monotonic accumulation models on experimental evolution data revolves around what is more similar to the systems we really care about \citep[see][and references therein]{weisberg2015}.

A related problem involves using single-cell data to validate inferences based on bulk sequencing data. We are not aware of this having been done, but could be worth addressing once single-cell data become more abundant. Two issues already raised are worth repeating here. First, ``validate bulk sequencing-based inferences with single-cell data'' is specific to the question asked, as discussed in \qref{sec:why-entities}: for example, are we referring to validating inferences about sign epistasis or validating a successful stratification of patients based on the dynamics of mutation acquisition? Second, and as mentioned in the paragraph above, validating specific inferences (e.g., reciprocal sign epistasis) in one tumour type is only indicative of likely success in other tumour types if the two tumour types are sufficiently similar in the features that matter \citep{weisberg2015}.

\subsection{Model fitting: probabilistic programming and simulation-based inference}\label{sec:simul_based_inf}

We have already mentioned (e.g., section \ref{sec:assumpt_stoch}) that changing assumptions or enhancing interpretation could be helped by using some of the existing models with specific modifications.  The available code for the methods discussed above has been carefully crafted for fast execution. Nonetheless, we might be willing to sacrifice some (or much) of that execution speed to easily explore models that would arise from relaxing or modifying some assumptions, or imposing other constraints. Some examples are: including reciprocal sign epistasis in deterministic models;  enforcing symmetry of pairwise interactions in stochastic models;  explicitly accounting for local fitness maxima\footnote{As mentioned in section \ref{sec:assumpt_stoch}, the new versions of HyperTraPS and HyperHMM will make this much easier by imposing priors on individual parameters.}; modelling the observation process to correct for observation biases \citep{schill2024a}. Using general purpose languages and libraries for probabilistic programming \citep{ronquist2021,vandemeent2021} might allow this  exploration.
Simulation-based inference will be even more important if the likelihood cannot be specified, for example if our models try to accommodate complex sampling schemes or population effects (e.g., as could arise from frequency-dependent fitness ---section \ref{sec:freq-dep}--- or  observation biases that arise from tumour sizes involving collections of different genotypes ---section \ref{sec:obs_bias_bulk}). In this case, Approximate Bayesian Computation and other simulation-based likelihood-free procedures might be necessary \citep[e.g.,][and references therein]{mo2023,craiu2023,sainsbury-dale2023,cranmer2020}. Indeed,  this is indeed almost what was done in \cite{williams2013}. To begin exploring these approaches, models with stochastic dependencies are probably much easier to play around with. In models with discrete dependencies there is a DAG and possibly a combination of AND/OR/XOR; this  forces us to use  different samplers for the discrete and continuous parameters, and require us to devise sensible procedures for moving over the DAGs \citep[e.g.,][]{Sakoparnig2012} and, depending on the inferential framework, specify appropriate priors. For some of these explorations, it might be possible to reuse building blocks (e.g., random walkers, in easily callable C/C++) from existing code, for example from HyperTraPS-CT (\url{https://github.com/StochasticBiology/hypertraps-ct}).

\section{Conclusion}
\label{sec:conclusion}

Cancer progression and monotonic accumulation models have provided new insights about dependencies in several dynamical systems, from cancer to HIV to malaria progression, and could be used as well to improve patient stratification and survival prediction. Because of their interest, for both basic and applied research, new methods continue to be developed.

Here, we have examined key evolutionary assumptions and interpretational issues when using these methods.  A recurring question in the manuscript is to which empirical entities the models apply. This underscores evaluating how the entities match the intended use of the inferences, and also has strong effects on how we should asses (e.g., via simulations) the performance of different approaches.  The focus on the empirical entities naturally led us to examine problems that arise from sampling (bulk sequencing) and deviations from the strong selection, weak mutation evolutionary model, and how models with stochastic vs.\ deterministic dependencies are affected differently by these issues.

Bulk sequencing, for example, sets limitations on the inferences that are possible and the extent of interpretations. As a  consequence, most interpretations in terms of within-cell epistatic interactions are only warranted under very strong  assumptions. Furthermore, interpretation can also hinge on thresholds for mutation detection/calling, requiring us to be explicit about their role; these thresholds can also affect how the probability of detection changes with tumour genotype or composition. The concepts of alternative evolutionary paths and mutual exclusivity are also discussed: in addition to often being used ambiguously in this literature, inferences about them are generally not possible with the data at hand. Nonetheless, many fruitful uses of these approaches, including patient stratification and improved survival analysis, are unlikely to be affected by these nuances.

A careful consideration of the empirical entities and the possible evolutionary assumptions will not only prevent us from endowing inferences with unsubstantiated implications, but also represents an opportunity for applying these models in novel scenarios. Four are worth considering, and highlight some research opportunities. First, bulk sequencing data makes it possible to attempt inferences about  dependencies when there is frequency-dependent selection, as we mentioned in section \ref{sec:freq-dep}. Second, these methods can be applied in other areas with a possibly better fit between empirical entities and model assumptions, as exemplified by the analysis of malaria progression in children (section \ref{sec:entities}), where many of the interpretation hurdles are absent. Third, data that are not independent can be analysed by these and closely related methods (we have mentioned HyperTraPS and HyperHMM as capable of analysing data with phylogenetic and longitudinal relationships, and TreeMHN and REVOLVER as methods that analyse within-patient phylogenetic trees ---see sections \ref{sec:assumptions} and \ref{sec:other-methods}); it will be important to clarify the relationship between these methods and the scope of their inferences, and issues about entities, evolutionary models, and sampling raised in this manuscript will be relevant for that clarification.  Finally, the rich variety of models available allows us to explore very different dependency structures between events; answering what is the ``best'' model (or combination of models and/or structures) will depend crucially on the appropriate validation procedure given the intended use of the model (section \ref{sec:model-choice}), and could involve the development of custom and expanded models (sections \ref{sec:simul_based_inf} and \ref{sec:heterogeneity}).

Finally, we said above ``could be used as well to improve patient stratification and survival prediction'': a pessimist might say that these approaches have had little translational or practical impact to date. This leaves open the question of “pathways to impact” of these approaches. Moving forward, tighter collaborations with experimental groups and clinical practise ---with a particular focus on how predicted interactions could be translated to actionable insights, and what predictive messages are most important in the clinic--- could strengthen these pathways and help realise the potential of these approaches.

\section{Acknowledgements}
Supported by grant PID2019-111256RB-I00 funded by MCIN/AEI/10.13039/501100011033 to Ramon Diaz-Uriarte.
This work was supported by the Trond Mohn Foundation [project HyperEvol under grant agreement No. TMS2021TMT09], through the Centre for Antimicrobial Resistance in Western Norway (CAMRIA) [TMS2020TMT11]. This project has received funding from the European Research Council (ERC) under the European Union’s Horizon 2020 research and innovation programme [grant agreement No. 805046 (EvoConBiO)].
We also thank Philip Nicol for answering questions about the interpretation of OncoBN and the DBN model, and Alex Graudenzi for answering questions about PMCE/H-ESBCN.

\bibliography{refs}
\bibliographystyle{natbib_noparen_date_urlbst}

\section{Appendix}

\subsection{Notation for genes, genotypes, DAGs, paths, landscapes}\label{sec:notation-genots}

In figures where we represent DAGs of restrictions and paths between mutational profiles (or tumour states) or genotypes side by side (like Fig \ref{fig:linear-deps} to \ref{fig:hesbcn-deps}), or figures where mutational profiles/genotypes are mentioned next to text (like the centre column of Fig.~\ref{fig:collapse}), we use a capital letter without curly brackets to denote a gene or event, and capital letters, between brackets, and separated by comma, to denote a mutational profile/genotype with those genes mutated. In other figures where it is clear we are showing genotypes (e.g., fitness landscapes plots or LOD, POMs, and tumour states as collections of genotypes ---Fig.~\ref{fig:lod_pom}) we do not use brackets or commas.

\subsection{What are the entities: quotations from the literature}\label{sec:quotes-entitites}

For easier reference, we indicate the method to which the comments directly apply:

\begin{description}
\item[OT] ``N independent specimens ("tumors") are obtained and the presence or absence of the alterations of interest is recorded as a binary vector''  and ``(...) mechanistic interpretation as the oncogenetic trees that we described'' \citep[][p.~2 and 22]{Szabo2008}.

  Interestingly, \citet[][pp.~220-221]{Szabo2002} write ``Some alterations present in the tumor might be missed because of the spatial heterogeneity'', which might suggest that for these authors the models are primarily about tumours, without stronger commitments (cells, SSWM, etc.).

  ``(...) the list of the copy number aberrations (CNAs) of a particular tumor'', ``They also imply that when certain combinations of vertices are reached, the abnormal cell would be sufficiently abnormal to be considered cancerous.'' ``It is extremely hard to figure out the cause and effect relationships in the laboratory, so a plausible model should be quite helpful to cancer geneticists.'' \citep[][pp.~38, 49, 49]{Desper1999JCB}

\item[OncoBN] ``(...) each directed path starting from the wild-type root in a DBN graph may be interpreted as a (sub)clone (Figure 2C), and each sample from the DBN graph can be thought of as an individual tumor consisting of (sub)clones'' \citep[][p.~3]{nicol2021}.  ``Each path of sampled graph forms a subclone'' \citep[][p.~3, legend to Fig.~2]{nicol2021}.  Therefore, necessarily, the mutational profile of a tumour would be the union of the mutations of all the individual clones. And this interpretation is also used when discussing some results of the application of their method to a specific dataset:  ``the BRAF and NRAS branches in Figure 6A do not share any child and therefore based on the phylogenetic interpretation of DBN (Figure 2C) although they can co-exist in a tumor, they may not co-occur at the cell level.'' \citep[][pp.~10-11]{nicol2021}.

  Further comments about the correct interpretation of the model are available from \url{https://github.com/phillipnicol/OncoBN/issues/8}. Briefly, the model parameters and predictions of probabilities of different states refer to the states of individual tumours (or mutational profiles), not clones or cells. However, the restrictions apply both within-cell, and for each tumour profile (or mutational profile). In particular (except for model violations accounted for by the error model), there can be no cells in the tumour that violate the restrictions implied by the DAG. The above holds both for the conjunctive (AND) and disjunctive (OR, the DBN \textit{sensu stricto}) versions.

  Moreover, we are implicitly excluding the ``until fixated'' assumption (in contrast to CBN) in so far as the mutational profile can show mutations that are only present in some of the subclones.

\item [CBN] \citet[][p.~2810]{Gerstung2009} mentions ``the genotype of the tumor'', which we understand is the same as the mutational profile. In this sense, this is similar to other authors \citep[e.g.,][]{nicol2021,angaroni2021} where it is clear that the data come from bulk sequencing and are not necessarily individual cell genotypes. However, we also read: ``mutations that need to be present before mutation \textit{i} can fixate'' and in \citet[][p.~8]{Gerstung2011} ``This waiting time may be interpreted as the time until the mutation has occurred and reached fixation in the population.''   These sentences would suggest that the description is about populations (the cell populations of the tumours) that fixate mutations successively, via clonal sweeps, similar to the SSWM regime. An explicit commitment to the SSWM is made in \citet[][p.~i390 ---see also their Fig.~1]{hosseini2019a} ``to systematically assess the validity of the CBN model for quantifying the predictability of cancer evolution, it is essential to establish that the CBN-based constraints on the ordering of mutations approximate well those based on the underlying fitness landscape and the SSWM assumption''.

  Interpreting restrictions in terms of epistatic interactions is explicit in  \citet[][p.~e15]{Beerenwinkel2015}\footnote{This paper is not exclusively about CBN, but it has been included under this item because some of their authors are the creators of CBN and because it seems to be coherent with other ideas mentioned in CBN papers.}: ``Direct and indirect interactions result from nonlinear, epistatic fitness landscapes underlying the evolutionary process and introduce statistical dependencies among genetic alterations.''.  Later \citealp[][p.~e17]{Beerenwinkel2015} we read ``(...) a genotype is defined by all mutations that have accumulated before a stopping time'', so  a genotype is not necessarily the observed genotype of a cell, but rather is defined based on accumulation. In an example that uses the HIV genome instead of cancer, \citet[][p.~i729]{montazeri_large-scale_2016} write ``(...) transition rates from each genotype to its successive genotypes''.

  So even if we understand that the data originate from bulk sequencing (and that we might not be measuring individual cells), by focusing on mutations that fixate in the complete population, the restrictions identified must correspond to within-cell restrictions, which justifies interpreting them as epistatic interactions. This within-cell interpretation would hold if we assume SSWM (as then the mutational profile would be the genotype of a cell), but would also hold if we define the ``tumour genotype'' as above: all mutations that have accumulated (fixated) by the time of sampling.

\item[H-ESBCN (PMCE)] ``(...) cancer progression models, even when inferred from low-resolution (binarized) mutational profiles of cross-sectional samples, may deliver accurate predictions on patients’ survival.'', ``Instead, PMCE allows one to infer a statistically robust population-level estimator of cancer evolution.'', ``A further pitfall of approaches processing low-resolution (binarized) data from bulk samples is due to the impact of intra-tumor heterogeneity,'' \citep[][pp.~760,, 761, and 761, respectively]{angaroni2021}. This makes it explicit that the model is to be interpreted as a model of the restrictions and probabilities of the bulk samples, not necessarily of the restrictions in individual cells within tumours.

  The authors also discuss heterogeneity, to argue for the OR and XOR in their models: ``the inference problem is complicated by the high levels of heterogeneity typically observed in most tumor types, which are due to the existence of multiple independent evolutionary trajectories'' \citep[][p.~754]{angaroni2021}. And  ``available methods can not infer logical formulas connecting events to represent alternative evolutionary routes or convergent evolution.''\citep[][p.~754, Abstract]{angaroni2021}. These should be interpreted as referring to heterogeneity and alternative evolutionary routes between patients; it is not  about within-tumour selection, competition, or cooperation between cells.

  The authors also indicate ``Driver (epi)genomic alterations underlie the positive selection of cancer subpopulations'' \citep[][p.~754, Abstract]{angaroni2021}. But since PMCE models bulk samples, patterns of selection between subpopulations   within a tumour cannot be discerned. In particular, the PMCE model need not reflect epistatic interactions leading to differences in fitness between cell types.

\item[MHN] ``Such progression events arise in individual tumour cells, but their effect on the reproductive fitness of this cell depends on earlier events'', ``assuming that the tumour genomes are observations from the same stochastic process, cancer progression models can infer dependencies between events from their co-occurrence patterns.'' \citep[][p.~241]{schill2020}. In their latest paper, \citet{schill2024a}, they refer to ``tumor genotype'' (p.~226) and explain that CPMs are used with ``bulk genotypes'' (p.~218).

\item[BML] ``BML first estimates the probability \(P(g)\) that a particular combination of mutations (denoted by genotype \(g\)) reaches fixation in a cell population that has evolved from a normal cell genotype and will eventually attain a tumor cell genotype'' and `` (...) the probabilities that we infer represent the chance of a combination of mutations reaching fixation in a cell population, as it evolves from a normal state'' \citep[][p.~2456, 2459]{Misra2014}.  The abstract also refers to cells, cell fitness, and epistasis: ``Cancer cell genomes acquire several genetic alterations during somatic evolution from a normal cell type. The relative order in which these mutations accumulate and contribute to cell fitness is affected by epistatic interactions.'' Furthermore, in p.~2456 they refer to ``Patterns of somatic mutations observed in tumor samples contain information, both about the evolutionary paths of cancer progression and the epistatic gene interactions that influence them'' and ``[BML] takes into account (...) and unknown epistatic interactions''. Thus, epistatic interactions and fixation of a genotype are explicit in BML.

\end{description}

\subsection{Software: main repositories}
\label{sec:software-repos}

\begin{description}
\item[OT] Oncotree R package: \url{https://CRAN.R-project.org/package=Oncotree}

\item[DBN (OncoBN)] GitHub repository for R package: \url{https://github.com/phillipnicol/OncoBN}

\item[CBN] Source code for h/ct-cbn (C program): \url{https://bsse.ethz.ch/cbg/software/ct-cbn.html}. GitHub repository for MC-CBN R package: \url{https://github.com/cbg-ethz/MC-CBN}. (EvAM-Tools includes a slightly modified version of H-CBN with a bug fix and output with lambdas and likelihood from the initial run and each of the iterations; available from \url{https://github.com/rdiaz02/EvAM-Tools#ct-cbn})

\item[H-ESBCN, PMCE] GitHub repository for H-ESBCN (C program): \url{https://github.com/danro9685/HESBCN}. GitHub repository for PMCE (includes H-ESBCN and R scripts): \url{https://github.com/BIMIB-DISCo/PMCE}.

\item[MHN] GitHub repository (R code that calls included C code): \url{https://github.com/RudiSchill/MHN}. A more recent Python version is available from \url{https://github.com/spang-lab/LearnMHN} and the code for the selection bias-correcting procedure from \url{https://github.com/cbg-ethz/ObservationMHN}.

\item[HyperTraPS]  GitHub repository (R code that calls included C++ code):
\url{https://github.com/StochasticBiology/HyperTraPS} (with a stripped-down version at \url{https://github.com/StochasticBiology/hypertraps-simple}); most recent version, including continuous time, at \url{https://github.com/StochasticBiology/hypertraps-ct}.

\item[HyperHMM]  GitHub repository (R code that calls included C++ code): \url{https://github.com/StochasticBiology/hypercube-hmm}.

\item[EvAM-Tools] EvAM-Tools is not a different method, but rather a web-based tool (\url{https://www.iib.uam.es/evamtools/}) and R package (GitHub repository with R and C code at \url{https://github.com/rdiaz02/EvAM-Tools#how-to-install-the-r-package}) that provides a unified interface to allow analysing data under most of the above methods, as well as simulating random dependency structures and data under the chosen models. It is under continuous development, and all of the above, with the notable exception of HyperHMM and HyperTraPS (which will be added soon), are available.

\item[BML] Code from \url{http://bml.molgen.mpg.de/}. An R  wrapper (using Rcpp) by Óscar Rodríguez available from \url{https://github.com/Deschain/BML}.

\end{description}

\subsection{Public domain silhouette images}\label{appendix:silho_images}

The silhouette images in Figure \ref{fig:cpm_cartoon} are in the public domain and are from \url{https://www.phylopic.org/}. These are the URLs of the images:

\url{https://www.phylopic.org/images/37536a9c-db56-42d3-8923-8c5ffdfa637b/homo-sapiens-sapiens}\\

\url{https://www.phylopic.org/images/b8c16fc6-d16b-4fac-8a04-67182448157e/homo-sapiens-sapiens}\\

\end{document}